\RequirePackage{lineno}
\newif\ifpreprint%
\preprintfalse%
\ifpreprint%
	\documentclass[preprint,english,aps,prl,floatfix,amssymb,superscriptaddress,a4paper]{revtex4-2}
\else%
	\documentclass[twocolumn,english,aps,prl,floatfix,amssymb,superscriptaddress,a4paper]{revtex4-2}
\fi%
\usepackage{amsmath}
\usepackage{physics}
\usepackage{chemformula}
\usepackage{dsfont}
\usepackage{color}
\usepackage{amssymb}
\usepackage{graphicx}
\usepackage{braket}
\usepackage{units}
\usepackage{lipsum}

\ifpreprint%
\else%
\fi%

\usepackage{hyperref}
\hypersetup{colorlinks = true, linkcolor=magenta,citecolor=blue, urlcolor=magenta, bookmarksnumbered =  true}
\newcommand{\ssm}{\scriptscriptstyle\rm}

\renewcommand{\theta}{\vartheta}
\renewcommand{\phi}{\varphi}

\begin{document}
\ifpreprint%
	\linenumbers%
\fi%

\title{Design and characterization of all two-dimensional fragile topological bands}

\author{
Samuel Bird}
\email{sabird@phys.ethz.ch}
\affiliation{
Institute for Theoretical Physics, ETH Zurich, 8093 Z\"urich, Switzerland
}

\author{
Chiara Devescovi}
\affiliation{
Institute for Theoretical Physics, ETH Zurich, 8093 Z\"urich, Switzerland
}
\affiliation{
    Donostia International Physics Center, Paseo Manuel de Lardizabal 4, 20018 Donostia-San Sebastián, Spain
}

\author{
Pascal Engeler}
\affiliation{
Institute for Theoretical Physics, ETH Zurich, 8093 Z\"urich, Switzerland
}

\author{
Agnes Valenti}
\affiliation{
Institute for Theoretical Physics, ETH Zurich, 8093 Z\"urich, Switzerland
}
\affiliation{Center for Computational Quantum Physics, Flatiron Institute, New York, NY, 10010, USA}

\author{
Doruk Efe Gökmen}
\affiliation{
Institute for Theoretical Physics, ETH Zurich, 8093 Z\"urich, Switzerland
}
\affiliation{
James Franck Institute and Department of Statistics, University of Chicago, Chicago, IL 60637, USA
}
\affiliation{
National Institute for Theory and Mathematics in Biology, Chicago, IL 60611, USA
}

\author{
Robin Worreby}
\affiliation{
Institute for Theoretical Physics, ETH Zurich, 8093 Z\"urich, Switzerland
}

\author{
Valerio Peri}
\affiliation{Department of Physics and Institute of Quantum Information and Matter, California Institute of Technology, Pasadena, CA 91125, USA
}
\affiliation{
Institute for Theoretical Physics, ETH Zurich, 8093 Z\"urich, Switzerland
}

\author{
Sebastian D. Huber}
\affiliation{
Institute for Theoretical Physics, ETH Zurich, 8093 Z\"urich, Switzerland
}

\begin{abstract}

Designing topological materials with specific topological indices is a complex inverse problem, traditionally tackled through manual, intuition-driven methods that are neither scalable nor efficient for exploring the vast space of possible material configurations. In this work, we develop an algorithm that leverages the covariance matrix adaptation evolution strategy to optimize the Fourier representation of the periodic functions shaping the designer material's characteristics. This includes mass profiles or dielectric tensors for phononic and photonic crystals, respectively, as much as synthetic potentials applicable to electronic and ultra-cold atomic systems. We demonstrate our methodology with a detailed characterization of a class of topological bands known as “fragile topological”, showcasing the algorithm’s capability to address both topological characteristics and spectral quality. This automation not only streamlines the design process but also significantly expands the potential for identifying and constructing high quality designer topological materials across the wide range of platforms, and is readily extendable to other setups, including higher-dimensional and non-linear systems.

\end{abstract}

\date{\today}

\maketitle

Since the advent of topology in condensed matter physics with the theoretical explanation of the Integer quantum Hall effect \cite{Klitzing80,Berry84,Simon83,Thouless82}, it has been used repeatedly, e.g., to construct field theories to explain the interacting version of the quantum Hall effects \cite{Tsui82,Hansson2017,Laughlin83,Zhang1989,Witten1989} or to categorize the electronic phases in all crystalline materials \cite{Zhang2019, Vergniory2019}. In the last example, topology led to predicted observations rather than explaining them in hindsight. With the recent advances  in {\em designer materials}, a new application of topology in physics emerged. In photonic \cite{Ozawa2019} and phononic materials \cite{Zhu2023,Huber16}, nano-structured electronic systems \cite{Gomes2012,Liu2020}, as well as in ensembles of cold atoms \cite{Mancini2015,Stuhl2015,Chalopin2020,Braun2023,Viebahn2023}, custom periodic structures are created to induce artificial Bloch bands with desired properties. The properties of these Bloch bands are then used to either achieve a sought-after functionality of the architectured material, e.g., a topological laser \cite{Bandres2018}, or to observe new physical phenomena \cite{Lu15, Nash15, Serra-Garcia18, Peri20}.

Encoding a functionality or a new physical phenomenon in terms of a topological index is a powerful tool for two reasons. First, such an index serves as a simple and stable objective function in the {\em design} of a periodic structure, may it be a specific optical lattice for cold atoms, an arrangement of gates defining a periodic potential for electrons in a two-dimensional electron gas, or an architectured photonic or phononic crystal. Second, the achieved functionality has the chance to {\em display some protection} against fabrication or implementation imperfections.
\begin{figure}[bth]
    \centering
    \includegraphics{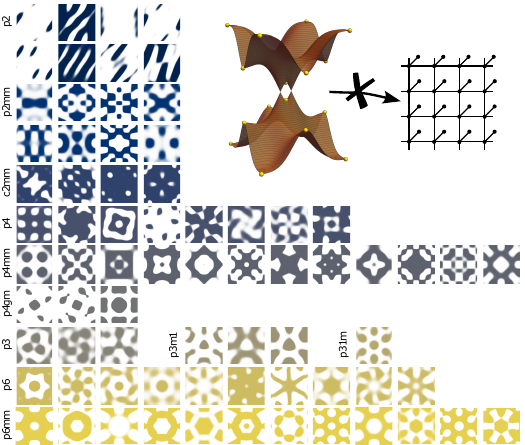}
    \caption{{\bf Complete catalog of samples with fragile bands.} Inset: Illustration of a topological set of bands that cannot be written in terms of a tight-binding model. The yellow dots indicate the high-symmetry points in the Brillouin zone where the eigenvalues of the space group symmetries determine the topology. Main panel: For each wallpaper group we present the unit cell of a two-dimensional structure which describes a mass profile $\sigma(x,y)$ or a periodic dielectric constant $\epsilon(x,y)$ that leads to fragile topological bands for the phonons or photons, respectively. We find structures for all the 79 distinct phases that have been theoretically predicted in \cite{Song2020}.}
    \label{fig:overview}
\end{figure}

Unfortunately, finding a periodic structure whose bands have a prescribed topological index is a complicated inverse problem with no generic solution. Despite this complication, the field of designer topological materials enjoyed remarkable success over the years \cite{Jotzu14, Gomes2012, Hafezi13, Susstrunk15,Serra-Garcia18,Peri20, Peterson18, He18}. However, in all the above examples a periodic structure had to be conceived and optimized by hand, typically guided by a simple discrete model. This approach is unsuitable for large-scale, high-throughput explorations, or for problems where either no simple discrete model is known or fabrication constraints severely obstruct the intuition-based workflow. Here we describe a systematic approach to this challenge and employ our methodology on one specific example, presenting an exhaustive characterization of a class of topological bands known as ``fragile topological'' in {\em all} the aforementioned platforms, an impossible endeavor without a fully automated algorithm. 

How are such topological bands typically found? For microscopic quantum materials, databases such as the  Inorganic Crystal Structure Database \cite{icsd} serve as a shopping list to identify new topological systems \cite{Vergniory2019}. Here we focus on the intermediate to large-scale designer materials described above. The Bloch bands in all of these systems are described by a partial differential equation where one of their coefficients $f(\vec r)$ is a periodic function of space. This can be a periodic potential $V(\vec r)$ for the Schrödinger equation for electrons or ultra-cold atoms \cite{SI}, a dielectric function $\epsilon(\vec r)$ in the Maxwell equations for photonic crystals \cite{SI}, or a mass distribution $\sigma(\vec r)$ for the Poisson equation describing the vibrations in thin membranes \cite{SI}. 

Designing intermediate to large-scale materials to showcase sought-after (topological) properties comes with advantages but also challenges compared to their microscopic counterparts. We are not constrained by combinations of atoms that form chemically stable compounds like in Ref.~\onlinecite{Vergniory2019}, but our design space is given by a set of reasonably well-behaved functions of the spatial coordinates. However, this design flexibility comes at a steep algorithmic price: Optimizing a structure $f(\vec r)$ with the goal to achieve some topological nature of the elementary excitations is suffering from all the challenges inherent to high-dimensional optimization routines. Here, we overcome this challenge using a modern evolution strategy in the form of the Covariance Matrix Adaptation Evolution Strategy (CMA-ES) \cite{Hansen2006, Hansen2016}.

We base our approach on topological phases that can be detected by eigenvalues of crystalline symmetries. There are a number of reasons for this choice. First, almost all known topological insulators have a crystalline counterpart, including Chern insulators \cite{Fu07} or $\mathbb Z_2$ \cite{Bradlyn17} insulators, to name just two examples. Second, for many designer materials, crystalline symmetries are a natural choice, as one typically replicates a local pattern periodically, with the local pattern obeying the symmetries of a point group e.g. rotations, mirrors, or glide reflections. 
Finally, eigenvalues of crystalline symmetries are straightforward to compute and hence serve as a computationally cheap input for a large scale search of topological bands. 
\begin{figure*}[bth]
    \centering
    \includegraphics{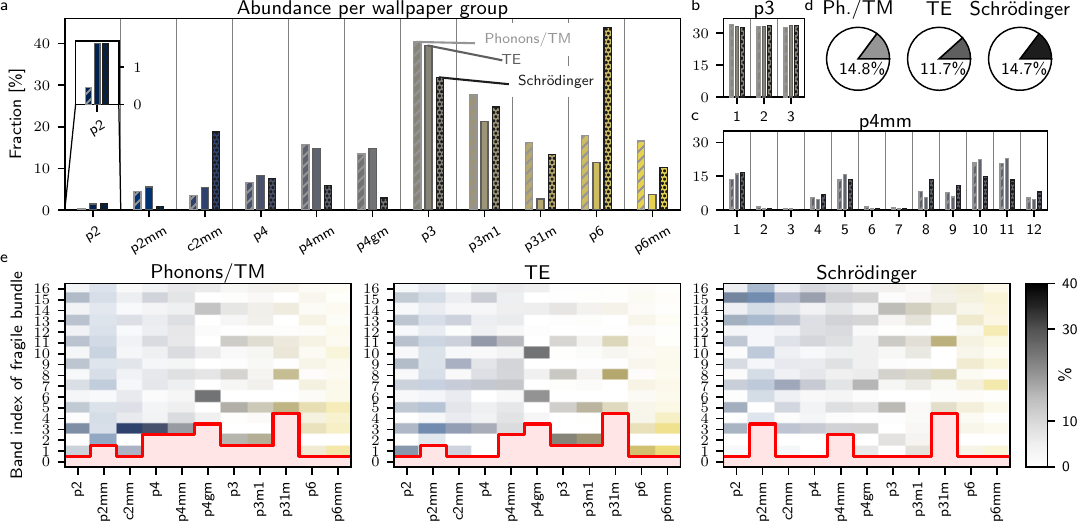}
    \caption{{\bf Statistical analysis of fragile topological bands.} (a) For each wallpaper group we show the likelihood that a random structure harbors a fragile topological band amongst its lowest 15 bands. The abundance of these topological bands are shown for phonons, photons, and the Schrödinger equation in a periodic potential. (b) For the group $p3$, all three distinct fragile bands are equally likely to occur, independent of the platform. (c) For $p4mm$, there are strong variations of the relative abundance. (d) Averaged over all wallpaper groups the three studied platforms show a similar likelihood that a random structure harbors a fragile topological bands bundle. (e) For each of the studied platform, we display how likely it is for a topological bundle to occur in the $n$'th band. The darker shade of the respective color indicates a higher likelihood. In the bands below the red line, no topological bundles were found. One observes that for some groups, e.g. in $p31m$ we cannot find any topological bundle among the five lowest bands.}
    \label{fig:statistics}
\end{figure*}

Concretely, we use the framework of topological quantum chemistry \cite{Bradlyn17, Bradlyn2018}: Sets of isolated bands are deemed topological if one cannot write them in terms of a basis of exponentially localized symmetric orbitals, see Fig.~\ref{fig:overview}. It turns out that there are two distinct ways to fail this test: Either bands are topological and can only be trivialized by adding another set of topological bands. All well-established topological insulators having a Chern number or a $\mathbb Z_2$ index fall into this category of ``stable topology''. However, there arises the possibility that bands are topological, but can be trivialized by a band that is itself trivial. This constitutes the class of ``fragile topology''. It turns out by simple inspection \cite{bilbao}, that in two dimensions without spin-orbit interaction, which for the chosen designer-materials platforms is the standard rather than the exception, fragile topology is the only possibility. Moreover, we will constrain ourselves to systems that do not break time-reversal symmetry for simplicity. 

All two-dimensional eigenvalue-indicated fragile phases have been tabulated according to their symmetry eigenvalues in Ref.~\onlinecite{Song2020}. Each phase can be labeled by the irreducible representations (irreps) of the wallpaper groups at high symmetry points in reciprocal space. For example, for the wallpaper group $p4mm$ with a four-fold rotation axis and two mirror planes, the possible irreps are $\Omega_1,\dots,\Omega_5$ for $\Omega=\Gamma=(0,0)$ and $\Omega=M=(1/2,1/2)$, respectively, as well as $X_1,X_2$ for the point $X=(1/2,0)$. The points in the Brillouin zone are written with respect to the reciprocal lattice vectors. An example of a set of connected fragile bands in the table of Ref.~\onlinecite{Song2020} is
\begin{equation}
    \label{eqn:root}
    \Gamma_1\Gamma_4,M_5,X_1X_2;
\end{equation}
where at the points $\Gamma$ and $X$ two singly degenerate bands are realized which are joined in the two-dimensional irrep $M_5$ at the $M$ point \cite{SI}. There are 79 different distinct sets of irreps, or roots, which describe fragile topological bands in 11 of the 17 wallpaper groups \cite{Hahn2005}. While some of these have been experimentally observed \cite{Peri20}, it is not known if all of them can occur in a realistic material. To find concrete material examples for all of them, to characterize those structures, and optimize them for scientific investigations or technological applications is the concrete challenge we want to meet with our proposed design algorithm.

\textit{Methodology. }Let us outline the algorithm. Our design space is given by two-dimensional periodic functions $f(\vec r)$. We parameterize these functions via their Fourier-coefficients $f(\vec r)= \sum_{n} u_{n} e^{i \vec k_n \cdot \vec r}$. The $\vec k_n$ are chosen to reflect the specific lattice in a given wallpaper group and the expansion coefficients $\{u_n\}$ fulfill the necessary relations for different $n$'s, such that all symmetries of the wallpaper group are realized \cite{Verberck2012, Hahn2005}. For some platforms we also discretize the profile by using $\tilde f(\vec r) = f_0+\frac{f_1-f_0}{2}\{1+\tanh[f(\vec r)/\xi]\}$ with $\xi\to 0$ to encode a system where $\tilde f$ only takes the two values $f_0$ and $f_1$. The set of independent $\{u_n\}$ make up the high-dimensional continuous search space for our algorithm.

We start the algorithm with a random set of $\{u_n\}$ and solve the partial differential equation (PDE) at the high-symmetry points in the Brillouin zone. We do this using a Finite Element Method (FEM). We design our own FEM meshes to ensure efficient enforcement of crystalline symmetries, which we then pass to the FEM package FEniCS, which generates the relevant FEM problem for each of the systems of interest \cite{AlnaesEtal2015,LoggEtal2012,LoggWells2010}. This formulates the PDE as a sparse matrix eigenvalue problem, that we diagonalize ourselves using LAPACK \cite{lapack99}. For each high symmetry point, the solutions form irreps of the respective little group, which can be labeled by the standard symbol of the high-symmetry point and a numerical label \cite{SI}. When ordering the symbols of a given solution by eigenvalues, these form words of the form (again on the example of $p4mm$)
\begin{equation}
    \Gamma_1\Gamma_1\Gamma_4\Gamma_4\dots, M_1M_1M_5M_4\dots, X_1X_2X_2X_1\dots.
\end{equation}
In a first step, one can {\em bundle these bands} into set of bands where symmetry compatibility relations require them to be {\em connected}
\begin{equation}
    \label{eqn:string}
    \Gamma_1^a\Gamma_1^b\Gamma_2^b\Gamma_4^c\dots, M_1^aM_5^bM_4^c\dots, X_1^aX_2^bX_2^bX_1^c\dots,
\end{equation}
where those irreps with the same superscript belong to a given bundle. 

In the next step, we can check if any of the bundles conform with the sought-after roots of Ref.~\onlinecite{Song2020}. If this is the case [in the example above the bundle labeled $b$ realizes the root of Eq.~(\ref{eqn:root})], one can store $\{u_n\}$ for later optimization. If none of the bundles correspond to the target, one move on to the next random set $\{u_n\}$. In CMA-ES, the parameters $\{u_n\}$ are drawn from a normal distribution $\mathcal N(\{\bar u_n\}; C_{nm})$, characterized by the means ${\bar u_n}$ and the covariance matrices $C_{nm}=\overline{(u_n-\bar u_n)(u_m-\bar u_m)}$. The coefficients ${\bar u_n}$ and $C_{nm}$ are then adapted in an evolution strategy to optimize a given cost function \cite{Hansen2006, Hansen2016}.

Finally, when a topological band is found, one can optimize further properties by using a refined cost function. A typical example would be to push for the maximum possible band gap without destroying the topological character. Such mixed-cost functions are straightforward to implement, and we will come back to this point below.

\textit{Results. }We now present the results achieved with this algorithm searching for all possible fragile bands. For each of the 79 roots, we display the corresponding structure in Fig.~\ref{fig:overview}. Shown are structures $f(\vec r)$ that lead to topological bands for electromagnetic TM and TE modes or phonons in thin membranes. Examples for the other platforms and the associated Bloch bands are shown in \cite{SI}. These structures represent the main result of this work where we establish that all fragile bands can in principle be realized. 

To highlight the power of the algorithm we move to a detailed statistical analysis of structures containing fragile roots. The most pressing questions one may have are: how likely is it to find a topological band? Does it depend on the wallpaper group? Are there classes of fragile bands that are harder to realize than others? And finally, can one target the lowest Bloch bands only, or does one have to consider highly excited bands? In Fig.~\ref{fig:statistics} we answer all the above questions. 

In each wallpaper group, we draw random structures $\{u_n\}$ until we obtain 10'000 samples with at least one fragile root amongst the 15 lowest bands. For each of the wallpaper groups and all the three studied platforms, we then report what fraction are topological. We observe that this depends strongly on the wallpaper group, and to a lesser extent, on the studied platform. One important observation is that for the group $p3$ for all platforms and for $p6$ for the Schrödinger equation, almost half of the random samples contain a fragile root. Hence, these groups may serve as an excellent starting point if one wants to optimize further properties beyond the presence of topological bands.

Next, we showcase for the two groups $p3$ and $p4mm$ how the relative abundance is distributed amongst the different roots. While for $p3$ all roots are equally likely, in $p4mm$ there is a significant variation between the different roots. This difference between $p3$ and $p4mm$ can be understood by the different structures of the specific roots \cite{SI}. Note, that there are also extreme outliers. For example, root \#3 in the group $p4gm$ is extremely hard to find. Only one out of $10^5$ random structures turn out to realize this root. We finally observe that overall fragile bands are more or less equally likely to occur in all the studied platforms.
\begin{figure}[t]
    \centering
    \includegraphics{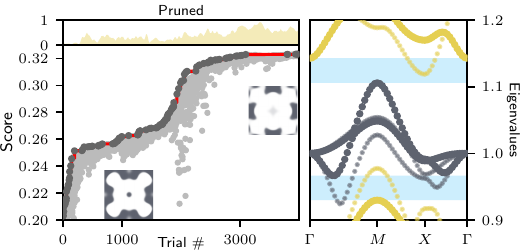}
    \caption{{\bf Gap optimization of a $p4mm$ sample.} Left panel: The Score $1/C$ indicates the size of the gap to the adjacent bands as a function of the number of optimization steps. The insets show the initial and the final real-space structure. Also shown is the fraction of samples that are pruned as they lost the initial topological bands. Right panel: Bloch bands along the high symmetry lines of $p4mm$. The small gray dots show the bands of the initial structure where the topological bands largely overlap with the bands below shown in yellow. The big dots represent the final, optimized spectrum with a full band gap to the bands above and below indicated by the blue areas.}
    \label{fig:optimization}|
\end{figure}

It is difficult to find fragile bands in the lowest bundle. In Fig.~\ref{fig:statistics} we show the relative abundance of fragile roots as a function of the band index of the fragile bundle per wallpaper group and platform. One can clearly observe that for some groups one needs to go to rather highly excited bands. This is mostly implied by the relatively complex irreps involved (the counterpart of $d$-, vs. $s$-wave orbitals), in particular the nature of the mechanism that binds the bands into the fragile 2-bundle: bundles that are bound by 2D or conjugate irreps are more common, whereas those bound by the fact that each individual band is necessarily a Chern band (and hence barred from isolation by time reversal symmetry) are less common. The bundling mechanism of each root is outlined in \cite{SI}.

In the presented statistical analysis we did not focus on the size of the band gaps separating the topological from its adjacent bands. This is, however, for most practical applications one of the most important quality measures. We show on the example of a specific root in group $p4mm$ how CMA-ES with an appropriate cost function can address this issue. Once the sought-after topological band is found in the form of a set of Fourier coefficients $\{u_n^0\}$, one can further optimize these coefficients for a large band gap. The evolution strategy of CMA-ES is optimally suited for this task: One generates additional samples by drawing new ones from $\mathcal N(\{u_n^0\};\lambda_n\delta_{nm})$, with $\lambda_n\ll u_n^0$. The cost function for the update strategy of $\bar u_n$ and $C_{nm}$ can be chosen to take the form
\begin{equation}
 C =  \frac{1}{\sum_{\vec k_i} \left(\epsilon_{n}({\vec k_i})-\epsilon_{n-1}(\vec k_i)+ \epsilon_{m+1}({\vec k_i})-\epsilon_{m}(\vec k_i)\right)},
\end{equation}
where $n$ ($m$) labels the eigenvalues in the lowest (highest) band in the bundle and the $\vec k_i$'s are chosen from a suitable set of reciprocal vectors. In order not to spoil the topology of the bundle in question, we can simply prune those structures $\{u_n^\alpha\}$ which do not conform with the root one optimizes from the population that is used in CMA-ES. In Fig.~\ref{fig:optimization} we show how a topological bundle that is initially overlapping with its neighboring band below, can be brought into a fully gapped system with a sizeable gap of more than five percent.

Two observations are worth pointing out. First, the fraction of pruned samples is growing at later stages of the optimization: At this point, improving the gap at all seems to be only possible by introducing unwanted band-inversion. Second, the initial real space structure $f_0(\vec r)$ and the final $f_{\ssm final}(\vec r)$ have very little in common with each other. This fact underlines why optimizations by hand or those parameterized by simple geometric motifs often do not yield satisfactory results.

\begin{figure}[t!bh]
    \centering
    \includegraphics{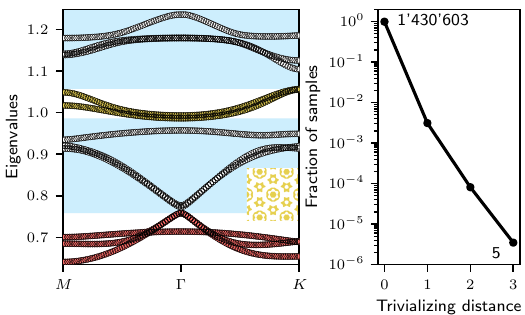}
    \caption{{\bf Trivializing distance}. Left panel: Dispersion relation of a sample (shown in the inset) hosting a topological bundle (yellow) where the trivializing bundle (red) is separated by two ``spectator'' bundles shown in white. For the purpose of topology, the effective gaps shown in blue are relevant. Right panel: Relative likelihood for a fragile topological bundle to be separated by 1,2, or 3 bundles from its trivializing one. The graph is based on more than 1.4 million samples, however, the overwhelming majority has its trivializing band direct adjacent.}
    \label{fig:trivializing}
\end{figure}

Another way to protect the topology is by separating the fragile bands as much as possible from the trivializing ones. Remember that fragile bundles can be trivialized by trivial bands. However, not any trivial band can do that job. By post-selecting those structures where the trivializing band is far away from the bundle in question, one can achieve the same goal as above without the need to optimize any gap. In Fig.~\ref{fig:trivializing} we show one example where the fragile bundle is surrounded by bundles that do not trivialize it, essentially inducing a protecting gap to the one that does trivialize the bundle. One may ask how likely such a situation is. In Fig.~\ref{fig:trivializing}.b we show the number of samples as a function of the distance to their trivializing band. While one can see that this distance is exponentially suppressed, we do find distances of up to three bundles. 

In summary, we presented an algorithm to find topological bands by using the covariance matrix adaptation evolution strategy that optimizes the Fourier representation of periodic functions entering a range of partial differential equations. We demonstrate the power of the approach by presenting a full catalog of all possible topological bands in two dimensions in the absence of spin-orbit coupling for systems of photons, phonons, electrons, and ultra-cold atoms. Our approach is by no means constrained to the presented application. Generalizations to higher dimensions, systems that break time-reversal symmetry, or further platforms, including those of non-linear systems, should be straightforward to implement. 

Author contributions: SB and CD performed most of the numerical simulations. PE, AV, DEG, and RW contributed to the algorithmic development. VP, SB, CD and SDH provided the background in topological quantum chemistry. SDH conceived the idea. All authors contributed to the writing of the manuscript.
  
\begin{acknowledgements}
We thank Bogdan A. Bernevig, Tena Dubcek, Eliska Greplova and Marc Serra-Garcia for valuable discussions. This work has received funding from the European Research Council under Grant agreement No. 771503. S.D.H. acknowledges support by the Benoziyo Endowment Fund for the Advancement of Science. The Flatiron Institute is a division of the Simons Foundation.
\end{acknowledgements}

\bibliographystyle{phd-url}
\bibliography{ref.bib}

\end{document}


\setcounter{equation}{0}
\setcounter{figure}{0}
\setcounter{page}{1}
\makeatletter
\renewcommand{\theequation}{S\arabic{equation}}
\renewcommand{\thefigure}{S\arabic{figure}}
\renewcommand{\thetable}{S\Roman{table}}
\renewcommand{\bibnumfmt}[1]{[S#1]}
\renewcommand{\citenumfont}[1]{S#1}
\renewcommand{\vec}[1]{\ensuremath{\boldsymbol{#1}}}

\title{Supplementary material for: Design and characterization of all two-dimensional fragile bands}

\author{
Samuel Bird}
\email{sabird@phys.ethz.ch}
\affiliation{
Institute for Theoretical Physics, ETH Zurich, 8093 Z\"urich, Switzerland
}

\author{
Chiara Devescovi}
\affiliation{
Institute for Theoretical Physics, ETH Zurich, 8093 Z\"urich, Switzerland
}
\affiliation{
    Donostia International Physics Center, Paseo Manuel de Lardizabal 4, 20018 Donostia-San Sebastián, Spain
}

\author{
Pascal Engeler}
\affiliation{
Institute for Theoretical Physics, ETH Zurich, 8093 Z\"urich, Switzerland
}

\author{
Agnes Valenti}
\affiliation{
Institute for Theoretical Physics, ETH Zurich, 8093 Z\"urich, Switzerland
}
\affiliation{Center for Computational Quantum Physics, Flatiron Institute, New York, NY, 10010, USA}

\author{
Doruk Efe Gökmen}
\affiliation{
Institute for Theoretical Physics, ETH Zurich, 8093 Z\"urich, Switzerland
}
\affiliation{
James Franck Institute and Department of Statistics, University of Chicago, Chicago, IL 60637, USA
}
\affiliation{
National Institute for Theory and Mathematics in Biology, Chicago, IL 60611, USA
}

\author{
Robin Worreby}
\affiliation{
Institute for Theoretical Physics, ETH Zurich, 8093 Z\"urich, Switzerland
}

\author{
Valerio Peri}
\affiliation{Department of Physics and Institute of Quantum Information and Matter, California Institute of Technology, Pasadena, CA 91125, USA}
\affiliation{
Institute for Theoretical Physics, ETH Zurich, 8093 Z\"urich, Switzerland
}

\author{
Sebastian D. Huber}
\affiliation{
Institute for Theoretical Physics, ETH Zurich, 8093 Z\"urich, Switzerland
}

\maketitle
\tableofcontents
\listoffigures
\listoftables

\section{Considered models}
In this section, we explain which models we consider in the statistical analysis of the fragile roots in the main text. All of them are described by scalar partial differential equations, where some coefficients are space-dependent with the symmetry given by one of the wallpaper groups. The Poisson equation either describes the transverse displacement of an elastic membrane or the $z$-component of the electric field for the transverse magnetic (TM) modes of the Maxwell equations in a slab geometry. For the transverse electric (TE) modes, the equation is slightly more complicated, see below. Finally, we consider the Schrödinger equation for a single electron confined to two dimensions exposed to an artificially structured potential.

\subsection{Mass patterned pre-stressed membrane}

We consider an elastic membrane with a mass profile $\sigma(x,y)$
\begin{equation}
    \label{eqn:poisson}
\nabla^2 \phi(x,y) = -\sigma(x,y)\omega^2\phi(x,y).
\end{equation}
We parameterize the mass profile $\sigma_{\ssm bare}(x,y)$ with Fourier coefficients $n_1 {\vec G}_1+n_2 {\vec G}_2$ ($n_1,n_2\in \mathbb Z$),
 using the reciprocal lattice vectors 
 \begin{equation}
    \vec{G}_i \cdot \vec{a}_j = 2 \pi \delta_{ij}
 \end{equation}
 of the corresponding lattice spanned by ${\vec a}_1$ and ${\vec a}_2$. 
 
 To be close to an easily manufacturable structure, e.g., a pre-stressed Silicon nitride membrane with gold patterning to achieve a periodic mass profile, we further use 
 \begin{equation}
    \sigma(x,y)=\frac{\sigma_{\ssm max}-\sigma_{\ssm min}}{1+e^{-\beta \sigma_{\ssm bare}(x,y)}}+\sigma_{\ssm min}.
 \end{equation}
 Unless otherwise stated we use $|n_i|\leq 4$, $\beta=2$, $\sigma_{\ssm min}=5$, and $\sigma_{\ssm max}=100$. Note,  that (\ref{eqn:poisson}) is scale invariant under $\sigma \to \lambda \sigma$, and hence the relevant parameter is the dynamical range $\eta=\sigma_{\ssm max}/\sigma_{\ssm min}=20$.

\subsection{Transverse magnetic modes in a two-dimensional photonic crystal}
TM modes two-dimensional photonic crystal are described by \cite{Joannopoulos2008}
\begin{equation}
\nabla^2 E_z(x,y) = -\epsilon(x,y)\omega^2E_z(x,y).
\end{equation}
This is formally the same equation as for the membranes above (\ref{eqn:poisson}) when identifying $\epsilon \equiv \sigma$.

\subsection{Transverse electric modes in a two-dimensional photonic crystal}

TE modes two-dimensional photonic crystal are described by \cite{Joannopoulos2008}
\begin{equation}
    -\nabla \log\epsilon(x,y)
    \cdot\nabla H_z(x,y)+
    \nabla^2 H_z(x,y)
    = -\epsilon(x,y)\omega^2H_z(x,y).
\end{equation}
We encode the dielectric constant as above.

\subsection{Two-dimensional Schrödinger equation}
The two-dimensional Schrödinger equation in a periodic potential is given by
\begin{equation}
    \label{eqn:2Dschroedinger}
    \nabla^2 \psi(x,y) = \left[V(x,y)+\omega\right]\psi(x,y).
\end{equation}
Note that the Schrödinger equation is {\em not} scale invariant. The typical  kinetic energy of a superstructure defined by $V(x,y)$ is given by
\begin{equation}
    E_{\ssm kin}=\frac{\hbar^2}{2m^*}\left(\frac{\pi}{a}\right)^2,
\end{equation}
where $a$ is the lattice constant, and $m^*$ the effective band mass. 

We highlight two specific applications of Eq.~(\ref{eqn:2Dschroedinger}). First electrons confined to two dimensions either in van der Waals materials or, e.g., on the (111) surface of Cu. Here, smaller $a$ are harder to fabricate, larger $a$ lower the temperature scale. For the purpose of this work we use $a\approx 10\,\mathrm{nm}$ and an effective mass of about a third of the electron mass, giving rise to a typical temperature scale of $E_{\ssm kin}/k_{\ssm B}=1\,{\mathrm K}$. Assuming the ability to expose the electron gas to a potential varying about an electronvolt, we use 
\begin{equation}
    V(x,y)=\frac{V_{0}}{1+e^{-\beta V_{\ssm bare}(x,y)}}
\end{equation}
with a $V_0 = 1000$. Note that written like this, we measure the strength of $V$ in units of $E_{\ssm natural}=E_{\ssm kin}/\pi^2$. These numbers are inspired, but not constrained to, carbon monoxide molecules on Cu(111) surfaces \cite{Gomes2012}. Another application are cold atoms in optical potentials. There, the established energy scale is 
\begin{equation}
    E_{\ssm recoil}=\frac{\hbar^2k^2}{2m}=16 E_{\ssm kin}=16\pi^2E_{\ssm natural},
\end{equation}
where $m$ now denotes the mass of the neutral atom and $k=2\pi/\lambda$, where $\lambda$ is the wave-length of the light responsible for the potential $V\propto \sin^2(k x)$ with a lattice constant $a=\lambda/2$. In other words, a $V_0=1000$ amounts to $\sim 6\,E_{\ssm recoil}$, corresponding to a moderate lattice depth.\footnote{The superfluid to Mott transition for Rubidium atoms in a square lattice occurs around $13\,E{\ssm recoil}$ \cite{Greiner02}.}

\section{The eleven wallpaper groups with fragile roots}

\subsection{Generic bundling strategy}

To detect fragile roots, we bundle bands into connected sets (all fragile roots contain at least two bands). When there are no high-symmetry {\em lines} present in the Brillouin zone, bands cannot generically cross. Hence, it suffices to make use of properties at high-symmetry {\em points} in the Brillouin zone. Three cases can be separated:
\begin{enumerate}
    \item Some of the high-symmetry points contain two-dimensional irreducible representations (irreps). By scanning through the irrep content at each high-symmetry point starting from the lowest bands allows us to uniquely determine the band bundles.
    \item Some of the high-symmetry points contain conjugate pairs of complex irreps. As we deal with time-reversal symmetric systems, these complex irreps have to come in pairs and hence are equivalent to two-dimensional irreps for the sake of bundling.
    \item No conjugate pairs or two-dimensional irreps are present at high-symmetry points. In that case, the irreps alone would only predict singly degenerate bands. However, in all of these cases, such bands would have a Chern number as shown below. Hence, these bands that would have opposite Chern number, if they were isolated, touch somewhere at arbitrary positions in the Brillouin zone. In these cases, one needs to check for all sets of neighboring bands, if by the irreps of the individual bands, one would conclude to have a Chern band, these Chern numbers are rendered zero by bundling them together.
\end{enumerate} 
We elaborate on the different scenarios below for each wallpaper group individually. In case there are additional high-symmetry lines induced by mirror or glide-symmetries, bands may be able to cross. When bundling in such groups, one has to take this into account and make sure compatibility relations are fulfilled. 

In the following subsections, we provide the symmetries and irreps for all eleven wallpaper groups that contain fragile roots. We discuss the different roots and show one example of a fragile band structure per distinct root.

\raggedbottom
\newpage

\subsection{$p2$}

\subsubsection{Basic group properties}

\begin{figure}[!h]
    \centering
    \includegraphics{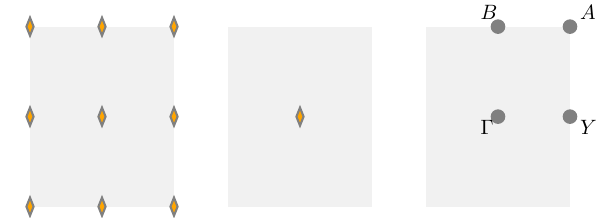}
    \caption[$p2$ unit cell.]{Left: Unit cell with the full symmetry of $p2$. Middle: General positions of $p2$. Right: Brillouin zone with the high-symmetry points and lines indicated.}
    \label{fig:p2-uc}
\end{figure}

The group $p2$ describes a rectangular lattice with lattice vectors $\vec a_1=(1,0)$ and $\vec a_2=(0,\beta)$. The corresponding space group is $P2$, (\#3) constrained to the $x$-$z$ plane. The group $p2$ contains the following group elements [\href{https://www.cryst.ehu.es/cgi-bin/plane/programs/nph-plane_getgen?what=gp&gnum=2}{retrieve from Bilbao server}]
\begin{align}
    &\{ 1|n_1 {\vec a}_1+n_2{\vec a}_2\}, \quad \mathrm{with} \quad n_1,n_2\, \in\, \mathbb Z\\
    &\{ 2|(0,0)\}.
\end{align}

\noindent
The relevant members of the little groups of the high-symmetry points and lines and their irreps are given in Tab.~\ref{tab:irreps-p2}. The full set can be obtained from the \href{https://www.cryst.ehu.es/cgi-bin/cryst/programs/representations_vec.pl?tipogroup=spg&super=3&complex=Submit}{Bilbao Server}.

\begin{table}[htb]
    \begin{subtable}[t]{0.22\textwidth}\centering
        \subcaption{$\Gamma$}
\begin{tabular}[b]{cc}
    \hline
    \hline\\[-10pt]
     &$\{2|(0,0)\}$ \\
    \hline\\[-10pt]
    $\Gamma_1$  &$\phantom{-}1$\\
    $\Gamma_2$  & $-1$ \\
    \hline
    \hline
\end{tabular}
\end{subtable}
\begin{subtable}[t]{0.22\textwidth}\centering
    \subcaption{$B$}
\begin{tabular}[b]{cc}
\hline
\hline\\[-10pt]
 &$\{2|(0,0)\}$ \\
\hline\\[-10pt]
$B_1$  &$\phantom{-}1$\\
$B_2$  & $-1$ \\
\hline
\hline
\end{tabular}
\end{subtable}
\begin{subtable}[t]{0.22\textwidth}\centering
    \subcaption{$Y$}
\begin{tabular}[b]{cc}
\hline
\hline\\[-10pt]
 &$\{2|(0,0)\}$ \\
\hline\\[-10pt]
$Y_1$  &$\phantom{-}1$\\
$Y_2$  & $-1$ \\
\hline
\hline
\end{tabular}
\end{subtable}
\begin{subtable}[t]{0.22\textwidth}\centering
    \subcaption{$A$}
\begin{tabular}[b]{cc}
\hline
\hline\\[-10pt]
 &$\{2|(0,0)\}$ \\
\hline\\[-10pt]
$A_1$  &$\phantom{-}1$\\
$A_2$  & $-1$ \\
\hline
\hline
\end{tabular}
\end{subtable}
\caption[$p2$ irreps.]{Irreducible representations of $p2$ at the high-symmetry points $\Gamma=(0,0)$, $B=(0,1/2)$, and $Y=(1/2,0)$  and $A=(1/2,1/2)$.}
\label{tab:irreps-p2}
\end{table}

\noindent
The fragile roots in $p2$ \cite{Song2020} are given in Tab.~\ref{tab:roots-p2}; the elementary band representations can be retrieved from the \href{https://www.cryst.ehu.es/cgi-bin/cryst/programs/bandrep.pl?super=3&elementaryTR=Elementary%20TR}{Bilbao server}.
\begin{table}[!htb]
    \begin{tabular}{c|l|c|l}
        \hline
        \# & root & \# of bands & type \\
        \hline\hline
        1 & $ 2 A_{2} + 2 B_{1} + 2 \Gamma_{2} + 2 Y_{2} $ &2& Chern\\
\hline
2 & $ 2A_{2} + 2 B_{1} + 2 \Gamma_{1} + 2 Y_{1} $&2& Chern\\
\hline
3 & $ 2 A_{1} + 2 B_{2} + 2 \Gamma_{2} + 2 Y_{2} $&2& Chern \\
\hline
4 & $ 2 A_{2} + 2 B_{2} + 2 \Gamma_{2} + 2 Y_{1} $ &2& Chern\\
\hline
5 & $ 2 A_{2} + 2 B_{2} + 2 \Gamma_{1} + 2 Y_{2} $ &2& Chern\\
\hline
6 & $ 2 A_{1} + 2 B_{1} + 2 \Gamma_{2} + 2 Y_{1} $ &2& Chern\\
\hline
7 & $ 2 A_{1} + 2 B_{2} + 2 \Gamma_{1} + 2 Y_{1} $ &2& Chern\\
\hline
8 & $ 2 A_{1} + 2 B_{1} + 2 \Gamma_{1} + 2 Y_{2} $ &2& Chern\\
\hline
        \end{tabular}
        \caption[$p2$ roots.]{Fragile roots and their types in wallpaper group $p2$}
        \label{tab:roots-p2}
 \end{table}

\subsubsection{Bundling strategy}

We analyze the structure of the roots in the wallpaper group $p2$ as shown in Tab.~\ref{tab:roots-p2}. As we will see, we need to invoke the vanishing Chern number argument to argue for the bundling of bands into fragile roots. For this, we state equation for the Chern number $C$ in $C_2$--symmetric systems \cite{Fu06,Fang2012}
\begin{equation}
    \label{eqn:fu-kane}
    (-1)^C=\prod_{i\,\in\, {\rm occ.}} \zeta_i(\Gamma)\zeta_i(B)\zeta_i(Y)\zeta_i(A),
\end{equation}    
where $\zeta$ are eigenvalues of $\{2|(0,0)\}$. Note, that there are no two-dimensional irreps at the high-symmetry points, nor conjugate pairs of complex irreps. Let us consider root \#1 as an example. It is easy to see, that if we were to have two disconnected sets of bands
\begin{equation}
    (A_2+B_1+\Gamma_2+Y_2) \,\oplus\,(A_2+B_1+\Gamma_2+Y_2),
\end{equation}
The Fu-Kane formula (\ref{eqn:fu-kane}) would tell us that we have two bands, each with an odd Chern number. Given that we do not break time-reversal symmetry, this cannot happen, and hence, these two sets of bands not touch somewhere off the high-symmetry points. The eight different roots in $p2$ are obtained by a permutation of the irreps. Finally, the types of the roots of $p2$ are indicated in the last column of Tab.~\ref{tab:roots-p2}.

\subsubsection{Examples}
\begin{figure}[h!bt]
    \centering
    \includegraphics{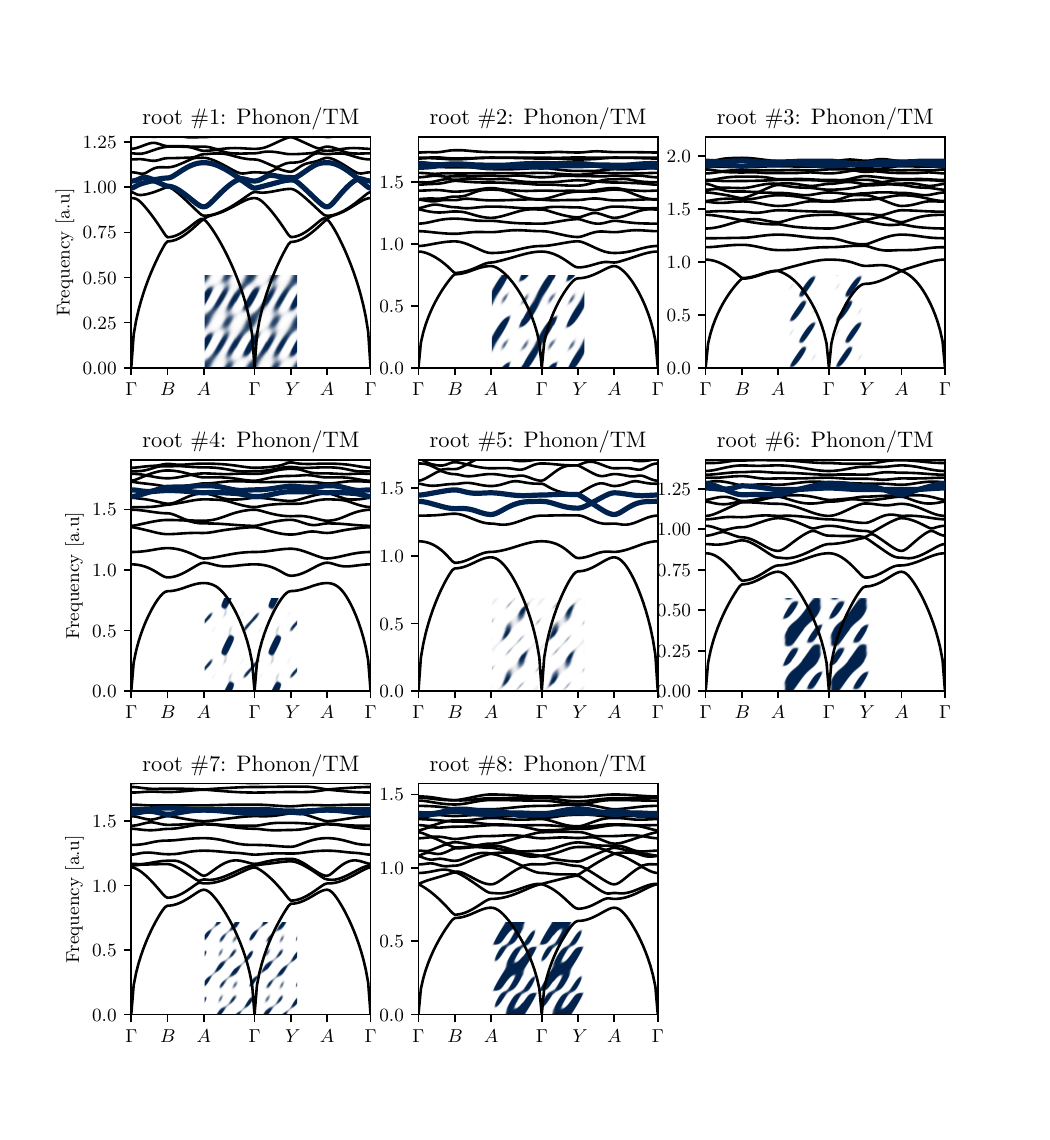}
    \caption[$p2$ phonons/TM examples.]{ One sample for each of the roots in $p2$ for phonons and TM photons.}
    \label{fig:p2-TM-examples}
\end{figure}

\begin{figure}[h!bt]
    \centering
    \includegraphics{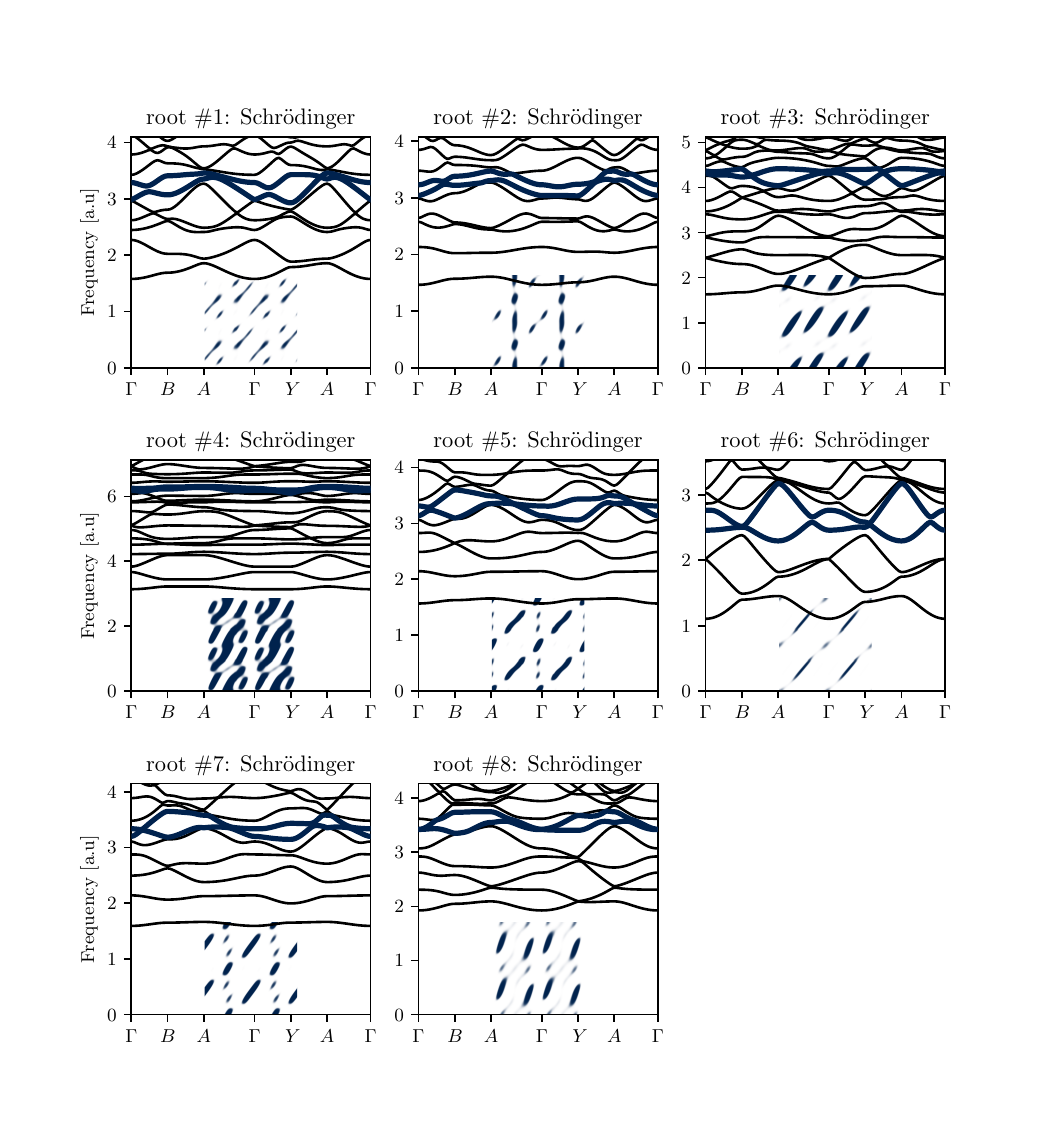}
    \caption[$p2$ Schrödinger examples. examples.]{One sample for each of the roots in $p2$ for systems described by the Schrödinger equation.}
    \label{fig:p2-Q-examples}
\end{figure}

\begin{figure}[h!bt]
    \centering
    \includegraphics{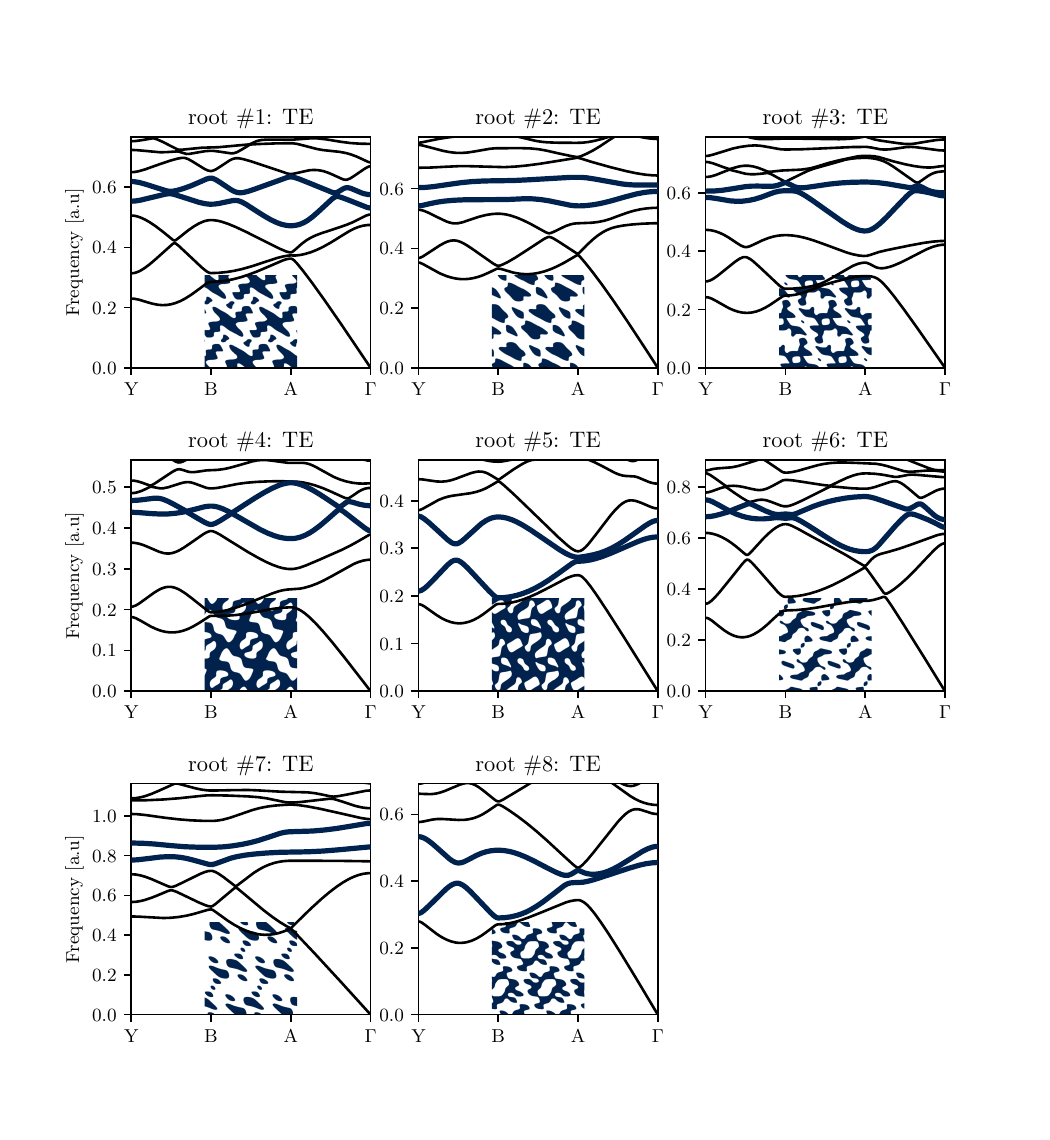}
    \caption[$p2$ TE examples.]{ One sample for each of the roots in $p2$ for TE photons.}
    \label{fig:p2-TE-examples}
\end{figure}

\raggedbottom
\pagebreak
\subsection{$p2mm$}

\subsubsection{Basic group properties}

\begin{figure}[h!bt]
    \centering
    \includegraphics{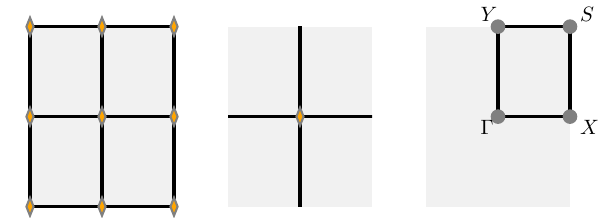}
    \caption[$p2mm$ unit cell.]{Left: Unit cell with the full symmetry of $p2mm$. Middle: General positions of $p2mm$. Right: Brillouin zone with the high-symmetry points and lines indicated.}
    \label{fig:p2mm-uc}
\end{figure}

The group $p2mm$ describes a rectangular lattice with lattice vectors $\vec a_1=(1,0)$ and $\vec a_2=(0,\beta)$. The corresponding space group is $P2mm$, (\#25) constrained to the $x$-$y$ plane. The group $p2mm$ contains the following group elements [\href{https://www.cryst.ehu.es/cgi-bin/plane/programs/nph-plane_getgen?what=gp&gnum=6}{retrieve from Bilbao server}]
\begin{align}
    &\{ 1|n_1 {\vec a}_1+n_2{\vec a}_2\}, \quad \mathrm{with} \quad n_1,n_2\, \in\, \mathbb Z\\
    &\{ 2|(0,0)\},\\
    &\{m_{10}|(0,0)\},\;\{m_{01}|(0,0)\}.
\end{align}

\noindent
The relevant members of the little groups of the high-symmetry points and lines and their irreps are given in Tab.~\ref{tab:irreps-p2mm}.  The full set can be obtained from the \href{https://www.cryst.ehu.es/cgi-bin/cryst/programs/representations_vec.pl?tipogroup=spg&super=25&complex=Submit}{Bilbao Server}.
\begin{table}[htb]
    \begin{subtable}[t]{0.32\textwidth}\centering
        \subcaption{$\Omega=\Gamma,\, X,\, Y,\, S$}
\begin{tabular}[b]{cccc}
    \hline
    \hline\\[-10pt]
     & $\{2|(0,0)\}$ & $\{m_{01}|(0,0)\}$& $\{m_{10}|(0,0)\}$\\
    \hline\\[-10pt]
    $\Omega_1$  &$\phantom{-}1$&$\phantom{-}1$&$\phantom{-}1$\\
    $\Omega_2$  & $\phantom{-}1$&$-1$&$-1$ \\
    $\Omega_3$  & $-1$&$-1$&$\phantom{-}1$ \\
    $\Omega_4$  & $-1$&$\phantom{-}1$&$-1$ \\
    \hline
    \hline
\end{tabular}
\end{subtable}
\begin{subtable}[t]{0.22\textwidth}\centering
    \subcaption{$\overline{\Gamma X},\,\overline{YS}$}
\begin{tabular}[b]{cc}
\hline
\hline\\[-10pt]
 &$\{m_{01}|(0,0)\}$ \\
\hline\\[-10pt]
$SM_1$  &$\phantom{-}1$\\
$SM_2$  & $-1$ \\
\hline
\hline
\end{tabular}
\end{subtable}
\begin{subtable}[t]{0.22\textwidth}\centering
    \subcaption{$\overline{\Gamma Y},\,\overline{YS}$}
\begin{tabular}[b]{cc}
\hline
\hline\\[-10pt]
 &$\{m_{10}|(0,0)\}$ \\
\hline\\[-10pt]
$DT_1$  &$\phantom{-}1$\\
$DT_2$  & $-1$ \\
\hline
\hline
\end{tabular}
\end{subtable}
\caption[$p4$ irreps.]{The relevant irreducible representations of $p2mm$ at the high-symmetry points $\Gamma=(0,0)$, $X=(1/2,0)$, and $Y=(0,1/2)$  and $S=(1/2,1/2)$ as well as along the lines $\overline{\Gamma X}$, $\overline{\Gamma Y}$, $\overline{XS}$, and $\overline{YS}$.}
\label{tab:irreps-p2mm}
\end{table}

\noindent
The fragile roots in $p4mm$ \cite{Song2020} are given in Tab.~\ref{tab:roots-p2mm}; the elementary band representations can be retrieved from the \href{https://www.cryst.ehu.es/cgi-bin/cryst/programs/bandrep.pl?super=25&elementaryTR=Elementary%20TR}{Bilbao server}.
\begin{table}[htb]
    \begin{tabular}{c|l|c|l}
        \hline
        \# & root & \# of bands & type \\
        \hline
        \hline
        1 & $ \Gamma_{1} + \Gamma_{2} + S_{1} + S_{2} + X_{3} + X_{4} + Y_{1} + Y_{2} $ & 2 & mirrors \\
\hline
2 & $ \Gamma_{1} + \Gamma_{2} + S_{3} + S_{4} + X_{1} + X_{2} + Y_{1} + Y_{2} $ & 2 & mirrors \\
\hline
3 & $ \Gamma_{3} + \Gamma_{4} + S_{3} + S_{4} + X_{3} + X_{4} + Y_{1} + Y_{2} $  & 2 & mirrors \\
\hline
4 & $ \Gamma_{1} + \Gamma_{2} + S_{3} + S_{4} + X_{3} + X_{4} + Y_{3} + Y_{4} $  & 2 & mirrors \\
\hline
5 & $ \Gamma_{1} + \Gamma_{2} + S_{1} + S_{2} + X_{1} + X_{2} + Y_{3} + Y_{4} $  & 2 & mirrors \\
\hline
6 & $ \Gamma_{3} + \Gamma_{4} + S_{3} + S_{4} + X_{1} + X_{2} + Y_{3} + Y_{4} $ & 2 & mirrors  \\
\hline
7 & $ \Gamma_{3} + \Gamma_{4} + S_{1} + S_{2} + X_{3} + X_{4} + Y_{3} + Y_{4} $  & 2 & mirrors \\
\hline
8 & $ \Gamma_{3} + \Gamma_{4} + S_{1} + S_{2} + X_{1} + X_{2} + Y_{1} + Y_{2} $  & 2 & mirrors \\
        \hline
        \end{tabular}
        \caption[$p2mm$ roots.]{Fragile roots and their types in wallpaper group $p2mm$}
        \label{tab:roots-p2mm}
 \end{table}

 \subsubsection{Bundling strategy}
 
 When consulting Tab.~\ref{tab:roots-p2mm}, we immediately observe that at each high-symmetry point, the irreps appear exclusively in the combination $\Omega_1+\Omega_2$ or $\Omega_3+\Omega_4$ (with $\Omega=\Gamma,\,X,\,Y,\,S$), i.e., with irreps of the same parity, see Tab.~\ref{tab:irreps-p2mm}. Using the same argument as for the group $p2$, we conclude that we again have a Chern number if we were to divide this root into two independent sets of bands. However, the mirrors give us a simpler handle on the bundling. 

 Let us consider root \#1. We start from $\Gamma_1$ and go along $\overline{\Gamma X}$, compatibility with the eigenvalues of $\{m_{01}|(0,0)\}$ require us to connect $\Gamma_1$--$SM_1$--$X_4$. Continuing along $\overline{XS}$ we find $X_4$--$DT_2$--$S_2$ due to the odd eigenvalues of $\{m_{10}|(0,0)\}$. Going along $\overline{YS}$ fixes $S_2$--$SM_2$--$Y_2$, and finally back to $\Gamma$ via $Y_2$--$DT_2$--$\Gamma_2$. In other words, we need to traverse the loop $\Gamma$--$X$--$S$--$Y$--$\Gamma$ twice to come back to the irrep $\Gamma_1$: The two Chern bands are linked through gap closings along the high-symmetry lines. This means, we can bundle by using the mirror eigenvalues along these lines and can then check for the irrep-content in the so-obtained bundle. There is no need to check the Chern number parity of adjacent bands in $p2mm$. All other roots are permutations of the irreps of root \#1.
 Finally, the types of the roots of $p2mm$ are indicated in the last column of Tab.~\ref{tab:roots-p2mm}.

\subsubsection{Examples}

\begin{figure}[h!bt]
    \centering
    \includegraphics{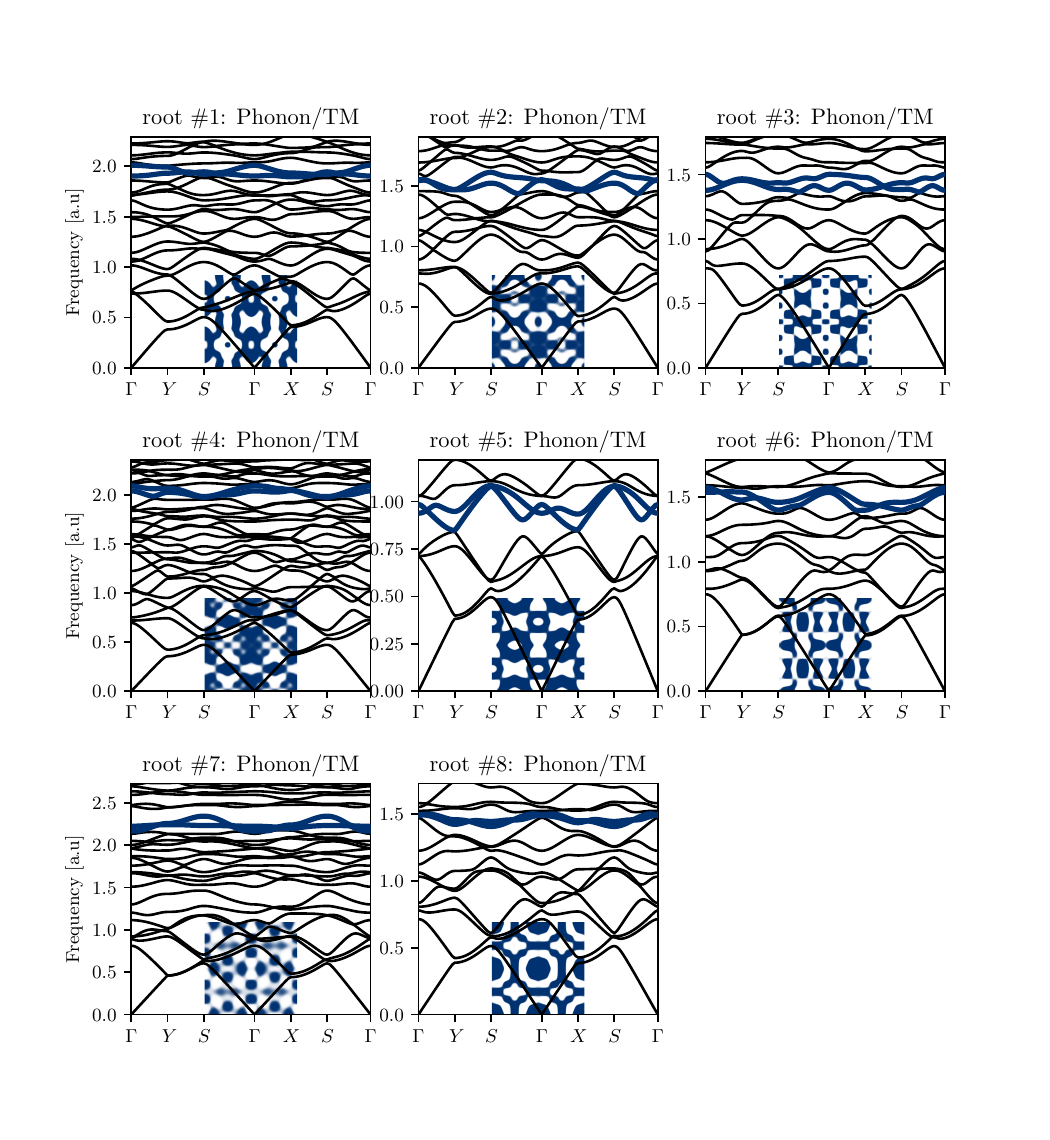}
    \caption[$p2mm$ phonons/TM examples.]{ One sample for each of the roots in $p2mm$ for phonons and TM photons.}
    \label{fig:p2mm-TM-examples}
\end{figure}

\begin{figure}[h!bt]
    \centering
    \includegraphics{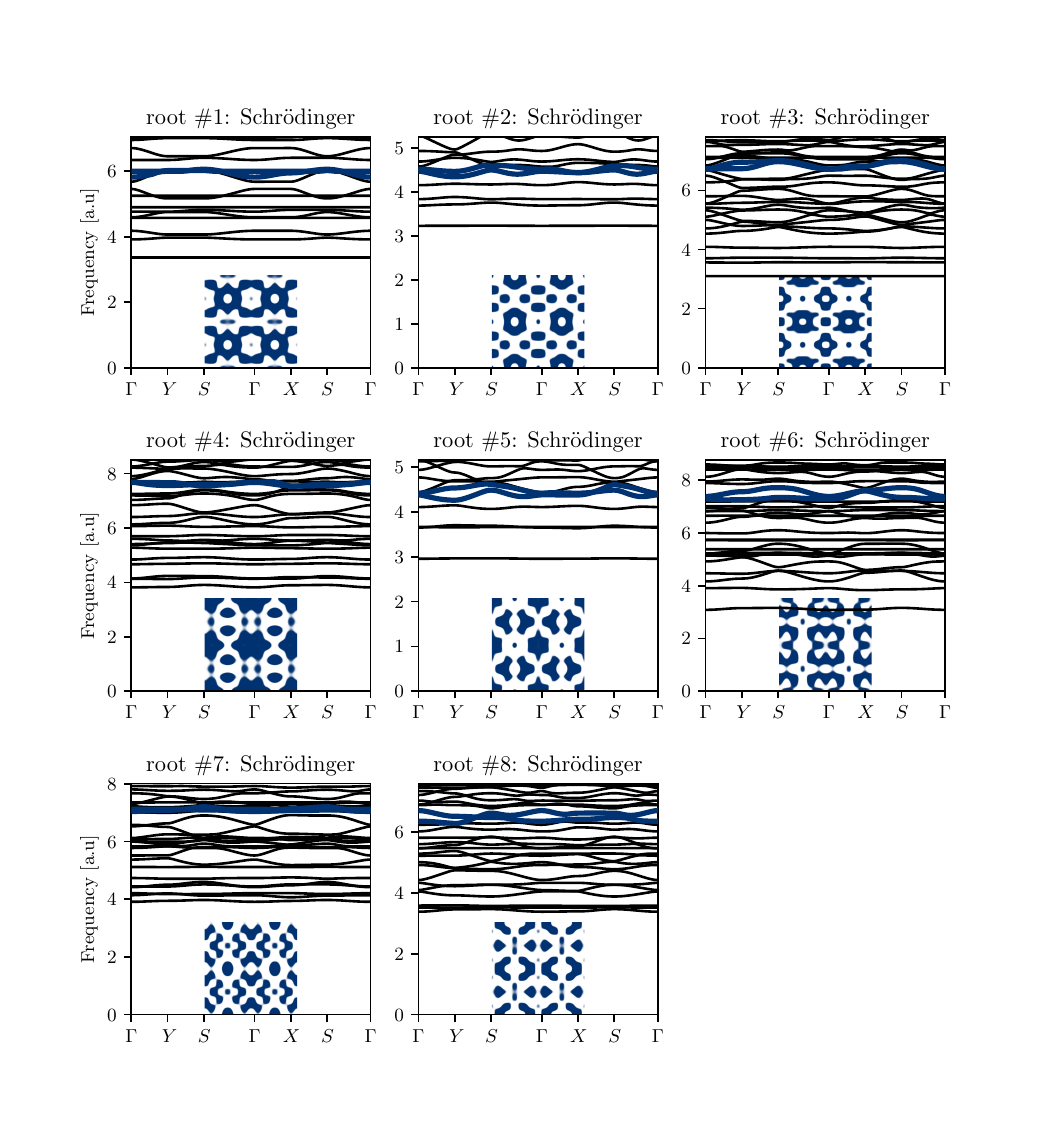}
    \caption[$p2mm$ Schrödinger examples. examples.]{One sample for each of the roots in $p2mm$ for systems described by the Schrödinger equation.}
    \label{fig:p2mm-Q-examples}
\end{figure}

\begin{figure}[h!bt]
    \centering
    \includegraphics{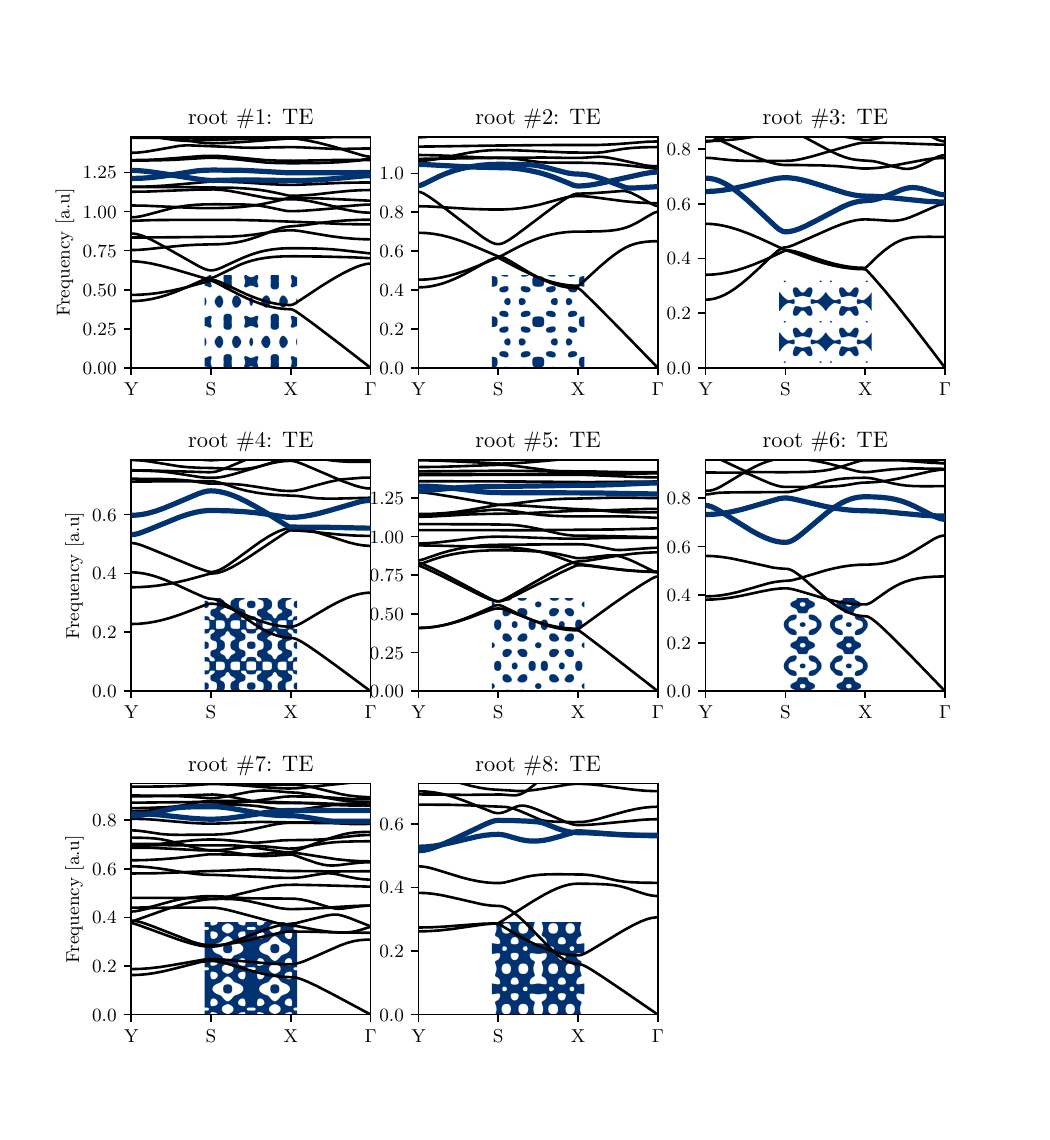}
    \caption[$p2mm$ TE examples.]{ One sample for each of the roots in $p2mm$ for TE photons.}
    \label{fig:p2mm-TE-examples}
\end{figure}

\raggedbottom
\newpage
\subsection{$c2mm$}

\subsubsection{Basic group properties}

\begin{figure}[!h]
    \centering
    \includegraphics{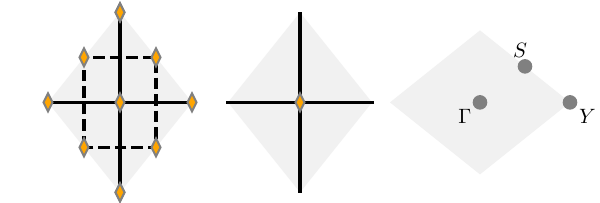}
    \caption[$c2mm$ unit cell.]{Left: Unit cell with the full symmetry of $c2mm$. Middle: General positions of $c2mm$. Right: Brillouin zone with the high-symmetry points and lines indicated.}
    \label{fig:c2mm-uc}
\end{figure}

The group $c2mm$ describes a rhombic lattice with lattice vectors $\vec a_1=(1,0)$ and $\vec a_2=(0,\beta)$. The corresponding space group is $Cmm2$, (\#35) constrained to the $x$-$y$ plane. The group $c2mm$ contains the following group elements [\href{https://www.cryst.ehu.es/cgi-bin/plane/programs/nph-plane_getgen?what=gp&gnum=9}{retrieve from Bilbao server}]
\begin{align}
    &\{ 1|n_1 {\vec a}_1+n_2{\vec a}_2\}, \quad \mathrm{with} \quad n_1,n_2\, \in\, \mathbb Z\\
    &\{ 2|(0,0)\},\\
    &\{m_{01} |(0,0\},\; \{m_{10} |(0,0)\}.
\end{align}

\noindent
The relevant members of the little groups of the high-symmetry points and lines and their irreps are given in Tab.~\ref{tab:irreps-c2mm}.  The full set can be obtained from the \href{https://www.cryst.ehu.es/cgi-bin/cryst/programs/representations_vec.pl?tipogroup=spg&super=35&complex=Submit}{Bilbao Server}.
\begin{table}[htb]
    \begin{subtable}[t]{0.48\textwidth}\centering
        \subcaption{$\Omega=\Gamma,\,Y$}
\begin{tabular}[b]{cccc}
    \hline
    \hline\\[-10pt]
    \multicolumn{1}{c}{} &$\{2|(0,0)\}$ & $\{m_{01}|(0,0)\}$& $\{m_{10}|(0,0)\}$ \\
    \hline\\[-10pt]
    $\Omega_1$  &$\phantom{-}1$ & $\phantom{-}1$  & $\phantom{-}1$ \\
    $\Omega_2$  & $\phantom{-}1$ & $-1$  &$-1$\\
    $\Omega_3$  & $-1$  & $-1$ &$\phantom{-}1$\\
    $\Omega_4$  & $-1$  & $\phantom{-}1$  &$-1$ \\
    \hline
    \hline
\end{tabular}
\end{subtable}
\begin{subtable}[t]{0.19\textwidth}\centering
    \subcaption{$S$}
\begin{tabular}[b]{cc}
\hline
\hline\\[-10pt]
\multicolumn{1}{c}{} & $\{2|(0,0)\}$ \\
\hline\\[-10pt]
$S_1$  &$\phantom{-}1$ \\
$S_2$  &$-1$ \\
\hline
\hline
\end{tabular}
\end{subtable}
\caption[$c2mm$ irreps.]{The relevant irreducible representations of $c2mm$ at the high-symmetry points $\Gamma=(0,0)$, $Y=(1,0)$, and $S=(1/2,1/2)$.}
\label{tab:irreps-c2mm}
\end{table}

\noindent
The fragile roots in $c2mm$ \cite{Song2020} are given in Tab.~\ref{tab:roots-c2mm}; the elementary band representations can be retrieved from the \href{https://www.cryst.ehu.es/cgi-bin/cryst/programs/bandrep.pl?super=35&elementaryTR=Elementary%20TR}{Bilbao server}.
\begin{table}[htb]
    \begin{tabular}{c|l|c|l}
        \hline
        \#& root &\# of bands & type\\
        \hline\hline
        1 & $ \Gamma_{1} + \Gamma_{2} + Y_{3} + Y_{4} + 2 S_{2} $ & 2 & Chern \\
        \hline
        2 & $ \Gamma_{1} + \Gamma_{2} + Y_{3} + Y_{4} + 2 S_{1} $ & 2 & Chern \\
        \hline
        3 & $ \Gamma_{3} + \Gamma_{4} + Y_{1} + Y_{2} + 2 S_{2} $ & 2 & Chern \\
        \hline
        4 & $ \Gamma_{3} + \Gamma_{4} + Y_{1} + Y_{2} + 2 S_{1} $ & 2 & Chern \\
        \hline
        \end{tabular}
        \caption[$c2mm$ roots.]{Fragile roots and their types in wallpaper group $c2mm$}
        \label{tab:roots-c2mm}
 \end{table}

\subsubsection{Bundling strategy}

As is obvious from the irreps at $\Gamma$, $Y$, and $S$, there are no 2D irreps, nor conjugate pairs. Moreover, no mirror symmetric lines are present in the Brillouin zone due to the rhombic lattice structure. We are therefore left with a Chern argument. 

Let us inspect root \#1: Given that $\Gamma_1$ and $\Gamma_2$ have the parity as much as $Y_3$ and $Y_4$, it does not matter which of the two we connect. The Fu-Kane formula (\ref{eqn:fu-kane}) immediately yields an odd Chern number: We need the two bands so constructed to cross somewhere in the Brillouin zone the yield a time-reversal symmetric bundle. Consequently, we apply the Chern bundling strategy outlined above. Finally, the types of the roots of $p2$ are indicated in the last column of Tab.~\ref{tab:roots-c2mm}.

 \subsubsection{Examples}
 \begin{figure}[h!bt]
    \centering
    \includegraphics{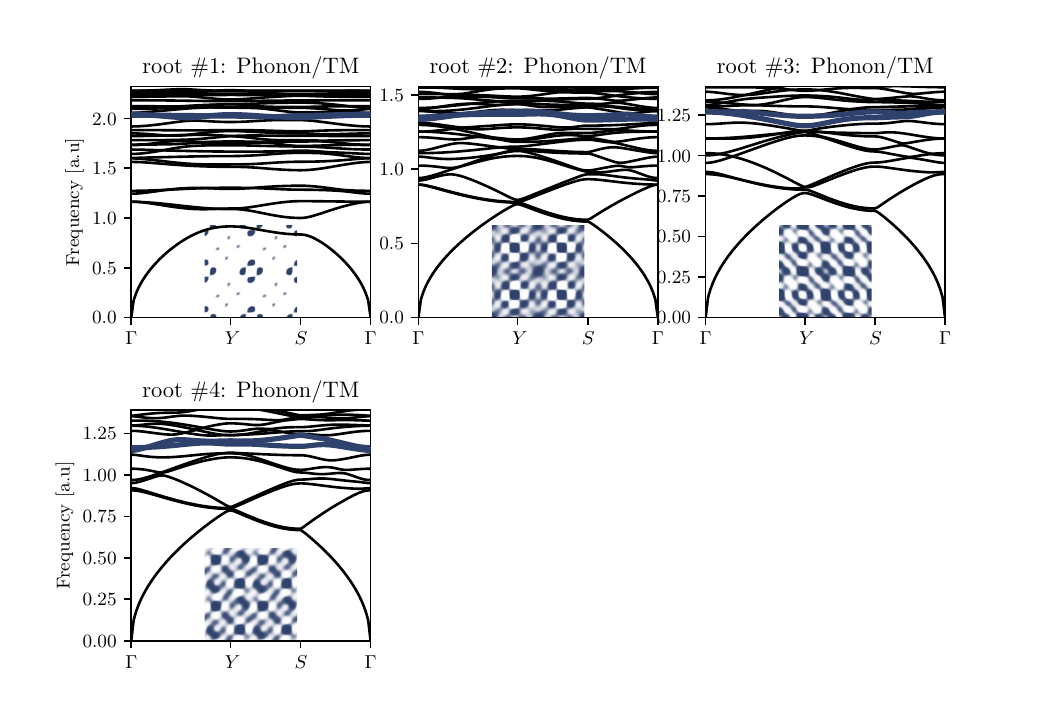}
    \caption[$c2mm$ phonons/TM examples.]{ One sample for each of the roots in $c2mm$ for phonons and TM photons.}
    \label{fig:c2mm-TM-examples}
\end{figure}

\begin{figure}[h!bt]
    \centering
    \includegraphics{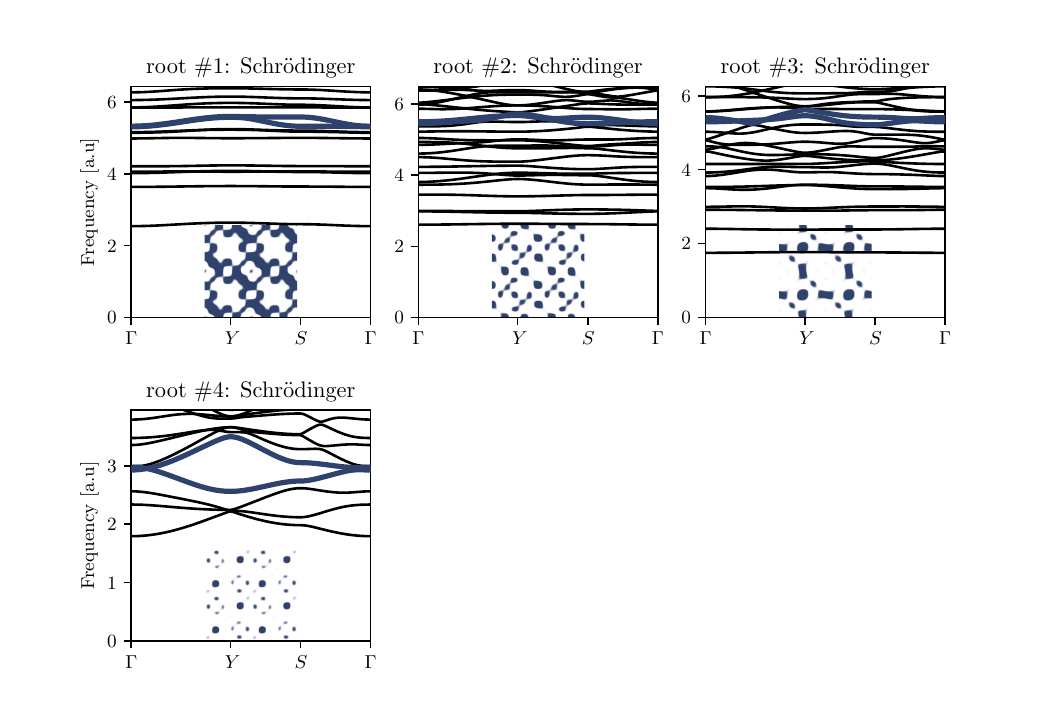}
    \caption[$c2mm$ Schrödinger examples.]{One sample for each of the roots in $c2mm$ for systems described by the Schrödinger equation.}
    \label{fig:c2mm-Q-examples}
\end{figure}

\begin{figure}[h!bt]
    \centering
    \includegraphics{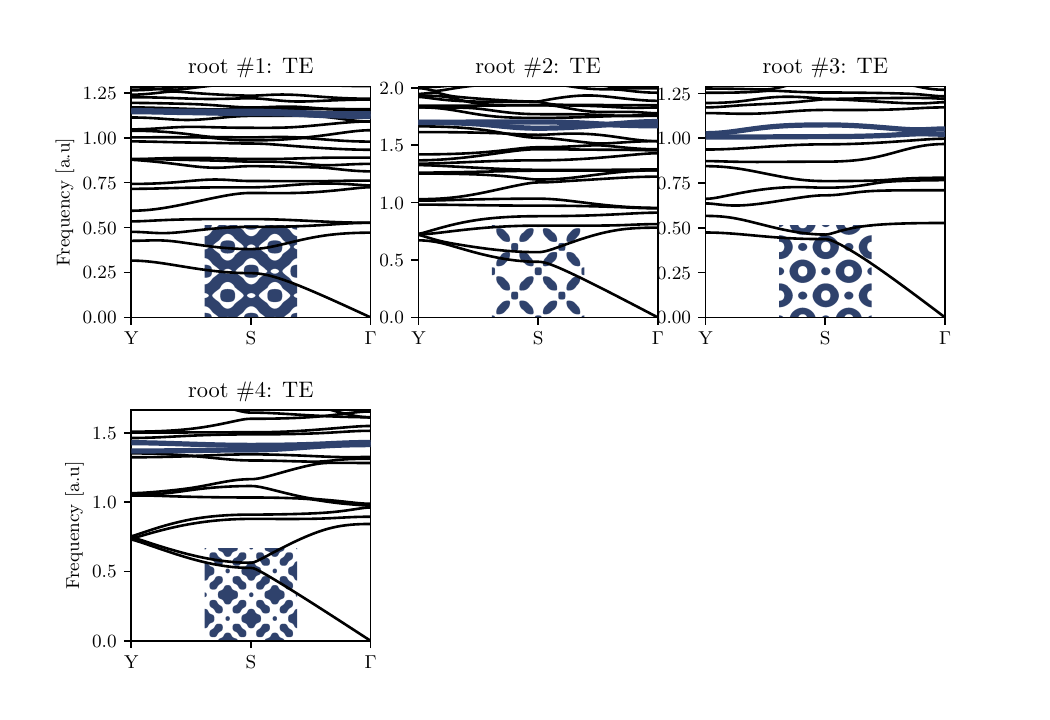}
    \caption[$c2mm$ TE examples.]{ One sample for each of the roots in $c2mm$ for TE photons.}
    \label{fig:c2mm-TE-examples}
\end{figure}

 \raggedbottom
 \newpage
 \subsection{$p4$}
 
 \subsubsection{Basic group properties}
 
 \begin{figure}[!h]
     \centering
     \includegraphics{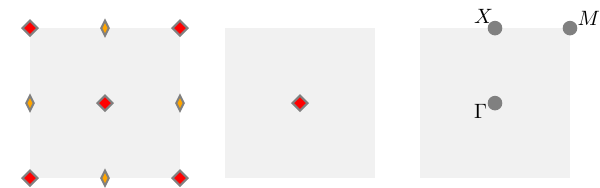}
     \caption[$p4$ unit cell.]{Left: Unit cell with the full symmetry of $p4$. Middle: General positions of $p4$. Right: Brillouin zone with the high-symmetry points and lines indicated.}
     \label{fig:p4-uc}
 \end{figure}
 
 The group $p4$ describes a square lattice with lattice vectors $\vec a_1=(1,0)$ and $\vec a_2=(0,1)$. The corresponding space group is $P4$, (\#75) constrained to the $x$-$y$ plane. The group $p4$ contains the following group elements [\href{https://www.cryst.ehu.es/cgi-bin/plane/programs/nph-plane_getgen?what=gp&gnum=10}{retrieve from Bilbao server}]
 \begin{align}
     &\{ 1|n_1 {\vec a}_1+n_2{\vec a}_2\}, \quad \mathrm{with} \quad n_1,n_2\, \in\, \mathbb Z\\
     &\{ 2|(0,0)\},\; \{4^+|(0,0)\},\; \{ 4^-|(0,0)\},.\\
 \end{align}
 
 \noindent
 The relevant members of the little groups of the high-symmetry points and lines and their irreps are given in Tab.~\ref{tab:irreps-p4}.  The full set can be obtained from the \href{https://www.cryst.ehu.es/cgi-bin/cryst/programs/representations_vec.pl?tipogroup=spg&super=75&complex=Submit}{Bilbao Server}.
 \begin{table}[htb]
    \begin{subtable}[t]{0.32\textwidth}\centering
        \subcaption{$\Gamma$}
\begin{tabular}[b]{ccc}
    \hline
    \hline\\[-10pt]
    \multicolumn{1}{c}{} &$\{2|(0,0)\}$ &$\{4^+|(0,0)\}$ \\
    \hline\\[-10pt]
    $\Gamma_1$  &$\phantom{-}1$ & $\phantom{-}1$ \\
    $\Gamma_2$  &$\phantom{-}1$ & $-1$ \\
    $\Gamma_3$  &$-1$ & $\phantom{-}\mathrm{i}$ \\
    $\Gamma_4$  &$-1$ & $-\mathrm{i}$ \\
    \hline
    \hline
\end{tabular}
\end{subtable}
\begin{subtable}[t]{0.32\textwidth}\centering
    \subcaption{$M$}
\begin{tabular}[b]{ccc}
\hline
\hline\\[-10pt]
\multicolumn{1}{c}{} &$\{2|(0,0)\}$ &$\{4^+|(0,0)\}$ \\
\hline\\[-10pt]
$M_1$  &$\phantom{-}1$ & $\phantom{-}1$ \\
$M_2$  &$\phantom{-}1$ & $-1$ \\
$M_3$  &$-1$ & $\phantom{-}\mathrm{i}$ \\
$M_4$  &$-1$ & $-\mathrm{i}$ \\
\hline
\hline
\end{tabular}
\vspace{6pt}
\end{subtable}
\begin{subtable}[t]{0.32\textwidth}\centering
    \subcaption{$X$}
\begin{tabular}[b]{cc}
\hline
\hline\\[-10pt]
\multicolumn{1}{c}{} &$\{2|(0,0)\}$ \\
\hline\\[-10pt]
$X_1$  &$\phantom{-}1$ \\
$X_2$  &$-1$ \\
\hline
\hline
\end{tabular}
\end{subtable}
\caption[$p4$ irreps.]{The relevant irreducible representations of $p4$ at the high-symmetry points $\Gamma=(0,0)$, $M=(1/2,1/2)$, and $X=(0,1/2)$.}
\label{tab:irreps-p4}
\end{table}

\noindent
The fragile roots in $p4$ \cite{Song2020} are given in Tab.~\ref{tab:roots-p4}; the elementary band representations can be retrieved from the \href{https://www.cryst.ehu.es/cgi-bin/cryst/programs/bandrep.pl?super=75&elementaryTR=Elementary%20TR}{Bilbao server}.
\begin{table}[htb]
    \begin{tabular}{c|l|c|l}
        \hline
        \#& root &\# of bands & type\\
        \hline\hline
        1 & $ 2 \Gamma_{2} + 2 M_{2} + 2 X_{2} $ & 2 & Chern \\
\hline
2 & $ 2 \Gamma_{1} + 2 M_{2} + 2 X_{1} $  & 2 & Chern\\
\hline
3 & $ 2 \Gamma_{2} + 2 M_{1} + 2 X_{1} $  & 2 & Chern\\
\hline
4 & $ 2 \Gamma_{1} + M_{3}M_{4} + 2 X_{2} $& 2 & conjugate pairs\\
\hline
5 & $ 2 \Gamma_{1} + 2 M_{1} + 2 X_{2} $ & 2 & Chern \\
\hline
6 & $ 2 \Gamma_{1} + M_{3}M_{4} + 2 X_{1} $ & 2 & conjugate pairs\\
\hline
7 & $ 2 \Gamma_{2} + M_{3}M_{4} + 2 X_{1} $& 2 & conjugate pairs \\
\hline
8 & $ 2 \Gamma_{2} + M_{3}M_{4} + 2 X_{2} $ & 2 & conjugate pairs\\
\hline
9 & $ \Gamma_{3}\Gamma_{4} + 2 M_{2} + 2 X_{2} $ & 2 & conjugate pairs\\
\hline
10 & $ \Gamma_{3}\Gamma_{4} + 2 M_{1} + 2 X_{1} $ & 2 & conjugate pairs\\
\hline
11 & $ \Gamma_{3}\Gamma_{4} + 2 M_{2} + 2 X_{1} $& 2 & conjugate pairs\\
\hline
12 & $ \Gamma_{3}\Gamma_{4} + 2 M_{1} + 2 X_{2} $ & 2 & conjugate pairs\\
\hline
        \end{tabular}
        \caption[$p4$ roots.]{Fragile roots and their types in wallpaper group $p4$.}
        \label{tab:roots-p4}
 \end{table}

 \subsubsection{Bundling strategy}

 We start by addressing roots \#1-\#3 and \#5 as these do not involve any conjugate pairs of irreps ($p4$ does not have any two-dimensional irreps). As in $p2$, we will rely on a Chern number argument. We state the following formula for the Chern number C in $C_4$--symmetric systems \cite{Fang2012}
\begin{equation}
    \label{eqn:chern4}
    \mathrm{i}^C = \prod_{i\,\in\, {\rm occ.}}(-1)^F \xi_i(\Gamma)\xi_i(M)\zeta_i({X}),
\end{equation}
where $F=2S$, with $S$ the total spin of the particles, $\xi$ and $\zeta$ the eigenvalues of $\{4^+|(0,0)\}$ and $\{2|(0,0)\}$, respectively. In the following we set $F=0$.

We see that root \#1 can be written as
\begin{equation}
    (\Gamma_2+M_2+X_2)\,\oplus\,(\Gamma_2+M_2+X_2).
\end{equation}
Using (\ref{eqn:chern4}), the above two isolated bands would each have a Chern number 
\begin{equation}
    C=2+4n\quad \mathrm{with} \quad n\in\mathbb Z.
\end{equation}
To be compliant with time-reversal symmetry, these two bands therefore need to be connected somewhere in the Brillouin zone. The remaining roots all contain a pair of conjugate irreps, which glues them into a bundle of two. Finally, the types of the roots of $p4$ are indicated in the last column of Tab.~\ref{tab:roots-p4}.

\subsubsection{Examples}

\begin{figure}[h!bt]
    \centering
    \includegraphics{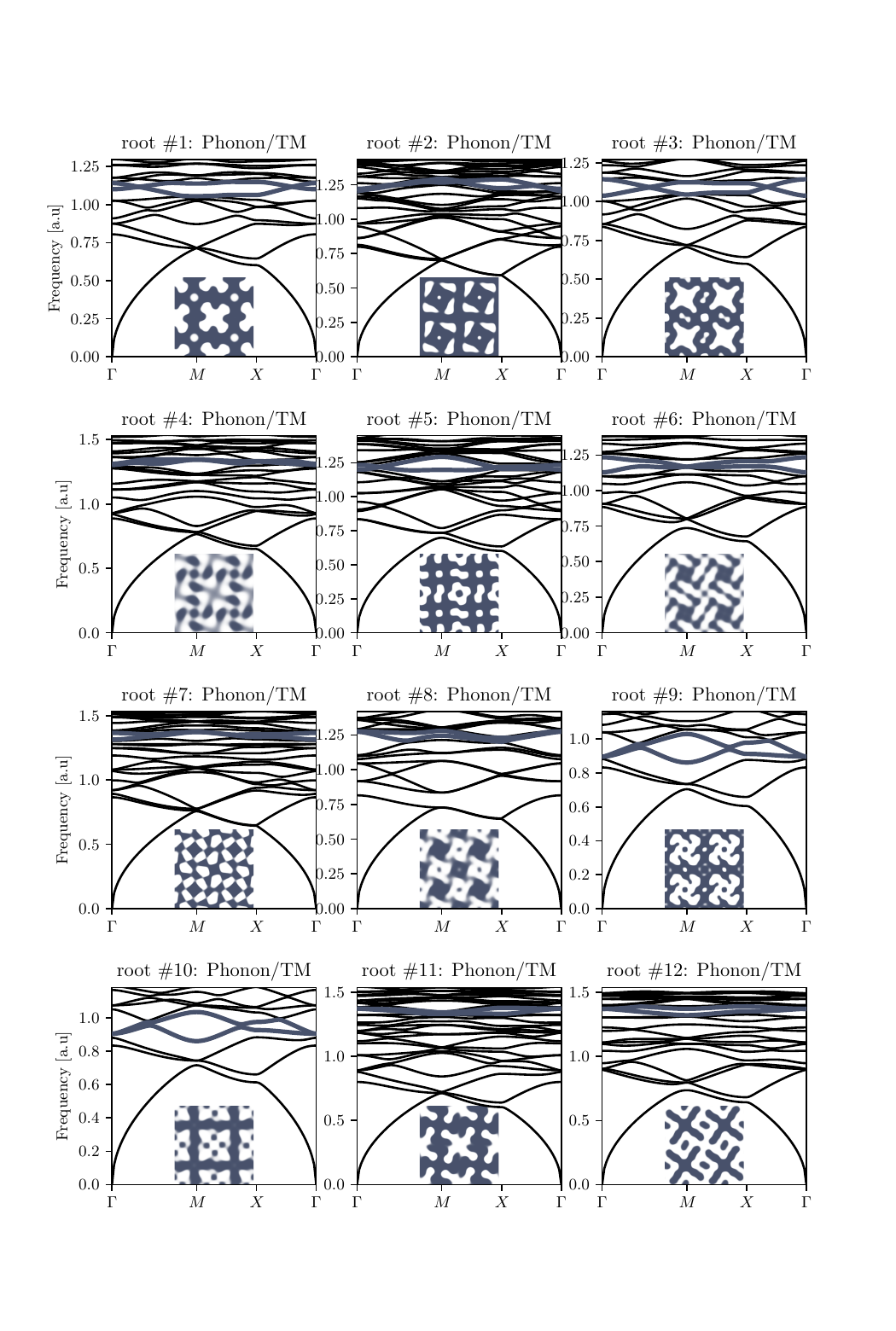}
    \caption[$p4$ phonons/TM examples.]{ One sample for each of the roots in $p4$ for phonons and TM photons.}
    \label{fig:p4-TM-examples}
\end{figure}

\begin{figure}[h!bt]
    \centering
    \includegraphics{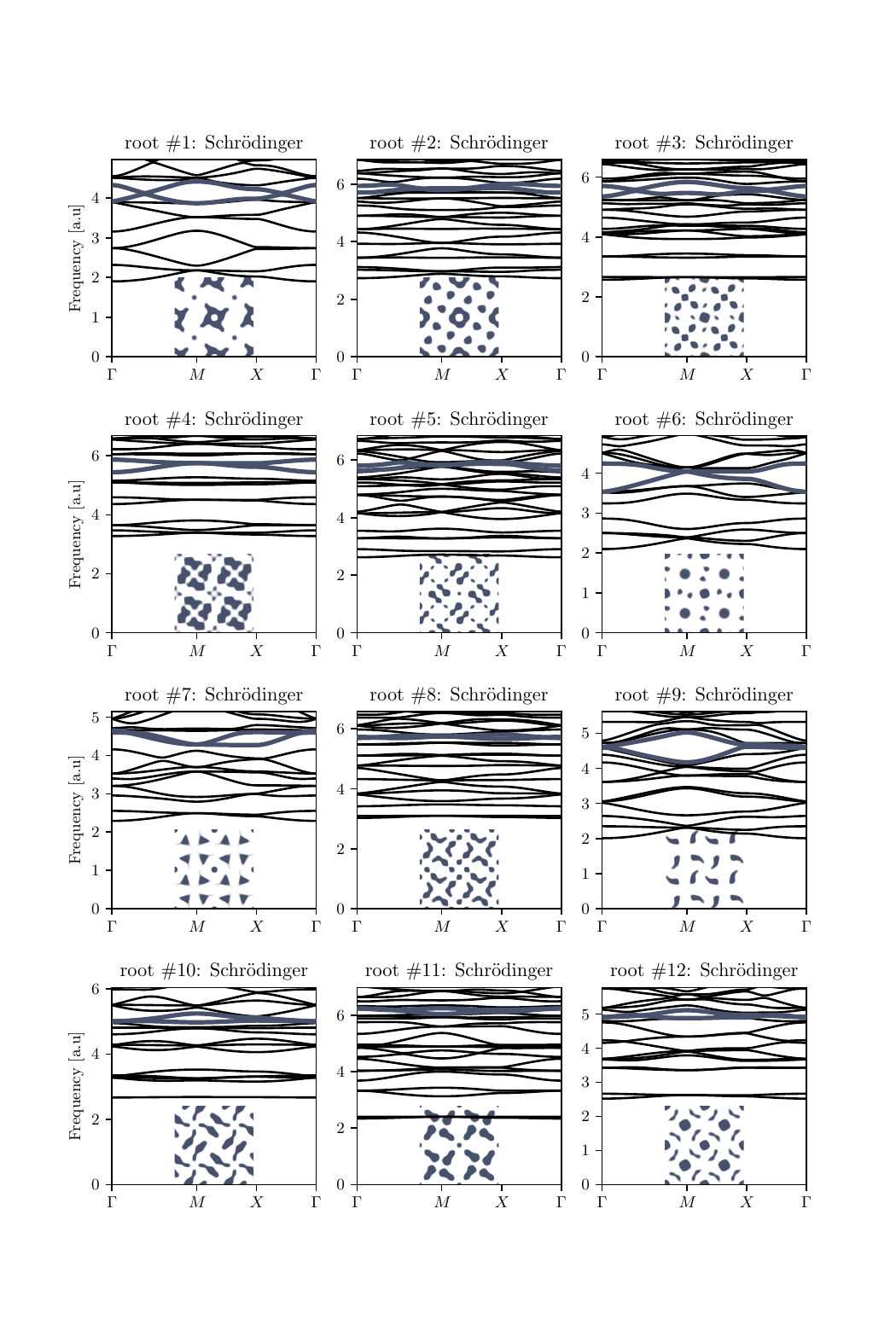}
    \caption[$p4$ Schrödinger examples.]{One sample for each of the roots in $p4$ for systems described by the Schrödinger equation.}
    \label{fig:p4-Q-examples}
\end{figure}

\begin{figure}[h!bt]
    \centering
    \includegraphics{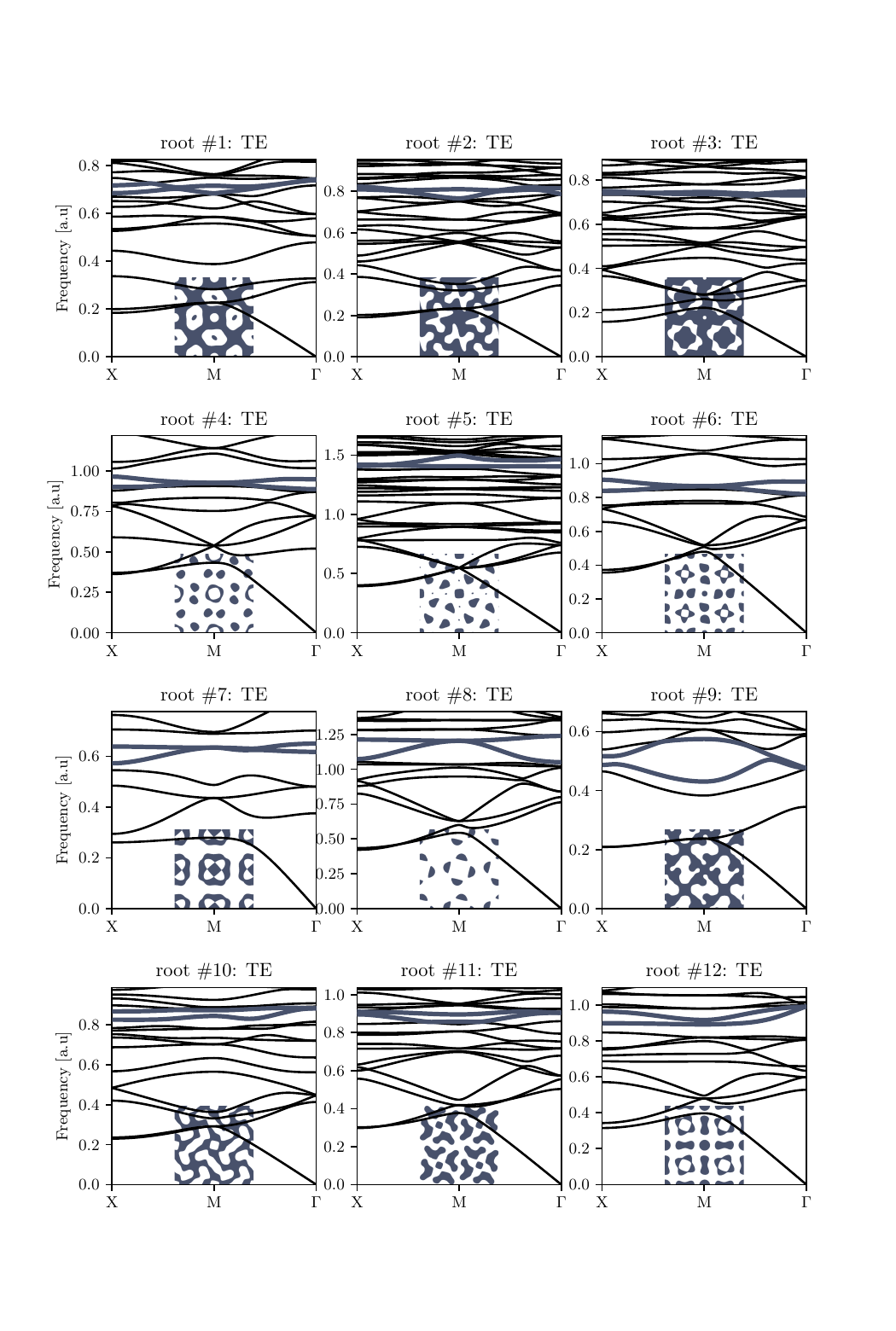}
    \caption[$p4$ TE examples.]{ One sample for each of the roots in $p4$ for TE photons.}
    \label{fig:p4-TE-examples}
\end{figure}

\raggedbottom
\newpage
\subsection{$p4mm$}

\subsubsection{Basic group properties}

\begin{figure}[!h]
    \centering
    \includegraphics{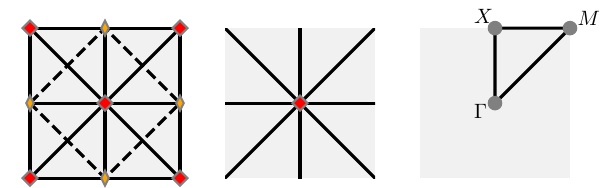}
    \caption[$p4gm$ unit cell.]{Left: Unit cell with the full symmetry of $p4mm$. Middle: General positions of $p4mm$. Right: Brillouin zone with the high-symmetry points and lines indicated.}
    \label{fig:p4mm-uc}
\end{figure}

The group $p4mm$ describes a square lattice with lattice vectors $\vec a_1=(1,0)$ and $\vec a_2=(0,1)$. The corresponding space group is $P4mm$, (\#99) constrained to the $x$-$y$ plane. The group $p4gm$ contains the following group elements [\href{https://www.cryst.ehu.es/cgi-bin/plane/programs/nph-plane_getgen?what=gp&gnum=11}{retrieve from Bilbao server}]
\begin{align}
    &\{ 1|n_1 {\vec a}_1+n_2{\vec a}_2\}, \quad \mathrm{with} \quad n_1,n_2\, \in\, \mathbb Z\\
    &\{ 2|(0,0)\},\; \{4^+|(0,0)\},\; \{ 4^-|(0,0)\},\\
    &\{m_{01} |(0,0)\},\; \{m_{10} |(0,0)\},\; \{m_{11} |(0,0)\},\; \{m_{1\bar1} |(0,0)\}.
\end{align}

\noindent
The relevant members of the little groups of the high-symmetry points and lines and their irreps are given in Tab.~\ref{tab:irreps-p4mm}. The full set can be obtained from the \href{https://www.cryst.ehu.es/cgi-bin/cryst/programs/representations_vec.pl?tipogroup=spg&super=99&complex=Submit}{Bilbao Server}.
\begin{table}[htb]
    \begin{subtable}[t]{0.48\textwidth}\centering
        \subcaption{$\Omega=\Gamma,\,M$}
\begin{tabular}[b]{cccccc}
    \hline
    \hline\\[-10pt]
    \multicolumn{1}{c}{} &$\{2|(0,0)\}$ &$\{4^+|(0,0)\}$ & $\{m_{10}|(0,0)\}$& $\{m_{01}|(0,0)\}$& $\{m_{1\bar1}|(0,0)\}$ \\
    \hline\\[-10pt]
    $\Omega_1$  &$\phantom{-}1$ & $\phantom{-}1$  & $\phantom{-}1$ & $\phantom{-}1$ & $\phantom{-}1$\\
    $\Omega_2$  & $\phantom{-}1$ & $-1$  & $\phantom{-}1$ & $\phantom{-}1$ &$-1$\\
    $\Omega_3$  & $\phantom{-}1$ & $-1$  & $-1$ & $-1$ &$\phantom{-}1$\\
    $\Omega_4$  & $\phantom{-}1$ & $\phantom{-}1$  & $-1$ & $-1$ & $-1$\\
    $\Omega_5$  & $-\mathds{1}$ & $-\mathrm{i}\sigma_y$  & $-\sigma_x$ & $\sigma_x$ &$-\sigma_z$\\
    \hline
    \hline
    \vspace{2pt}
\end{tabular}
\end{subtable}
\begin{subtable}[t]{0.48\textwidth}\centering
    \subcaption{$X$}
\begin{tabular}[b]{cccc}
\hline
\hline\\[-10pt]
\multicolumn{1}{c}{} &$\{2|(0,0)\}$ & $\{m_{10}|(0,0)\}$& $\{m_{01}|(0,0)\}$ \\
\hline\\[-10pt]
$X_1$  &$\phantom{-}1$ & $\phantom{-}1$ & $\phantom{-}1$\\
$X_2$  & $\phantom{-}1$ & $-1$  & $-1$ \\
$X_3$  & $-1$ & $\phantom{-}1$  &$-1$\\
$X_4$  & $-1$ & $-1$  & $\phantom{-}1$\\
\hline
\hline
\end{tabular}
\end{subtable}

\begin{subtable}[t]{0.19\textwidth}\centering
    \subcaption{$\overline{\Gamma X}$}
\begin{tabular}[b]{cc}
\hline
\hline\\[-10pt]
\multicolumn{1}{c}{} & $\{m_{10}|(0,0)\}$ \\
\hline\\[-10pt]
$DT_1$  &$\phantom{-}1$ \\
$DT_2$  &$-1$ \\
\hline
\hline
\end{tabular}
\end{subtable}
\begin{subtable}[t]{0.19\textwidth}\centering
    \subcaption{$\overline{\Gamma M}$}
\begin{tabular}[b]{cc}
\hline
\hline\\[-10pt]
\multicolumn{1}{c}{} & $\{m_{1\bar1}|(0,0)\}$ \\
\hline\\[-10pt]
$SM_1$  &$\phantom{-}1$ \\
$SM_2$  &$-1$ \\
\hline
\hline
\end{tabular}
\end{subtable}
\begin{subtable}[t]{0.19\textwidth}\centering
    \subcaption{$\overline{X M}$}
\begin{tabular}[b]{cc}
\hline
\hline\\[-10pt]
\multicolumn{1}{c}{} & $\{m_{01}|(0,0)\}$ \\
\hline\\[-10pt]
$Y_1$  &$\phantom{-}1$ \\
$Y_2$  &$-1$ \\
\hline
\hline
\end{tabular}
\end{subtable}
\caption[$p4mm$ irreps.]{The relevant irreducible representations of $p4mm$ at the high-symmetry points $\Gamma=(0,0)$, $M=(1/2,1/2)$, and $X=(0,1/2)$ as well as along the lines $\overline{\Gamma X}$, $\overline{\Gamma M}$, and $\overline{XM}$.}
\label{tab:irreps-p4mm}
\end{table}

\noindent
The fragile roots in $p4mm$ \cite{Song2020} are given in Tab.~\ref{tab:roots-p4mm}; the elementary band representations can be retrieved from the \href{https://www.cryst.ehu.es/cgi-bin/cryst/programs/bandrep.pl?super=99&elementaryTR=Elementary%20TR}{Bilbao server}.
\begin{table}[htb]
    \begin{tabular}{c|l|c|l}
        \hline
        \#& root &\# of bands & type\\
        \hline\hline
        1 & $ \Gamma_{2} + \Gamma_{3} + M_{5} + X_{3} + X_{4} $ 2D &2&2D irrep \& mirrors \\
        \hline
        2 & $ \Gamma_{2} + \Gamma_{3} + M_{2} + M_{3} + X_{3} + X_{4} $ & 2 & mirrors\\
        \hline
        3 & $ \Gamma_{1} + \Gamma_{4} + M_{1} + M_{4} + X_{3} + X_{4} $ & 2 & mirrors\\
        \hline
        4 & $ \Gamma_{1} + \Gamma_{4} + M_{5} + X_{1} + X_{2} $ &2&2D irrep \& mirrors \\
        \hline
        5 & $ \Gamma_{2} + \Gamma_{3} + M_{5} + X_{1} + X_{2} $ &2&2D irrep \& mirrors \\
        \hline
        6 & $ \Gamma_{2} + \Gamma_{3} + M_{1} + M_{4} + X_{1} + X_{2} $ & 2 & mirrors\\
        \hline
        7 & $ \Gamma_{1} + \Gamma_{4} + M_{2} + M_{3} + X_{1} + X_{2} $ & 2 & mirrors\\
        \hline
        8 & $ \Gamma_{5} + M_{1} + M_{4} + X_{3} + X_{4} $ &2&2D irrep \& mirrors \\
        \hline
        9 & $ \Gamma_{5} + M_{2} + M_{3} + X_{1} + X_{2} $ &2&2D irrep \& mirrors \\
        \hline
        10 & $ \Gamma_{5} + M_{2} + M_{3} + X_{3} + X_{4} $ &2&2D irrep \& mirrors \\
        \hline
        11 & $ \Gamma_{5} + M_{1} + M_{4} + X_{1} + X_{2} $ &2&2D irrep \& mirrors \\
        \hline
        12 & $ \Gamma_{1} + \Gamma_{4} + M_{5} + X_{3} + X_{4} $ &2&2D irrep \& mirrors \\
\hline
        \end{tabular}
        \caption[$p4mm$ roots.]{Fragile roots and their types in wallpaper group $p4mm$.}
        \label{tab:roots-p4mm}
 \end{table}

 \subsubsection{Bundling strategy}

 We start with the roots \#1, \#4, \#5, \#8--\#12. All of those contain either $\Gamma_5$ or $M_5$, i.e., a two-dimensional irrep. The mirror eigenvalues along the lines connecting these points to their partners fully determine the bundling. 

 For the remaining root \#2, \#3, \#6, \#7 the situation is the same as of $p2mm$: The Chern number formula (\ref{eqn:chern4}) suggests that if we were to divide these two sets of bands into disconnected sets, we were to have a Chern number. However, here again, mirrors come to our rescue: Starting from $\Gamma_2$ of root \#2, we need to traverse the loop $\Gamma$--$M$--$X$--$\Gamma$ twice to come back to $\Gamma_2$. In other words, the mirror eigenvalues force a bundle of two bands. Finally, the types of the roots of $p4mm$ are indicated in the last column of Tab.~\ref{tab:roots-p4mm}.

 \subsubsection{Examples}
 \begin{figure}[h!bt]
    \centering
    \includegraphics{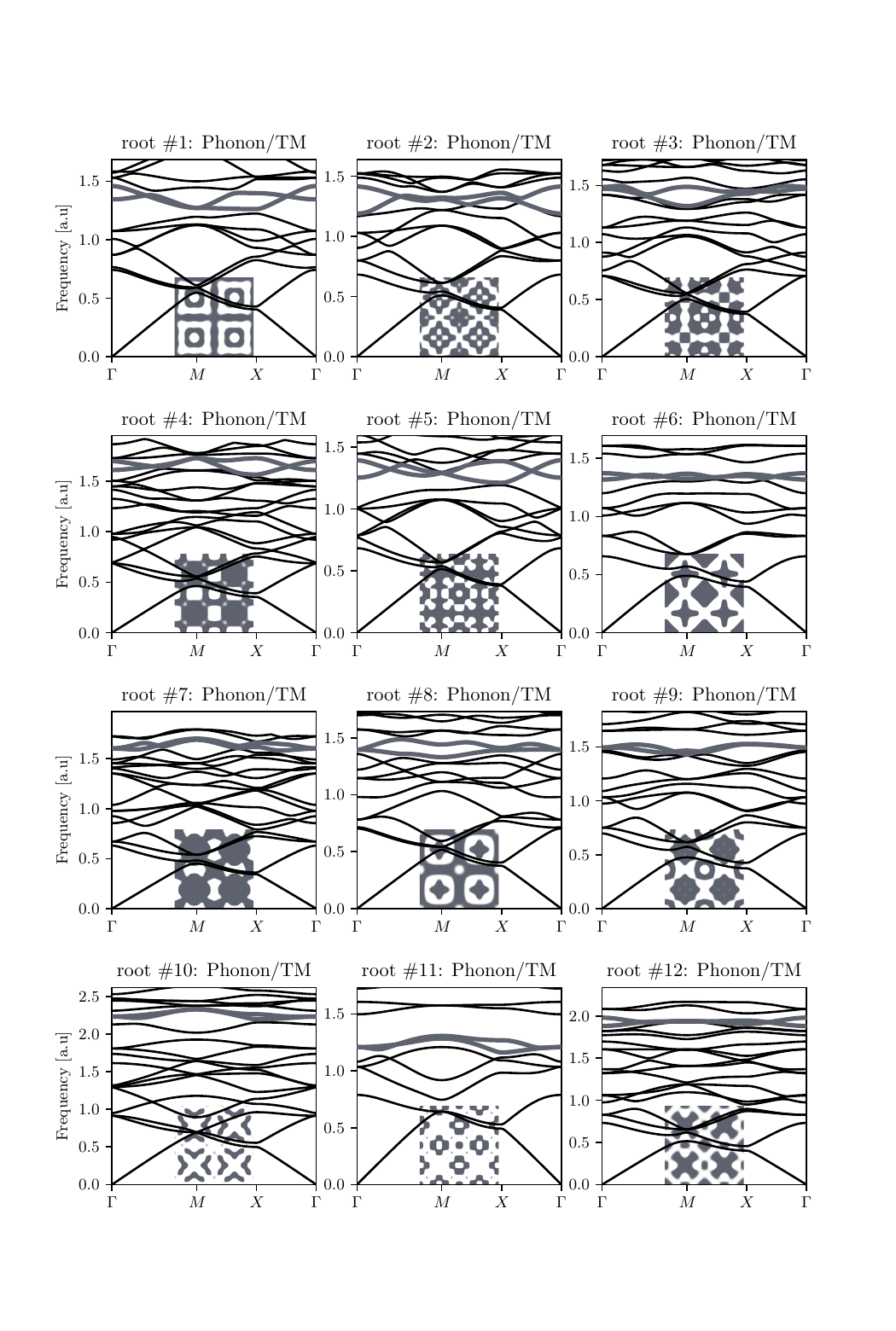}
    \caption[$p4mm$ phonons/TM examples.]{ One sample for each of the roots in $p4mm$ for phonons and TM photons.}
    \label{fig:p4mm-TM-examples}
\end{figure}

\begin{figure}[h!bt]
    \centering
    \includegraphics{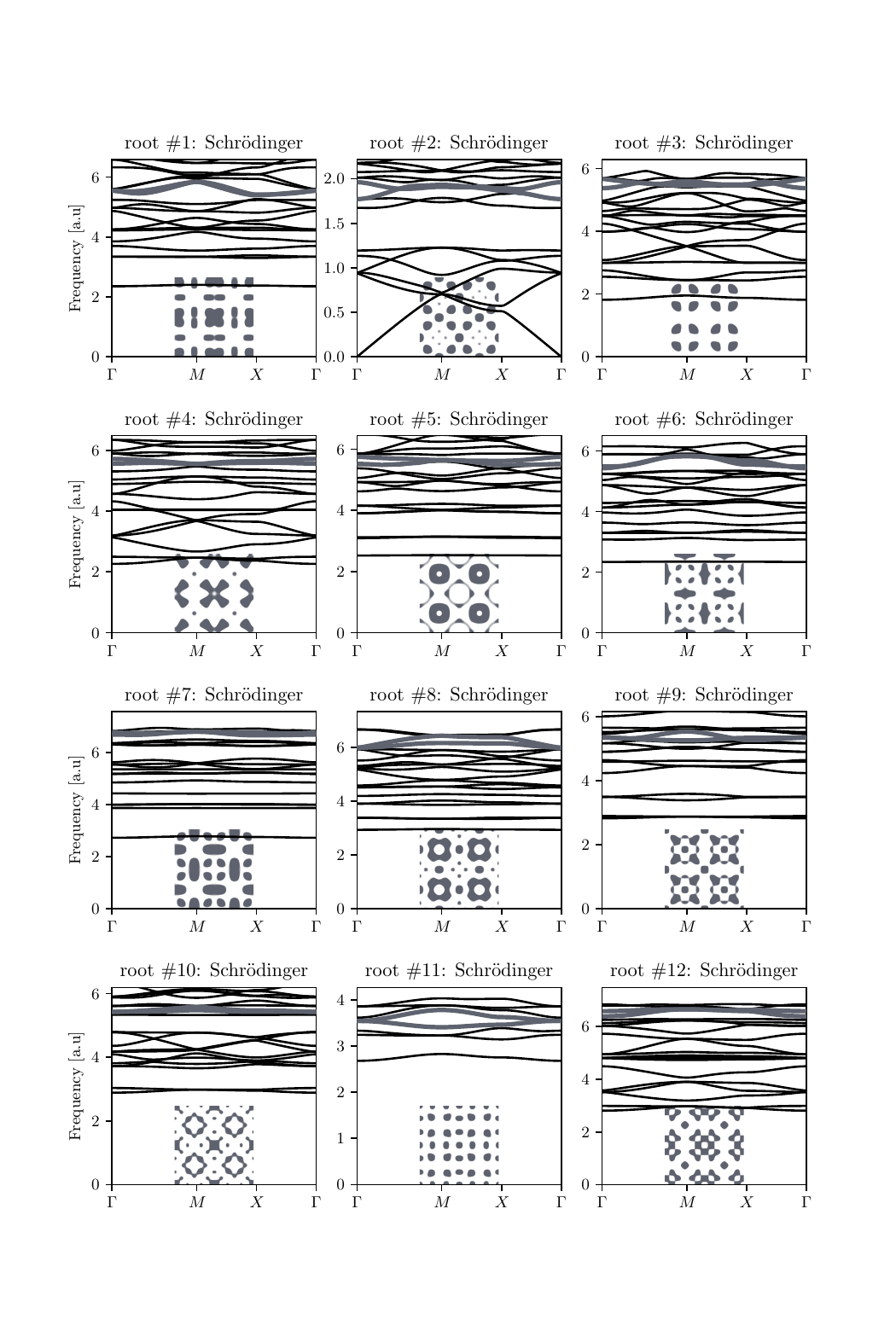}
    \caption[$p4mm$ Schrödinger examples.]{One sample for each of the roots in $p4mm$ for systems described by the Schrödinger equation.}
    \label{fig:p4mm-Q-examples}
\end{figure}

\begin{figure}[h!bt]
    \centering
    \includegraphics{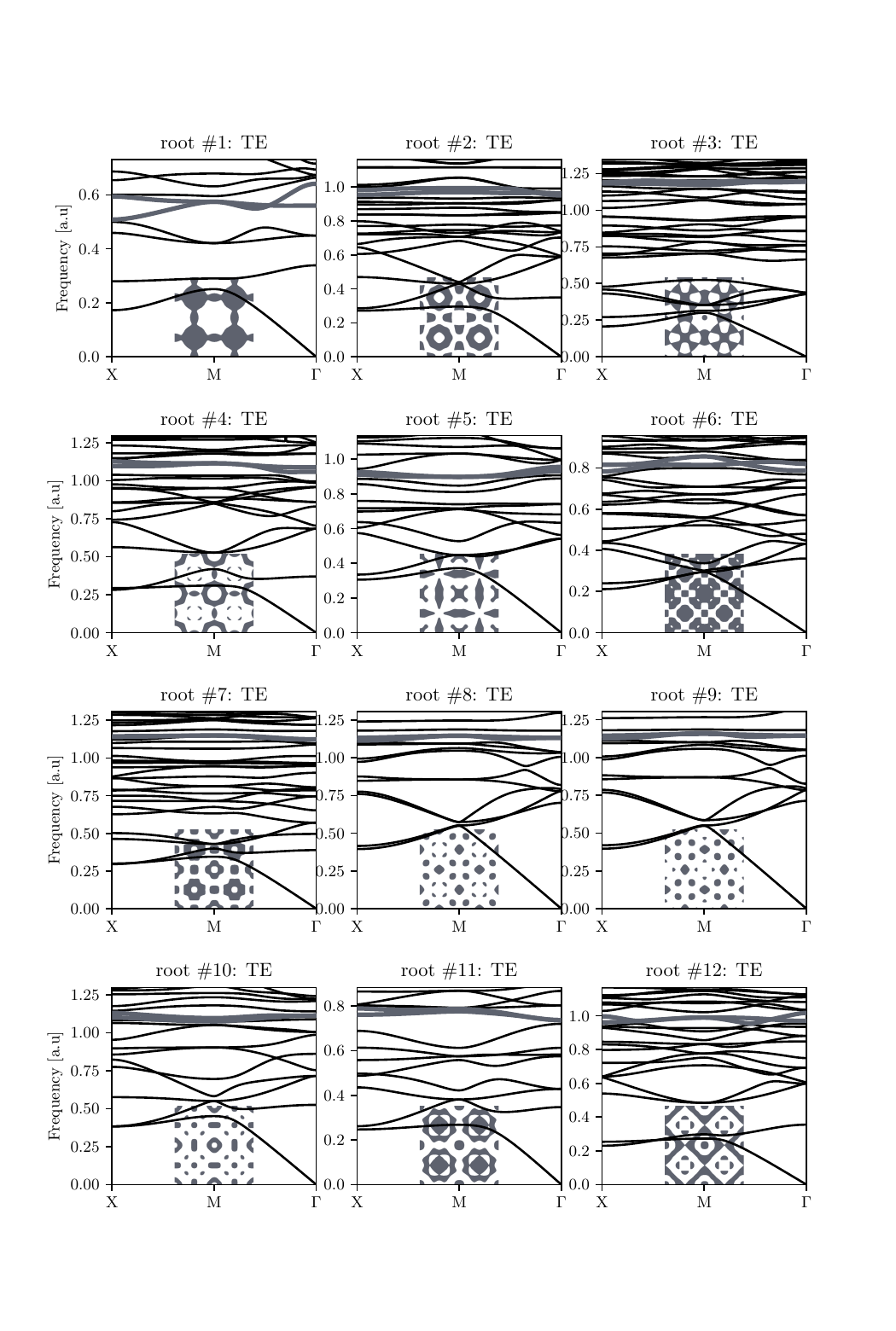}
    \caption[$p4mm$ TE examples.]{ One sample for each of the roots in $p4mm$ for TE photons.}
    \label{fig:p4mm-TE-examples}
\end{figure}

\raggedbottom
\newpage
\subsection{$p4gm$}

\subsubsection{Basic group properties}

\begin{figure}[!h]
    \centering
    \includegraphics{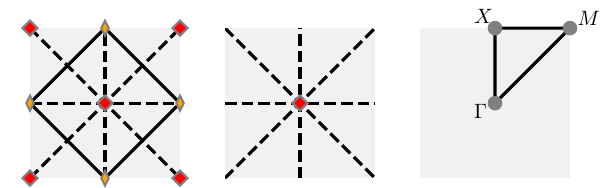}
    \caption[$p4gm$ unit cell.]{Left: Unit cell with the full symmetry of $p4gm$. Middle: General positions of $p4gm$. Right: Brillouin zone with the high-symmetry points and lines indicated.}
    \label{fig:p4gm-uc}
\end{figure}

The group $p4gm$ describes a square lattice with lattice vectors $\vec a_1=(1,0)$ and $\vec a_2=(0,1)$. The corresponding space group is $P4bm$, (\#100) constrained to the $x$-$y$ plane. The group $p4gm$ contains the following group elements [\href{https://www.cryst.ehu.es/cgi-bin/plane/programs/nph-plane_getgen?what=gp&gnum=12}{retrieve from Bilbao server}]
\begin{align}
    &\{ 1|n_1 {\vec a}_1+n_2{\vec a}_2\}, \quad \mathrm{with} \quad n_1,n_2\, \in\, \mathbb Z\\
    &\{ 2|(0,0)\},\; \{4^+|(0,0)\},\; \{ 4^-|(0,0)\},\\
    &\{m_{01} |(1/2,1/2)\},\; \{m_{10} |(1/2,1/2)\},\; \{m_{11} |(1/2,1/2)\},\; \{m_{1\bar1} |(1/2,1/2)\}.
\end{align}

\noindent
The relevant members of the little groups of the high-symmetry points and lines and their irreps are given in Tab.~\ref{tab:irreps-p4gm}. The full set can be obtained from the \href{https://www.cryst.ehu.es/cgi-bin/cryst/programs/representations_vec.pl?tipogroup=spg&super=100&complex=Submit}{Bilbao Server}.
\begin{table}[htb]
    \begin{subtable}[t]{0.48\textwidth}\centering
        \subcaption{$\Gamma$}
\begin{tabular}[b]{ccccc}
    \hline
    \hline\\[-10pt]
    \multicolumn{1}{c}{} &$\{2|(0,0)\}$ &$\{4^+|(0,0)\}$ & $\{m_{01}|(1/2,1/2)\}$& $\{m_{1\bar1}|(1/2,1/2)\}$ \\
    \hline\\[-10pt]
    $\Gamma_1$  &$\phantom{-}1$ & $\phantom{-}1$  & $\phantom{-}1$ & $\phantom{-}1$\\
    $\Gamma_2$  & $\phantom{-}1$ & $-1$  & $\phantom{-}1$ &$-1$\\
    $\Gamma_3$  & $\phantom{-}1$ & $-1$  & $-1$ &$\phantom{-}1$\\
    $\Gamma_4$  & $\phantom{-}1$ & $\phantom{-}1$  & $-1$ &$-1$\\
    $\Gamma_5$  & $-\mathds{1}$ & $-\mathrm{i}\sigma_y$  & $\sigma_x$ &$-\sigma_z$\\
    \hline
    \hline
\end{tabular}
\end{subtable}
\begin{subtable}[t]{0.48\textwidth}\centering
    \subcaption{$M$}
\begin{tabular}[b]{ccccc}
\hline
\hline\\[-10pt]
\multicolumn{1}{c}{} &$\{2|(0,0)\}$ &$\{4^+|(0,0)\}$ & $\{m_{01}|(1/2,1/2)\}$& $\{m_{1\bar1}|(1/2,1/2)\}$ \\
\hline\\[-10pt]
$M_1$  &$-1$ & $\phantom{-}\mathrm{i}$  & $-\mathrm{i}$ & $-1$\\
$M_2$  & $-1$ & $-\mathrm{i}$  & $-\mathrm{i}$ &$\phantom{-}1$\\
$M_3$  & $-1$ & $-\mathrm{i}$  & $\phantom{-}\mathrm{i}$ &$-1$\\
$M_4$  & $-1$ & $\phantom{-}\mathrm{i}$  & $\phantom{-}\mathrm{i}$ &$\phantom{-}1$\\
$M_5$  & $\phantom{-}\mathds{1}$ & $\sigma_x$  & $\mathrm{i}\sigma_y$ &$\sigma_z$\\
\hline
\hline
\end{tabular}
\vspace{6pt}
\end{subtable}
\begin{subtable}[t]{0.4\textwidth}\centering
    \subcaption{$X$}
\begin{tabular}[b]{cccc}
\hline
\hline\\[-10pt]
\multicolumn{1}{c}{} &$\{2|(0,0)\}$ & $\{m_{01}|(1/2,1/2)\}$& $\{m_{1\bar1}|(1/2,1/2)\}$ \\
\hline\\[-10pt]
$X_1$  &$\sigma_x$ & $\sigma_y$  & $-\mathrm{i}\sigma_z$ \\
\hline
\hline
\end{tabular}
\end{subtable}
\begin{subtable}[t]{0.19\textwidth}\centering
    \subcaption{$\overline{\Gamma X}$}
\begin{tabular}[b]{cc}
\hline
\hline\\[-10pt]
\multicolumn{1}{c}{} & $\{m_{10}|(1/2,1/2)\}$ \\
\hline\\[-10pt]
$DT_1$  &$\exp(\pi  u \mathrm{i})$ \\
$DT_2$  &$\exp( \pi [1+ u] \mathrm{i})$ \\
\hline
\hline
\end{tabular}
\end{subtable}
\begin{subtable}[t]{0.19\textwidth}\centering
    \subcaption{$\overline{\Gamma M}$}
\begin{tabular}[b]{cc}
\hline
\hline\\[-10pt]
\multicolumn{1}{c}{} & $\{m_{1\bar1}|(1/2,1/2)\}$ \\
\hline\\[-10pt]
$SM_1$  &$\exp(2\pi  u\mathrm{i})$ \\
$SM_2$  &$\exp(\pi [1+2u]\mathrm{i})$ \\
\hline
\hline
\end{tabular}
\end{subtable}
\begin{subtable}[t]{0.19\textwidth}\centering
    \subcaption{$\overline{X M}$}
\begin{tabular}[b]{cc}
\hline
\hline\\[-10pt]
\multicolumn{1}{c}{} & $\{m_{01}|(1/2,1/2)\}$ \\
\hline\\[-10pt]
$Y_1$  &$\exp(\pi   u\mathrm{i})$ \\
$Y_2$  &$\exp( \pi  [1+u]\mathrm{i})$ \\
\hline
\hline
\end{tabular}
\end{subtable}
\caption[$p4gm$ irreps.]{The relevant irreducible representations of $p4gm$ at the high-symmetry points $\Gamma=(0,0)$, $M=(1/2,1/2)$, and $X=(0,1/2)$ as well as along the lines $\overline{\Gamma X}=(0,u)$, $\overline{\Gamma M}=(u,u)$, and $\overline{XM}=(u,1/2)$.}
\label{tab:irreps-p4gm}
\end{table}

\noindent
The fragile roots in $p4gm$ \cite{Song2020} are given in Tab.~\ref{tab:roots-p4gm}; the elementary band representations can be retrieved from the \href{https://www.cryst.ehu.es/cgi-bin/cryst/programs/bandrep.pl?super=100&elementaryTR=Elementary%20TR}{Bilbao server}.
\begin{table}[htb]
    \begin{tabular}{c|l|c|l}
        \hline
        \#& root &\# of bands & type\\
        \hline\hline
        1 & $ 2 \Gamma_{2} + 2 \Gamma_{3} + M_{1}M_{3} + M_{2}M_{4} + 2 X_{1} $&4& conjugate pairs \& Chern\\
        \hline
        2 & $ 2 \Gamma_{1} + 2 \Gamma_{4} + M_{1}M_{3} + M_{2}M_{4} + 2 X_{1} $ &4&conjugate pairs \& Chern\\
        \hline
        3 & $ \Gamma_{1} + \Gamma_{4} + \Gamma_{5} + M_{1}M_{3} + M_{2}M_{4} + 2 X_{1} $&4&conjugate pairs \& 2D irrep \\
        \hline
        4 & $ \Gamma_{2} + \Gamma_{3} + \Gamma_{5} + M_{1}M_{3} + M_{2}M_{4} + 2 X_{1} $ &4&conjugate pairs \& 2D irrep\\
        \hline
        \end{tabular}
        \caption[$p4gm$ roots.]{Fragile roots and their types in wallpaper group $p4gm$.}
        \label{tab:roots-p4gm}
 \end{table}

\subsubsection{Bundling strategy}
We analyze the structure of the roots in the wallpaper group $p4gm$ shown in Tab.~\ref{tab:roots-p4gm}. As we will see, we need to invoke the vanishing Chern number argument to argue for the bundling of bands into fragile roots. We start with root $\#1$ 
\begin{equation}
    2 \Gamma_{2} + 2 \Gamma_{3} + M_{1}M_{3} + M_{2}M_{4} + 2 X_{1}.
\end{equation}
Let us see who we can link up with whom. For this, we consult the tables with the irreducible representations at $\Gamma$, $M$, $X$, and along the line $\overline{\Gamma X}=(u,u)$ in Tab.~\ref{tab:irreps-p4gm}.

Let us start with the conjugate pair $M_1M_3$ which both are odd under $\{m_{1\bar1}|(1/2,1/2)\}$ and, accordingly, are compatible with $SM_1$, which in turn forces us to connect $M_1M_3$ to $2\Gamma_3$. In the other pair $M_2M_4$, both transform evenly under $\{m_{1\bar1}|(1/2,1/2)\}$ and are therefore via $SM_2$ connected to $2\Gamma_2$. Along $\overline{XM}$, the irreps $Y_1$ and $Y_2$ turn the eigenvalues $\pm i$ of $\{m_{01}|(1/2,1/2)\}$ of $M_2M_4$ and $M_1M_3$ into $\pm 1$ of $X_1$, cf. Tab.~\ref{tab:irreps-p4gm}. Moreover, the non-symmorphic $\{m_{01}|(1/2,1/2)\}$ together with the time-reversal symmetry $T=K$, with $K$ denoting complex conjugation, gives rise to effective Kramers pair. In other words, we find ourselves in the typical situation of band-stickings along the zone boundary.

Hence, from compatibility relations alone, we would bundle 
\begin{equation}
    2 \Gamma_{2} + 2 \Gamma_{3} + M_{1}M_{3} + M_{2}M_{4} + 2 X_{1}=
    (2\Gamma_2+M_{2}M_{4}+X_1) \oplus (2\Gamma_3+M_{1}M_{3}+X_1)
\end{equation}
Using formula (\ref{eqn:chern4}) we conclude that we have
\begin{align}
    (2\Gamma_2+M_{2}M_{4}+X_1):\quad \mathrm{i}^C&=(-1)^2\times -\mathrm{i} \times \mathrm{i} \times -1 \times 1  = -1
    \quad \rightarrow\quad C = 2+4 n \;\;\mathrm{with}\;\; n\in \mathbb Z\\
    (2\Gamma_3+M_{1}M_{2}+X_1):\quad \mathrm{i}^C&=(-1)^2\times \mathrm{i} \times -\mathrm{i} \times -1 \times 1  = -1
    \quad \rightarrow\quad C = 2+4 n \;\;\mathrm{with}\;\; n\in \mathbb Z
\end{align}
From this we read that (i) for one of the two $n=0$ and for the other $n=-1$, such that by having (ii) a gap-closing between the two sets of bands, we deal with a {\em bundle of four bands} with zero Chern number.

We now consider the next root $\#2$
\begin{equation}
    2 \Gamma_{1} + 2 \Gamma_{4} + M_{1}M_{3} + M_{2}M_{4} + 2 X_{1}
\end{equation}
The same argument as above leads us to the conclusion that we can write 
\begin{equation}
    2 \Gamma_{1} + 2 \Gamma_{4} + M_{1}M_{3} + M_{2}M_{4} + 2 X_{1}=
    (2\Gamma_4+M_{2}M_{4}+X_1) \oplus (2\Gamma_1+M_{1}M_{3}+X_1)
\end{equation}
And again, through the argument that the Chern number has to be zero, we conclude that we have a bundle of four bands with band touchings off high-symmetry lines or points.
\begin{figure}[bt]
    \centering 
    \includegraphics{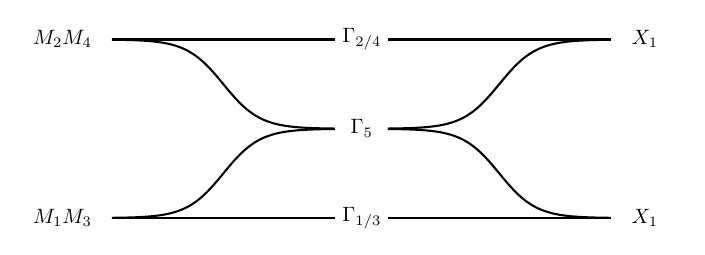}
    \caption[$p4gm$ bundle linkage.]{Linkage of bands for the fragile roots $\#3$ and $\#4$ of $p4gm$.}
    \label{fig:linkage-p4gm}
\end{figure}

For the two remaining roots of this group
\begin{eqnarray}
    \Gamma_{1}+ \Gamma_{4} + \Gamma_{5} + M_{1}M_{3} + M_{2}M_{4} + 2 X_{1} \\
 \Gamma_{2} + \Gamma_{3}+ \Gamma_{5} + M_{1}M_{3} + M_{2}M_{4} + 2 X_{1} 
\end{eqnarray}
we see that of the two $\{m_{1\bar1}|(1/2,1/2)\}$-even (odd) lines emerging from $M_{2}M_{4}$ ($M_1M_3$), which along $\overline{\Gamma M}$ turn  $\{m_{1\bar1}|(1/2,1/2)\}$-odd (even), one has to be absorbed by $\Gamma_{4/2}$ ($\Gamma_{1/3}$) and one by $\Gamma_5$, respectively. In other words: The four bands are linked together at high-symmetry points as in Fig.~\ref{fig:linkage-p4gm}.

Finally, the types of the roots of $p4gm$ are indicated in the last column of Tab.~\ref{tab:roots-p4gm}.

\subsubsection{Examples}

\begin{figure}[h!bt]
    \centering
    \includegraphics{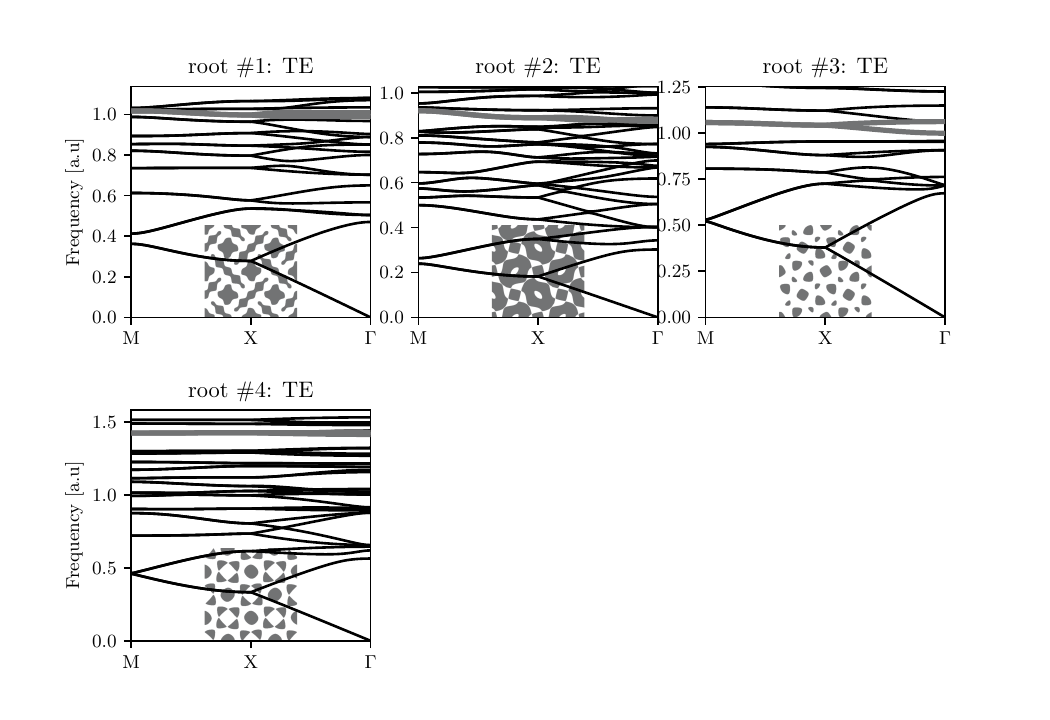}
    \caption[$p4gm$ TE examples.]{ One sample for each of the roots in $p4gm$ for TE photons.}
    \label{fig:p4gm-TE-examples}
\end{figure}

\raggedbottom
\newpage

\subsection{$p3$}

\subsubsection{Basic group properties}

\begin{figure}[!h]
    \centering
    \includegraphics{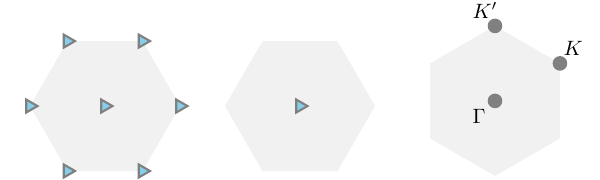}
    \caption[$p3$ unit cell.]{Left: Unit cell with the full symmetry of $p3$. Middle: General positions of $p3$. Right: Brillouin zone with the high-symmetry points and lines indicated.}
    \label{fig:p3-uc}
\end{figure}

The group $p3$ describes a hexagonal lattice with lattice vectors $\vec a_1=(1,0)$ and $\vec a_2=(-\sqrt{3}/2,1/2)$. The corresponding space group is $P3$, (\#143) constrained to the $x$-$y$ plane. The group $p3$ contains the following group elements [\href{https://www.cryst.ehu.es/cgi-bin/plane/programs/nph-plane_getgen?what=gp&gnum=13}{retrieve from Bilbao server}]
\begin{align}
    &\{ 1|n_1 {\vec a}_1+n_2{\vec a}_2\}, \quad \mathrm{with} \quad n_1,n_2\, \in\, \mathbb Z\\
    &\{3^+|(0,0)\},\; \{ 3^-|(0,0)\}.
\end{align}

The relevant members of the little groups of the high-symmetry points and lines and their irreps are given in Tab.~\ref{tab:irreps-p3}. The full set can be obtained from the \href{https://www.cryst.ehu.es/cgi-bin/cryst/programs/representations_vec.pl?tipogroup=spg&super=143&complex=Submit}{Bilbao Server}.
\begin{table}[htb]
    \begin{subtable}[t]{0.15\textwidth}\centering
        \subcaption{$\Omega=\Gamma,\,K$}
\begin{tabular}[b]{cc}
    \hline
    \hline\\[-10pt]
    \multicolumn{1}{c}{} &$\{3^+|(0,0)\}$ \\
    \hline\\[-10pt]
    $\Omega_1$  &$1$\\
    $\Omega_2$  & $z^2$ \\
    $\Omega_3$  & $\bar z^2$ \\
    \hline
    \hline
    \vspace{6pt}
\end{tabular}
\end{subtable}
    \begin{subtable}[t]{0.15\textwidth}\centering
        \subcaption{$K'$}
\begin{tabular}[b]{cc}
    \hline
    \hline\\[-10pt]
    \multicolumn{1}{c}{} &$\{3^+|(0,0)\}$ \\
    \hline\\[-10pt]
    $KA_1$  &$1$\\
    $KA_2$  & $\bar z^2$ \\
    $KA_3$  & $z^2$ \\
    \hline
    \hline
    \vspace{6pt}
\end{tabular}
\end{subtable}
\caption[$p3$ irreps.]{The relevant irreducible representations of $p3$ at the high-symmetry points $\Gamma=(0,0)$, $K=(1/3,1/3)$, and $K'=(-1/3,-1/3)$. Note that $z=\exp(\mathrm{i}\pi/3)$.}
\label{tab:irreps-p3}
\end{table}

\noindent
The fragile roots in $p3$ \cite{Song2020} are given in Tab.~\ref{tab:roots-p3}; the elementary band representations can be retrieved from the \href{https://www.cryst.ehu.es/cgi-bin/cryst/programs/bandrep.pl?super=143&elementaryTR=Elementary%20TR}{Bilbao server}.
\begin{table}[htb]
    \begin{tabular}{c|l|c|l}
        \hline
        \#& root &\# of bands & type\\
        \hline\hline
        1 & $ \Gamma_{2}\Gamma_{3} + 2 K_{2} + 2 KA_{2} + 2 M_{1} $ & 2& conjugate pairs \\
\hline
2 & $ \Gamma_{2}\Gamma_{3} + 2 K_{1} + 2 KA_{1} + 2 M_{1} $ & 2& conjugate pairs\\
\hline
3 & $ \Gamma_{2}\Gamma_{3} + 2 K_{3} + 2 KA_{3} + 2 M_{1} $ & 2& conjugate pairs\\
        \hline
        \end{tabular}
        \caption[$p3$ roots.]{Fragile roots and their types in wallpaper group $p3$.}
        \label{tab:roots-p3}
 \end{table}

 \subsubsection{Bundling strategy}

 In all fragile roots of Tab.~\ref{tab:irreps-p3}, only the conjugate pairs $\Gamma_2\Gamma_3$ appear at the $\Gamma$-point. Hence, a bundling according to these is sufficient as shown in the last column of Tab.~\ref{tab:roots-p3}.

 \subsubsection{Examples}

 \begin{figure}[h!bt]
    \centering
    \includegraphics{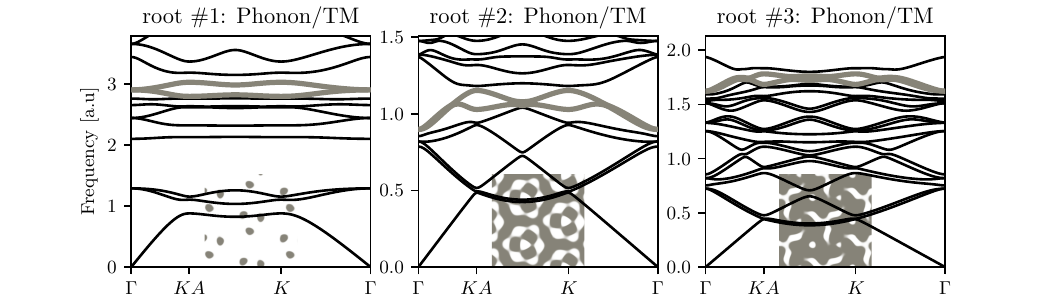}
    \caption[$p3$ phonons/TM examples.]{ One sample for each of the roots in $p3$ for phonons and TM photons.}
    \label{fig:p3-TM-examples}
\end{figure}

\begin{figure}[h!bt]
    \centering
    \includegraphics{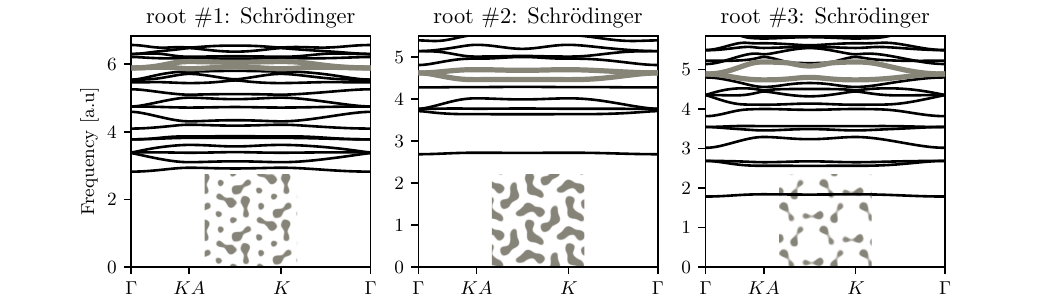}
    \caption[$p3$ Schrödinger examples.]{One sample for each of the roots in $p3$ for systems described by the Schrödinger equation.}
    \label{fig:p3-Q-examples}
\end{figure}

\begin{figure}[h!bt]
    \centering
    \includegraphics{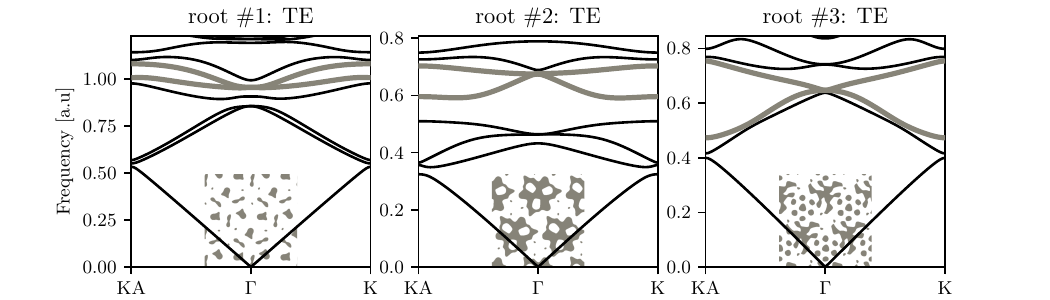}
    \caption[$p3$ TE examples.]{ One sample for each of the roots in $p3$ for TE photons.}
    \label{fig:p3-TE-examples}
\end{figure}

\raggedbottom
\newpage

\subsection{$p3m1$}

\subsubsection{Basic group properties}

\begin{figure}[!h]
    \centering
    \includegraphics{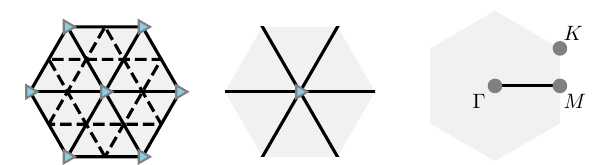}
    \caption[$p3m1$ unit cell.]{Left: Unit cell with the full symmetry of $p3m1$. Middle: General positions of $p3m1$. Right: Brillouin zone with the high-symmetry points and lines indicated.}
    \label{fig:p3m1-uc}
\end{figure}

The group $p3m1$ describes a hexagonal lattice with lattice vectors $\vec a_1=(1,0)$ and $\vec a_2=(-\sqrt{3}/2,1/2)$. The corresponding space group is $P3m1$, (\#156) constrained to the $x$-$y$ plane. The group $p3m1$ contains the following group elements [\href{https://www.cryst.ehu.es/cgi-bin/plane/programs/nph-plane_getgen?what=gp&gnum=14}{retrieve from Bilbao server}]
\begin{align}
    &\{ 1|n_1 {\vec a}_1+n_2{\vec a}_2\}, \quad \mathrm{with} \quad n_1,n_2\, \in\, \mathbb Z\\
    & \{3^+|(0,0)\},\; \{ 3^-|(0,0)\}, \\
    & \{m_{11}|(0,0)\},\;\{m_{10}|(0,0)\},\;\{m_{01}|(0,0)\}.
\end{align}

The relevant members of the little groups of the high-symmetry points and lines and their irreps are given in Tab.~\ref{tab:irreps-p3m1}. The full set can be obtained from the \href{https://www.cryst.ehu.es/cgi-bin/cryst/programs/representations_vec.pl?tipogroup=spg&super=156&complex=Submit}{Bilbao Server}.
\begin{table}[htb]
    \begin{subtable}[t]{0.5\textwidth}\centering
        \subcaption{$\Gamma$}
\begin{tabular}[b]{ccccc}
    \hline
    \hline\\[-10pt]
    \multicolumn{1}{c}{} &$\{3^+|(0,0)\}$ & $\{m_{11}|(0,0)\}$&$\{m_{10}|(0,0)\}$&$\{m_{01}|(0,0)\}$\\
    \hline\\[-10pt]
    $\Gamma_1$  &$\phantom{-}1$&$\phantom{-}1$&$\phantom{-}1$&$\phantom{-}1$\\
    $\Gamma_2$  & $\phantom{-}1$&$-1$&$-1$&$-1$ \\
    $\Gamma_3$  & $\phantom{-}\begin{pmatrix}
        z^2&0\\0&\bar z^2
    \end{pmatrix}$&$\phantom{-}\sigma_x$&$\begin{pmatrix}
        0&\bar z^2\\z^2&0
    \end{pmatrix}$ & $\begin{pmatrix}
        0&z^2\\ \bar z^2&0
    \end{pmatrix}$\\[8pt]
    \hline
    \hline
    \vspace{6pt}
\end{tabular}
\end{subtable}
\begin{subtable}[t]{0.15\textwidth}\centering
    \subcaption{$K$}
\begin{tabular}[b]{cc}
\hline
\hline\\[-10pt]
\multicolumn{1}{c}{} &$\{3^+|(0,0)\}$\\
\hline\\[-10pt]
$K_1$  & $1$ \\
$K_2$  & $z^2$\\
$K_3$  & $\bar z^2$\\
\hline
\hline
\end{tabular}
\end{subtable}
\begin{subtable}[t]{0.15\textwidth}\centering
    \subcaption{$M$}
\begin{tabular}[b]{cc}
\hline
\hline\\[-10pt]
\multicolumn{1}{c}{} &$\{m_{01}|(0,0)\}$ \\
\hline\\[-10pt]
$M_1$  & $\phantom{-}1$ \\
$M_2$  & $-1$ \\
\hline
\hline
\end{tabular}
\end{subtable}
\begin{subtable}[t]{0.15\textwidth}\centering
    \subcaption{$\overline{\Gamma M}$}
\begin{tabular}[b]{cc}
\hline
\hline\\[-10pt]
\multicolumn{1}{c}{} &$\{m_{01}|(0,0)\}$ \\
\hline\\[-10pt]
$SM_1$  & $\phantom{-}1$ \\
$SM_2$  & $-1$ \\
\hline
\hline
\end{tabular}
\end{subtable}
\caption[$p3m1$ irreps.]{The relevant irreducible representations of $p3m1$ at the high-symmetry points $\Gamma=(0,0)$, $M=(1/2,0)$, and $K=(1/3,1/3)$. Note that $z=\exp(\mathrm{i}\pi/3)$.}
\label{tab:irreps-p3m1}
\end{table}

\noindent
The fragile roots in $p3m1$ \cite{Song2020} are given in Tab.~\ref{tab:roots-p3m1}; the elementary band representations can be retrieved from the \href{https://www.cryst.ehu.es/cgi-bin/cryst/programs/bandrep.pl?super=156&elementaryTR=Elementary%20TR}{Bilbao server}.
\begin{table}[htb]
    \begin{tabular}{c|l|c|l}
        \hline
        \#& root &\# of bands & type\\
        \hline\hline
        1 & $ \Gamma_{3} + 2 K_{2} + M_{1} + M_{2} $ &2& 2D irrep\\
\hline
2 & $ \Gamma_{3} + 2 K_{1} + M_{1} + M_{2} $&2& 2D irrep \\
\hline
3 & $ \Gamma_{3} + 2 K_{3} + M_{1} + M_{2} $ &2& 2D irrep\\
        \hline
        \end{tabular}
        \caption[$p3m1$ roots.]{Fragile roots and their types in wallpaper group $p3m1$.}
        \label{tab:roots-p3m1}
 \end{table}

\subsubsection{Bundling strategy}

In all three roots, the two-dimensional irrep $\Gamma_3$ is appearing. Hence, the bands are glued together at the origin of the Brillouin zone as indicated in the last column of Tab.~\ref{tab:roots-p3m1}

\subsubsection{Examples}

\begin{figure}[h!bt]
    \centering
    \includegraphics{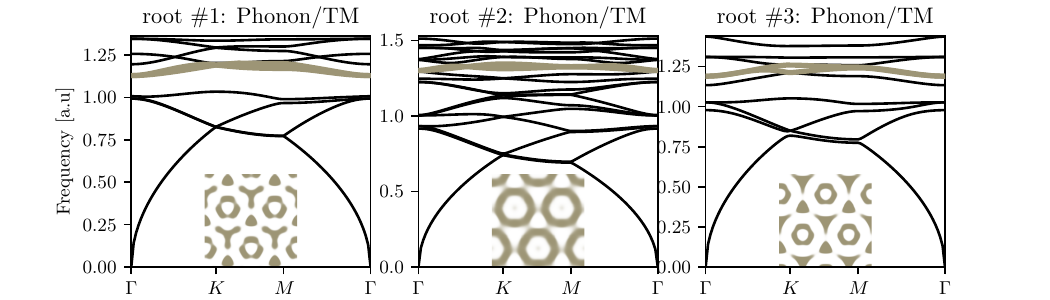}
    \caption[$p3m1$ phonons/TM examples.]{ One sample for each of the roots in $p3m1$ for phonons and TM photons.}
    \label{fig:p3m1-TM-examples}
\end{figure}

\begin{figure}[h!bt]
    \centering
    \includegraphics{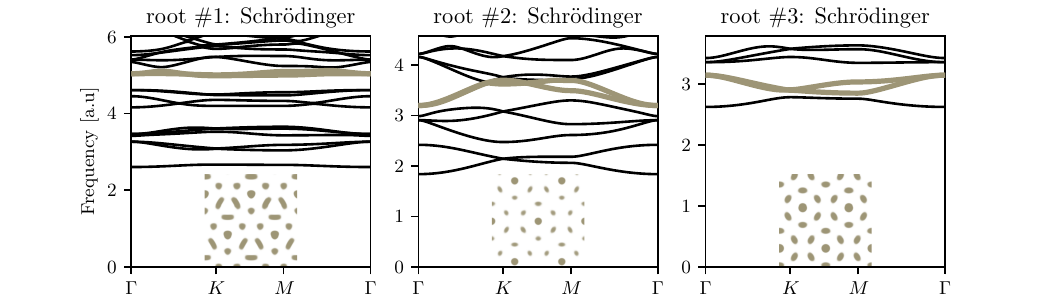}
    \caption[$p3m1$ Schrödinger examples.]{One sample for each of the roots in $p3m1$ for systems described by the Schrödinger equation.}
    \label{fig:p3m1-Q-examples}
\end{figure}

\begin{figure}[h!bt]
    \centering
    \includegraphics{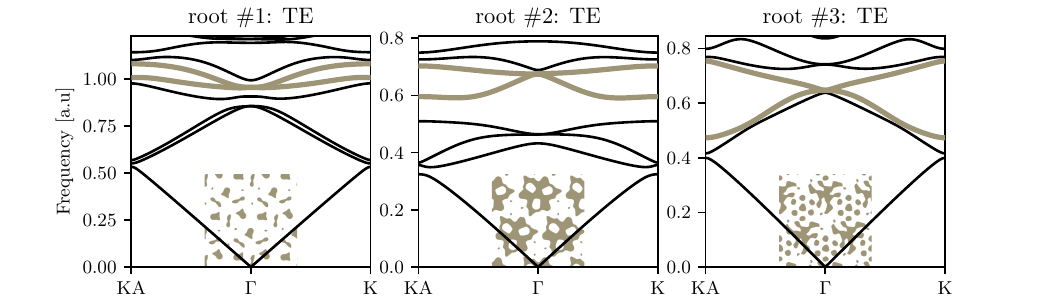}
    \caption[$p3m1$ TE examples.]{ One sample for each of the roots in $p3m1$ for TE photons.}
    \label{fig:p3m1-TE-examples}
\end{figure}

\raggedbottom
\newpage

\subsection{$p31m$}

\subsubsection{Basic group properties}

\begin{figure}[!h]
    \centering
    \includegraphics{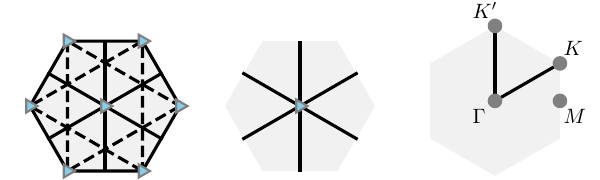}
    \caption[$p31m$ unit cell.]{Left: Unit cell with the full symmetry of $p31m$. Middle: General positions of $p31m$. Right: Brillouin zone with the high-symmetry points and lines indicated.}
    \label{fig:p6-uc}
\end{figure}

The group $p31m$ describes a hexagonal lattice with lattice vectors $\vec a_1=(1,0)$ and $\vec a_2=(-\sqrt{3}/2,1/2)$. The corresponding space group is $P31m$, (\#157) constrained to the $x$-$y$ plane. The group $p31m$ contains the following group elements [\href{https://www.cryst.ehu.es/cgi-bin/plane/programs/nph-plane_getgen?what=gp&gnum=15}{retrieve from Bilbao server}]
\begin{align}
    &\{ 1|n_1 {\vec a}_1+n_2{\vec a}_2\}, \quad \mathrm{with} \quad n_1,n_2\, \in\, \mathbb Z\\
    & \{3^+|(0,0)\},\; \{ 3^-|(0,0)\}, \\
    & \{m_{1\bar 1}|(0,0)\},\;\{m_{12}|(0,0)\},\;\{m_{21}|(0,0)\}.
\end{align}

The relevant members of the little groups of the high-symmetry points and lines and their irreps are given in Tab.~\ref{tab:irreps-p31m}. The full set can be obtained from the \href{https://www.cryst.ehu.es/cgi-bin/cryst/programs/representations_vec.pl?tipogroup=spg&super=157&complex=Submit}{Bilbao Server}.
\begin{table}[htb]
    \begin{subtable}[t]{0.45\textwidth}\centering
        \subcaption{$\Omega=\Gamma,\,K$}
\begin{tabular}[b]{ccccc}
    \hline
    \hline\\[-10pt]
    \multicolumn{1}{c}{} &$\{3^+|(0,0)\}$ & $\{m_{1\bar 1}|(0,0)\}$&$\{m_{12}|(0,0)\}$&$\{m_{2  1}|(0,0)\}$\\
    \hline\\[-10pt]
    $\Omega_1$  &$\phantom{-}1$&$\phantom{-}1$&$\phantom{-}1$&$\phantom{-}1$\\
    $\Omega_2$  & $\phantom{-}1$&$-1$&$-1$&$-1$ \\
    $\Omega_3$  & $\phantom{-}\begin{pmatrix}
        z^2&0\\0&\bar z^2
    \end{pmatrix}$&$\phantom{-}\sigma_x$&$\begin{pmatrix}
        0&\bar z^2\\z^2&0
    \end{pmatrix}$ & $\begin{pmatrix}
        0&z^2\\ \bar z^2&0
    \end{pmatrix}$\\[8pt]
    \hline
    \hline
    \vspace{6pt}
\end{tabular}
\end{subtable}
\begin{subtable}[t]{0.45\textwidth}\centering
    \subcaption{$K'$}
\begin{tabular}[b]{ccccc}
\hline
\hline\\[-10pt]
\multicolumn{1}{c}{} &$\{3^+|(0,0)\}$ & $\{m_{1\bar 1}|(0,0)\}$&$\{m_{12}|(0,0)\}$&$\{m_{2  1}|(0,0)\}$\\
\hline\\[-10pt]
$KA_1$  &$\phantom{-}1$&$\phantom{-}1$&$\phantom{-}1$&$\phantom{-}1$\\
$KA_2$  & $\phantom{-}1$&$-1$&$-1$&$-1$ \\
$KA_3$  & $\phantom{-}\begin{pmatrix}
    \bar z^2&0\\0& z^2
\end{pmatrix}$&$\phantom{-}\sigma_x$&$\begin{pmatrix}
    0& z^2\\ \bar z^2&0
\end{pmatrix}$ & $\begin{pmatrix}
    0&\bar z^2\\  z^2&0
\end{pmatrix}$\\[8pt]
\hline
\hline
\vspace{6pt}
\end{tabular}
\end{subtable}

\begin{subtable}[t]{0.45\textwidth}\centering
    \subcaption{$M$}
\begin{tabular}[b]{cc}
\hline
\hline\\[-10pt]
\multicolumn{1}{c}{} &$\{m_{21}|(0,0)\}$ \\
\hline\\[-10pt]
$M_1$  & $\phantom{-}1$ \\
$M_2$  & $-1$ \\
\hline
\hline
\end{tabular}
\end{subtable}
\begin{subtable}[t]{0.45\textwidth}\centering
    \subcaption{$\overline{K' \Gamma K}$}
\begin{tabular}[b]{cc}
\hline
\hline\\[-10pt]
\multicolumn{1}{c}{} &$\{m_{1\bar 1}|(0,0)\}$ \\
\hline\\[-10pt]
$LD_1$  & $\phantom{-}1$ \\
$LD_2$  & $-1$ \\
\hline
\hline
\end{tabular}
\end{subtable}
\caption[$p31m$ irreps.]{The relevant irreducible representations of $p31m$ at the high-symmetry points $\Gamma=(0,0)$, $M=(1/2,0)$, $K=(1/3,1/3)$ and $K'=(-1/3,-1/3)$ as wall as on the line $\overline{K'\Gamma K}$. Note that $z=\exp(\mathrm{i}\pi/3)$.}
\label{tab:irreps-p31m}
\end{table}

\noindent
The fragile root in $p31m$ \cite{Song2020} is given in Tab.~\ref{tab:roots-p31m}; the elementary band representations can be retrieved from the \href{https://www.cryst.ehu.es/cgi-bin/cryst/programs/bandrep.pl?super=157&elementaryTR=Elementary%20TR}{Bilbao server}.
\begin{table}[htb]
    \begin{tabular}{c|l|c|l}
        \hline
        \#& root &\# of bands & type\\
        \hline\hline
        1 & $ \Gamma_{3} + K_{1} + K_{2} + KA_{1} + KA_{2} + M_{1} + M_{2} $ &2& 2D irrep\\
        \hline
        \end{tabular}
        \caption[$p31m$ root.]{The fragile roots and its type in wallpaper group $p31m$.}
        \label{tab:roots-p31m}
 \end{table}

 \subsubsection{Bundling strategy}
 
 The two-dimensional irrep $\Gamma_3$ is appearing in the only fragile root of $p31m$. Therefore, bundling can be trivially achieved using this double degeneracy as shown in the last column of Tab.~\ref{tab:roots-p31m}.

 \subsubsection{Examples}

 \begin{figure}[h!bt]
    \centering
    \includegraphics{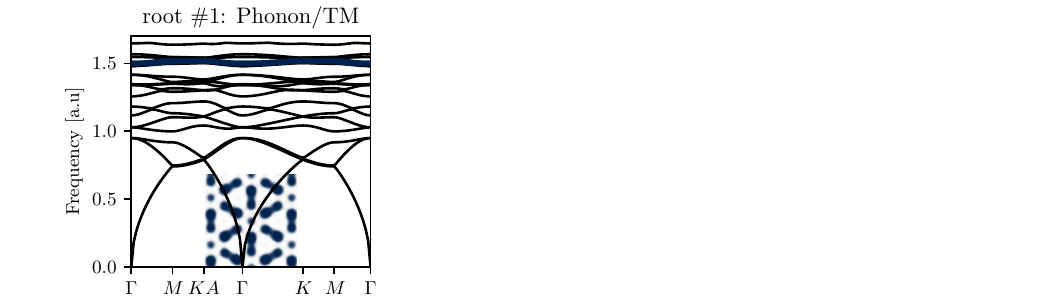}
    \caption[$p31m$ phonons/TM examples.]{ One sample for each of the roots in $p31m$ for phonons and TM photons.}
    \label{fig:p31m-TM-examples}
\end{figure}

\begin{figure}[h!bt]
    \centering
    \includegraphics{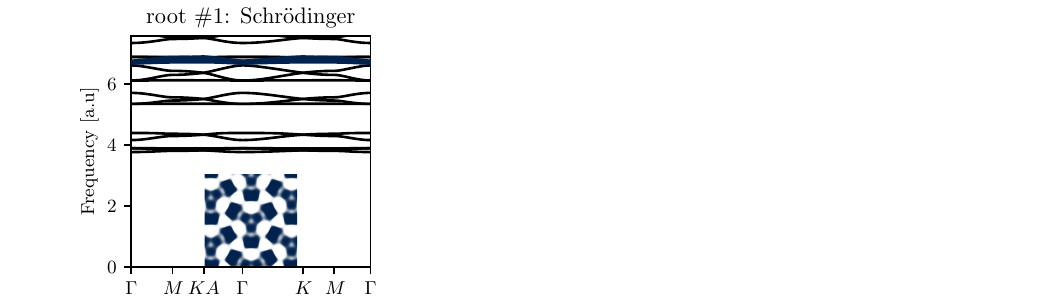}
    \caption[$p31m$ Schrödinger examples.]{One sample for each of the roots in $p31m$ for systems described by the Schrödinger equation.}
    \label{fig:p31m-Q-examples}
\end{figure}

\begin{figure}[h!bt]
    \centering
    \includegraphics{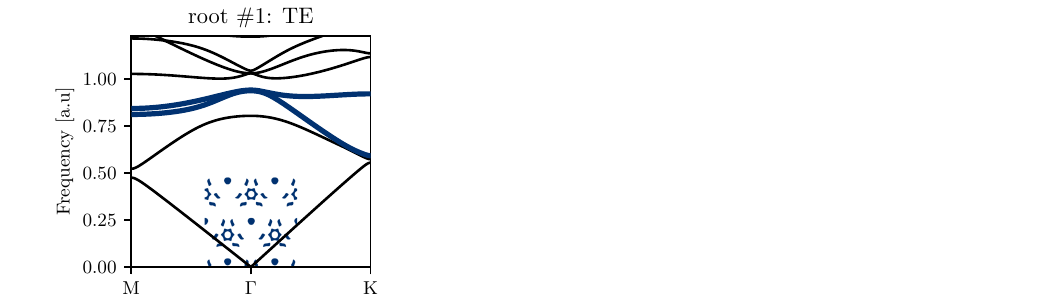}
    \caption[$p31m$ TE examples.]{ One sample for each of the roots in $p31m$ for TE photons.}
    \label{fig:p31m-TE-examples}
\end{figure}

\raggedbottom
\newpage

\subsection{$p6$}

\subsubsection{Basic group properties}

\begin{figure}[!h]
    \centering
    \includegraphics{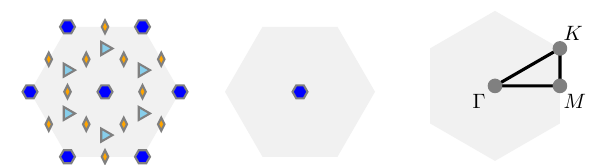}
    \caption[$p6$ unit cell.]{Left: Unit cell with the full symmetry of $p6$. Middle: General positions of $p6$. Right: Brillouin zone with the high-symmetry points and lines indicated.}
    \label{fig:p6-uc}
\end{figure}

The group $p6$ describes a hexagonal lattice with lattice vectors $\vec a_1=(1,0)$ and $\vec a_2=(-\sqrt{3}/2,1/2)$. The corresponding space group is $P6$, (\#168) constrained to the $x$-$y$ plane. The group $p6mm$ contains the following group elements [\href{https://www.cryst.ehu.es/cgi-bin/plane/programs/nph-plane_getgen?what=gp&gnum=16}{retrieve from Bilbao server}]
\begin{align}
    &\{ 1|n_1 {\vec a}_1+n_2{\vec a}_2\}, \quad \mathrm{with} \quad n_1,n_2\, \in\, \mathbb Z\\
    &\{ 2|(0,0)\},\; \{3^+|(0,0)\},\; \{ 3^-|(0,0)\},\; \{6^+|(0,0)\},\; \{ 6^-|(0,0)\}.
\end{align}

The relevant members of the little groups of the high-symmetry points and lines and their irreps are given in Tab.~\ref{tab:irreps-p6mm}. The full set can be obtained from the \href{https://www.cryst.ehu.es/cgi-bin/cryst/programs/representations_vec.pl?tipogroup=spg&super=168&complex=Submit}{Bilbao Server}.
\begin{table}[htb]
    \begin{subtable}[t]{0.15\textwidth}\centering
        \subcaption{$\Gamma$}
\begin{tabular}[b]{cc}
    \hline
    \hline\\[-10pt]
    \multicolumn{1}{c}{} &$\{6^+|(0,0)\}$ \\
    \hline\\[-10pt]
    $\Gamma_1$  &$\phantom{-}1$\\
    $\Gamma_2$  & $-1$ \\
    $\Gamma_3$  & $\phantom{-}\bar z^2$ \\
    $\Gamma_4$  & $\phantom{-}z$ \\
    $\Gamma_5$  & $\phantom{-}z^2$\\
    $\Gamma_6$  & $\phantom{-}\bar z$\\
    \hline
    \hline
    \vspace{6pt}
\end{tabular}
\end{subtable}
\begin{subtable}[t]{0.15\textwidth}\centering
    \subcaption{$K$}
\begin{tabular}[b]{cc}
\hline
\hline\\[-10pt]
\multicolumn{1}{c}{} &$\{3^+|(0,0)\}$\\
\hline\\[-10pt]
$K_1$  & $1$ \\
$K_2$  & $z^2$\\
$K_3$  & $\bar z^2$\\
\hline
\hline
\end{tabular}
\end{subtable}
\begin{subtable}[t]{0.15\textwidth}\centering
    \subcaption{$M$}
\begin{tabular}[b]{cc}
\hline
\hline\\[-10pt]
\multicolumn{1}{c}{} &$\{2|(0,0)\}$ \\
\hline\\[-10pt]
$M_1$  & $\phantom{-}1$ \\
$M_2$  & $-1$ \\
\hline
\hline
\end{tabular}
\end{subtable}
\caption[$p6$ irreps.]{The relevant irreducible representations of $p6$ at the high-symmetry points $\Gamma=(0,0)$, $M=(1/2,0)$, and $K=(1/3,1/3)$. Note that $z=\exp(\mathrm{i}\pi/3)$.}
\label{tab:irreps-p6}
\end{table}

\noindent
The fragile roots in $p6$ \cite{Song2020} are given in Tab.~\ref{tab:roots-p6}; the elementary band representations can be retrieved from the \href{https://www.cryst.ehu.es/cgi-bin/cryst/programs/bandrep.pl?super=168&elementaryTR=Elementary%20TR}{Bilbao server}.
\begin{table}[htb]
    \begin{tabular}{c|l|c|l}
        \hline
        \#& root &\# of bands & type\\
        \hline\hline
        1 & $ 2 \Gamma_{2} + 2 K_{1} + 2 M_{1} $ & 2 & Chern \\
        \hline
        2 & $ 2 \Gamma_{2} + K_{2}K_{3} + 2 M_{1} $ & 2 & conjugate pairs\\
        \hline
        3 & $ 2 \Gamma_{2} + K_{2}K_{3} + 2 M_{2} $ & 2 & conjugate pairs\\
        \hline
        4 & $ 2 \Gamma_{1} + 2 K_{1} + 2 M_{2} $  & 2 & Chern\\
        \hline
        5 & $ 2 \Gamma_{1} + K_{2}K_{3} + 2 M_{2} $ & 2 & conjugate pairs\\
        \hline
        6 & $ 2 \Gamma_{1} + K_{2}K_{3} + 2 M_{1} $ & 2 & conjugate pairs\\
        \hline
        7 & $ \Gamma_{4}\Gamma_{6} + 2 K_{1} + 2 M_{1} $ & 2 & conjugate pairs\\
        \hline
        8 & $ \Gamma_{3}\Gamma_{5} + 2 K_{1} + 2 M_{1} $ & 2 & conjugate pairs\\
        \hline
        9 & $ \Gamma_{3}\Gamma_{5} + K_{2}K_{3} + 2 M_{2} $ & 2 & conjugate pairs\\
        \hline
        10 & $ \Gamma_{4}\Gamma_{6} + K_{2}K_{3} + 2 M_{1} $ & 2 & conjugate pairs\\
        \hline
        11 & $ \Gamma_{4}\Gamma_{6} + 2 K_{1} + 2 M_{2} $ & 2 & conjugate pairs\\
        \hline
        12 & $ \Gamma_{3}\Gamma_{5} + 2 K_{1} + 2 M_{2} $ & 2 & conjugate pairs\\
        \hline
        \end{tabular}
        \caption[$p6$ roots.]{Fragile roots and their types in wallpaper group $p6$.}
        \label{tab:roots-p6}
 \end{table}

 \subsubsection{Bundling strategies}

 Let us start with the first two roots \#1 and \#4, where only one dimensional irreps with real characters are involved. Moreover, no mirrors help us in further constraining the bundling. This means, in principle, we could write for \#1
 \begin{equation}
    (\Gamma_2+K_1+M_1)\,\oplus\,(\Gamma_2+K_1+M_1).
 \end{equation}
We now invoke the usual Chern argument by writing \cite{Fang2012}
\begin{equation}
    e^{\mathrm{i}\pi C/3} = \prod_{i\,\in\,\mathrm{occ.}}(-1)^F \eta_i(\Gamma)\vartheta_i(K)\zeta_i(M),
\end{equation}
where $F=2S$, with $S$ the total spin of the particles, $\eta$, $\vartheta$, and $\zeta$ are the eigenvalues of $\{6^+|(0,0)\}$ and $\{3^+|(0,0)\}$, and $\{2|(0,0)\}$ respectively. In the following we set $F=0$. For the bands above, this means, the Chern number C of these bands is given by 
\begin{equation}
    C=3+6n\quad\mathrm{with}\quad n \,\in\, \mathbb Z.
\end{equation}
The only way this is compatible with time reversal symmetry is by having a band touching between the sets of bands. The same holds for root \#4.

For all remaining roots, there is either at $K$ or at $\Gamma$ a pair of conjugate irreps which glues the bands together. Finally, the types of the roots of $p6$ are indicated in the last column of Tab.~\ref{tab:roots-p6}.

\subsubsection{Examples}

\begin{figure}[h!bt]
    \centering
    \includegraphics{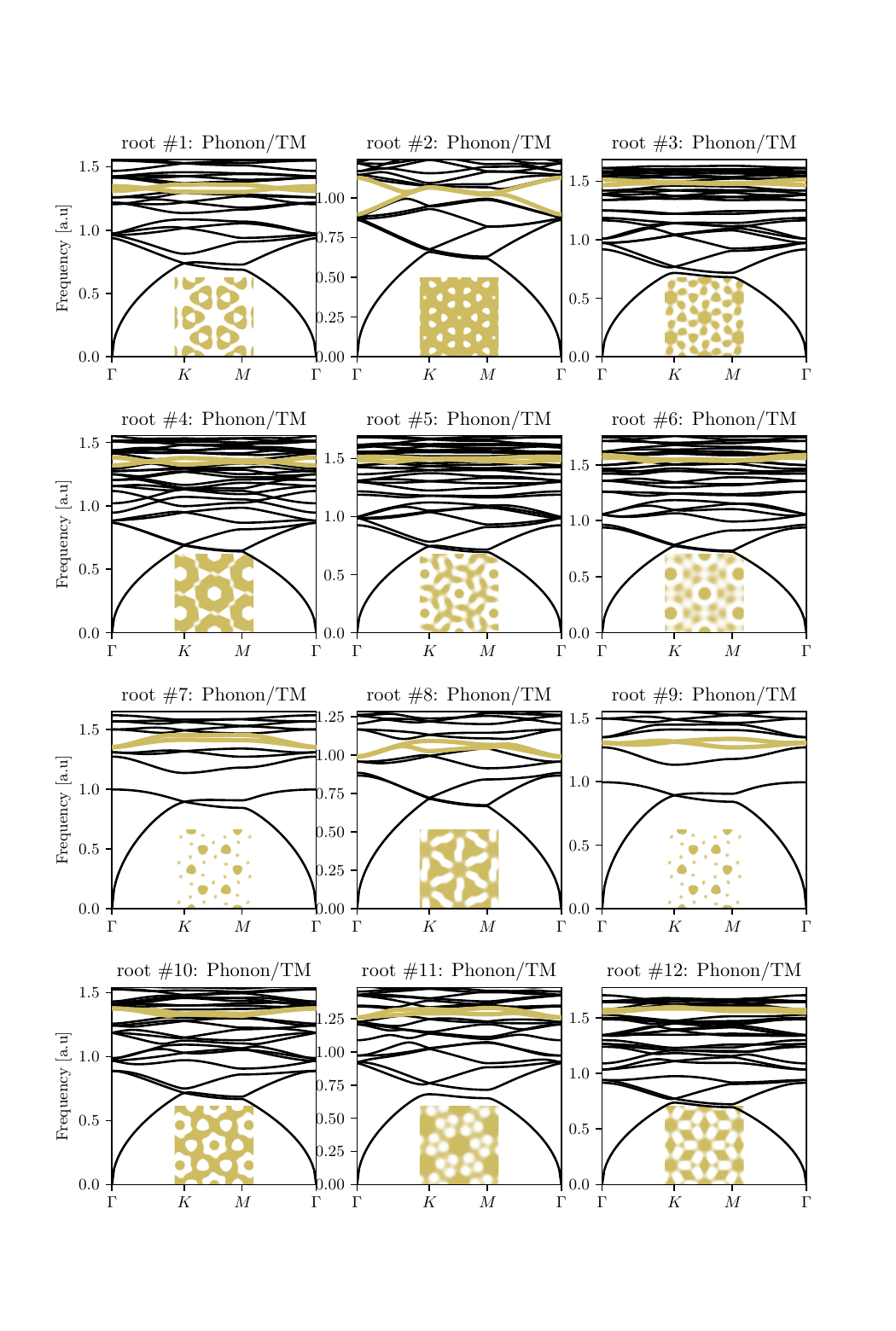}
    \caption[$p6$ phonons/TM examples.]{ One sample for each of the roots in $p6$ for phonons and TM photons.}
    \label{fig:p6-TM-examples}
\end{figure}

\begin{figure}[h!bt]
    \centering
    \includegraphics{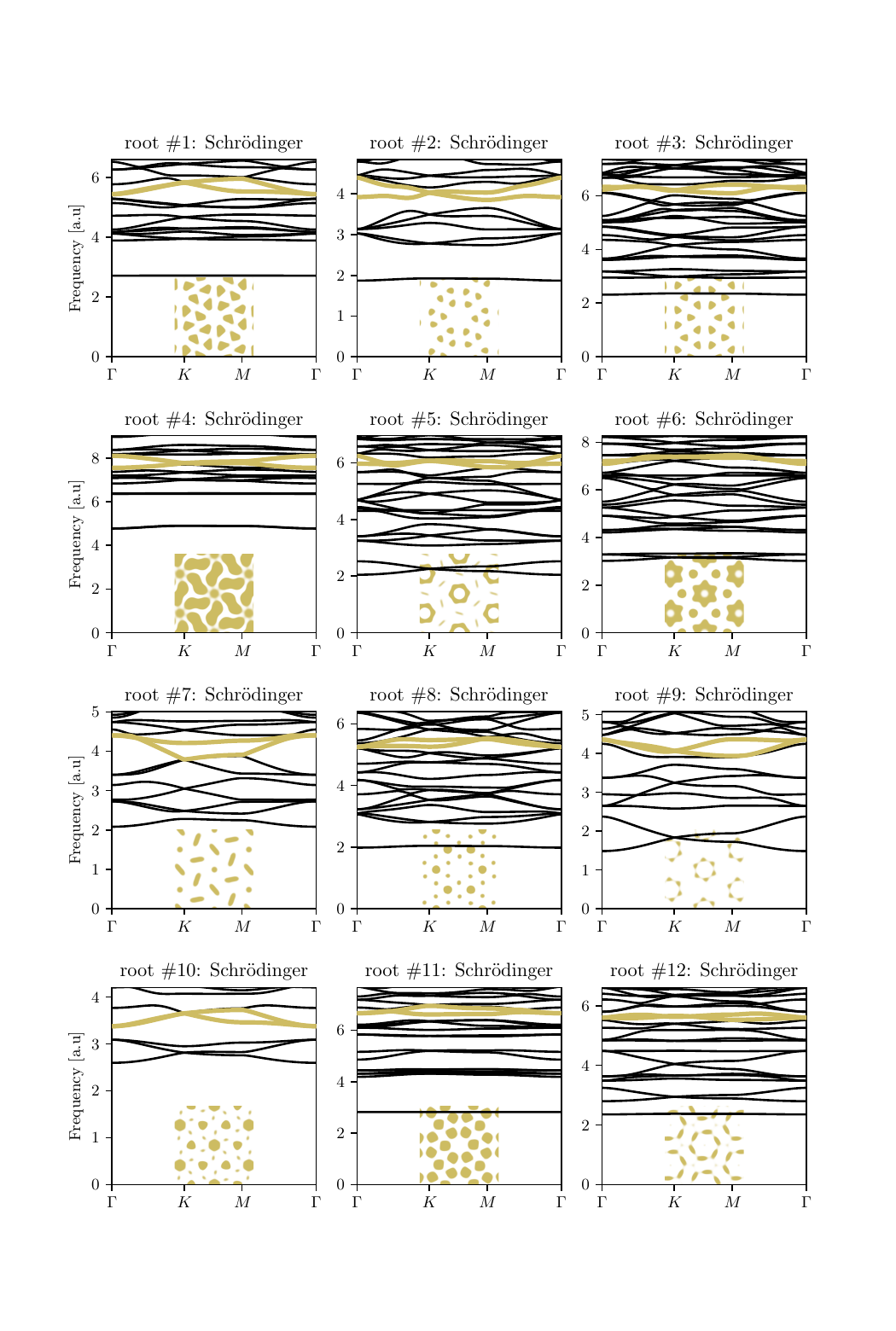}
    \caption[$p6$ Schrödinger examples.]{One sample for each of the roots in $p6$ for systems described by the Schrödinger equation.}
    \label{fig:p6-Q-examples}
\end{figure}

\begin{figure}[h!bt]
    \centering
    \includegraphics{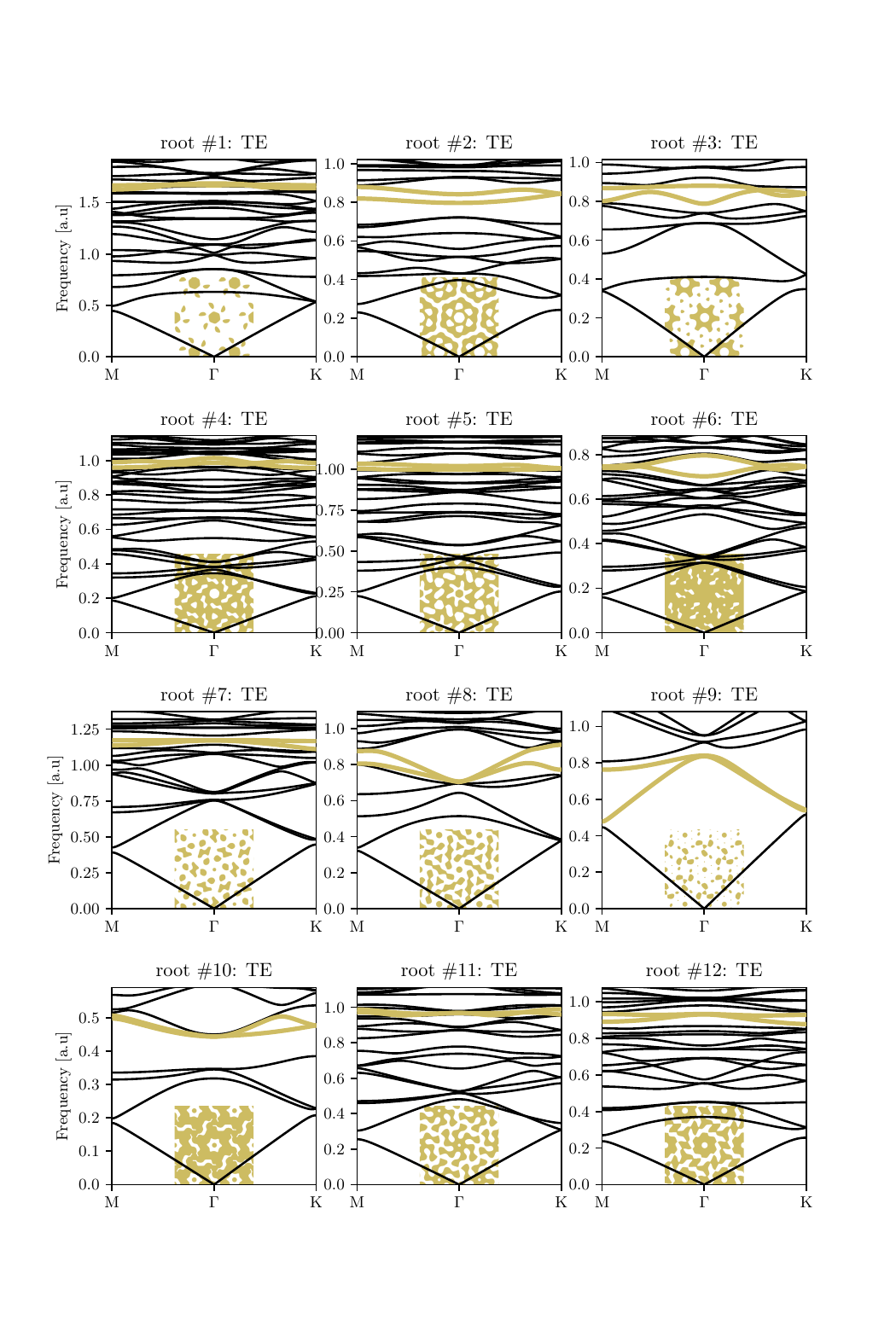}
    \caption[$p6$ TE examples.]{ One sample for each of the roots in $p6$ for TE photons.}
    \label{fig:p6-TE-examples}
\end{figure}

\raggedbottom
\newpage
\subsection{$p6mm$}

\subsubsection{Basic group properties}

\begin{figure}[!h]
    \centering
    \includegraphics{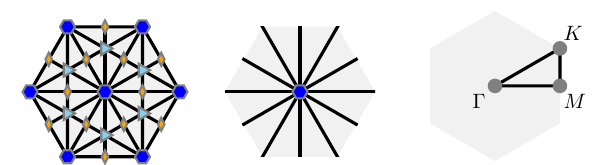}
    \caption[$p6mm$ unit cell.]{Left: Unit cell with the full symmetry of $p6mm$. (There are additional glide planes parallel to the mirror planes, always halfway in between the mirror planes. They are not shown for simplicity). Middle: General positions of $p6mm$. Right: Brillouin zone with the high-symmetry points and lines indicated.}
    \label{fig:p6mm-uc}
\end{figure}

The group $p6mm$ describes a hexagonal lattice with lattice vectors $\vec a_1=(1,0)$ and $\vec a_2=(-\sqrt{3}/2,1/2)$. The corresponding space group is $P6mm$, (\#183) constrained to the $x$-$y$ plane. The group $p6mm$ contains the following group elements [\href{https://www.cryst.ehu.es/cgi-bin/plane/programs/nph-plane_getgen?what=gp&gnum=17}{retrieve from Bilbao server}]
\begin{align}
    &\{ 1|n_1 {\vec a}_1+n_2{\vec a}_2\}, \quad \mathrm{with} \quad n_1,n_2\, \in\, \mathbb Z\\
    &\{ 2|(0,0)\},\; \{3^+|(0,0)\},\; \{ 3^-|(0,0)\},\; \{6^+|(0,0)\},\; \{ 6^-|(0,0)\}\\
    &\{m_{11} |(0,0)\},\; \{m_{10} |(0,0)\},\; \{m_{01} |(0,0)\},\; \{m_{1\bar1} |(0,0)\},\; \{m_{12} |(0,0)\},\; \{m_{21} |(0,0)\}.
\end{align}

The relevant members of the little groups of the high-symmetry points and lines and their irreps are given in Tab.~\ref{tab:irreps-p6mm}. The full set can be obtained from the \href{https://www.cryst.ehu.es/cgi-bin/cryst/programs/representations_vec.pl?tipogroup=spg&super=183&complex=Submit}{Bilbao Server}.
\begin{table}[htb]
    \begin{subtable}[t]{0.35\textwidth}\centering
        \subcaption{$\Gamma$}
\begin{tabular}[b]{cccc}
    \hline
    \hline\\[-10pt]
    \multicolumn{1}{c}{} &$\{6^+|(0,0)\}$ & $\{m_{01}|(0,0)\}$& $\{m_{1\bar1}|(0,0)\}$ \\
    \hline\\[-10pt]
    $\Gamma_1$  &$\phantom{-}1$ & $\phantom{-}1$  & $\phantom{-}1$ \\
    $\Gamma_2$  & $\phantom{-}1$ & $-1$  & $-1$ \\
    $\Gamma_3$  & $-1$ & $-1$  & $\phantom{-}1$ \\
    $\Gamma_4$  & $-1$ & $\phantom{-}1$  & $-1$ \\
    $\Gamma_5$  & $\phantom{-}\begin{pmatrix}\bar z^2&0\\0&z^2\end{pmatrix}$ & $\phantom{-}\begin{pmatrix}0&\bar z^2\\z^2&0\end{pmatrix}$&$\phantom{-}\sigma_x$\\[10pt]
    $\Gamma_6$  & $\phantom{-}\begin{pmatrix} z&0\\0&\bar z\end{pmatrix}$ & $\phantom{-}\begin{pmatrix}0&z^2\\\bar z^2&0\end{pmatrix}$&$-\sigma_x$\\[8pt]
    \hline
    \hline
    \vspace{6pt}
\end{tabular}
\end{subtable}
\begin{subtable}[t]{0.32\textwidth}\centering
    \subcaption{$K$}
\begin{tabular}[b]{cccc}
\hline
\hline\\[-10pt]
\multicolumn{1}{c}{} &$\{3^+|(0,0)\}$ &$\{m_{1\bar 1}|(0,0)\}$ &$\{m_{21}|(0,0)\}$\\
\hline\\[-10pt]
$K_1$  & $\phantom{-}1$  & $\phantom{-}1$&$\phantom{-}1$\\
$K_2$  & $\phantom{-}1$  & $-1$&$-1$\\
$K_3$  &$\phantom{-}\begin{pmatrix}\bar z^2&0\\0&z^2\end{pmatrix}$&$\phantom{-}\sigma_x$&$\begin{pmatrix}
    0&z^2\\\bar z^2 & 0
\end{pmatrix}$\\[8pt]
\hline
\hline
\end{tabular}
\end{subtable}
\begin{subtable}[t]{0.32\textwidth}\centering
    \subcaption{$M$}
\begin{tabular}[b]{cccc}
\hline
\hline\\[-10pt]
\multicolumn{1}{c}{} &$\{2|(0,0)\}$ &$\{m_{01}|(0,0)\}$ &$\{m_{21}|(0,0)\}$ \\
\hline\\[-10pt]
$M_1$  & $\phantom{-}1$  & $\phantom{-}1$&$\phantom{-}1$\\
$M_2$  & $\phantom{-}1$  & $-1$&$-1$\\
$M_3$  & $-1$  & $-1$&$\phantom{-}1$\\
$M_4$  & $-1$  & $\phantom{-}1$&$-1$\\
\hline
\hline
\end{tabular}
\end{subtable}
\begin{subtable}[t]{0.19\textwidth}\centering
    \subcaption{$\overline{\Gamma M}$}
\begin{tabular}[b]{cc}
\hline
\hline\\[-10pt]
\multicolumn{1}{c}{} & $\{m_{01}|(0,0)\}$ \\
\hline\\[-10pt]
$SM_1$  &$\phantom{-}1$ \\
$SM_2$  &$-1$ \\
\hline
\hline
\end{tabular}
\end{subtable}
\begin{subtable}[t]{0.19\textwidth}\centering
    \subcaption{$\overline{\Gamma K}$}
\begin{tabular}[b]{cc}
\hline
\hline\\[-10pt]
\multicolumn{1}{c}{} & $\{m_{1\bar1}|(0,0)\}$ \\
\hline\\[-10pt]
$LD_1$  &$\phantom{-}1$  \\
$LD_2$  &$-1$  \\
\hline
\hline
\end{tabular}
\end{subtable}
\begin{subtable}[t]{0.19\textwidth}\centering
    \subcaption{$\overline{K M}$}
\begin{tabular}[b]{cc}
\hline
\hline\\[-10pt]
\multicolumn{1}{c}{} & $\{m_{21}|(0,0)\}$ \\
\hline\\[-10pt]
$Y_1$  &$\phantom{-}1$  \\
$Y_2$  &$-1$ \\
\hline
\hline
\end{tabular}
\end{subtable}
\caption[$p6mm$ irreps.]{The relevant irreducible representations of $p6mm$ at the high-symmetry points $\Gamma=(0,0)$, $M=(1/2,0)$, and $K=(1/3,1/3)$, as well as along the lines $\overline{\Gamma M}$, $\overline{\Gamma K}$, and $\overline{KM}$. Note that $z=\exp(\mathrm{i}\pi/3)$.}
\label{tab:irreps-p6mm}
\end{table}

\noindent
The fragile roots in $p6mm$ \cite{Song2020} are given in Tab.~\ref{tab:roots-p6mm}; the elementary band representations can be retrieved from the \href{https://www.cryst.ehu.es/cgi-bin/cryst/programs/bandrep.pl?super=183&elementaryTR=Elementary%20TR}{Bilbao server}.
\begin{table}[htb]
    \begin{tabular}{c|l|c|l}
        \hline
        \#& root &\# of bands & type\\
        \hline\hline
        1 & $ \Gamma_{1} + \Gamma_{2} + K_{1} + K_{2} + M_{3} + M_{4} $ & 2 & mirrors\\
        \hline
        2 & $ \Gamma_{3} + \Gamma_{4} + K_{1} + K_{2} + M_{1} + M_{2} $ & 2 & mirrors\\
        \hline
        3 & $ \Gamma_{3} + \Gamma_{4} + K_{3} + M_{1} + M_{2} $ & 2 & 2D irrep \& mirrors\\
        \hline
        4 & $ \Gamma_{3} + \Gamma_{4} + K_{3} + M_{3} + M_{4} $ & 2 & 2D irrep \& mirrors\\
        \hline
        5 & $ \Gamma_{1} + \Gamma_{2} + K_{3} + M_{3} + M_{4} $ & 2 & 2D irrep \& mirrors\\
        \hline
        6 & $ \Gamma_{1} + \Gamma_{2} + K_{3} + M_{1} + M_{2} $ & 2 & 2D irrep \& mirrors\\
        \hline
        7 & $ \Gamma_{6} + K_{1} + K_{2} + M_{1} + M_{2} $ & 2 & 2D irrep \& mirrors\\
        \hline
        8 & $ \Gamma_{5} + K_{1} + K_{2} + M_{1} + M_{2} $ & 2 & 2D irrep \& mirrors\\
        \hline
        9 & $ \Gamma_{5} + K_{3} + M_{3} + M_{4} $ & 2 & 2D irrep (twice) \& mirrors\\
        \hline
        10 & $ \Gamma_{6} + K_{3} + M_{1} + M_{2} $& 2 & 2D irrep (twice) \& mirrors\\
        \hline
        11 & $ \Gamma_{6} + K_{1} + K_{2} + M_{3} + M_{4} $ & 2 & 2D irrep \& mirrors\\
        \hline
        12 & $ \Gamma_{5} + K_{1} + K_{2} + M_{3} + M_{4} $ & 2 & 2D irrep \& mirrors\\
        \hline
        \end{tabular}
        \caption[$p6mm$ roots.]{Fragile roots and their types in wallpaper group $p6mm$.}
        \label{tab:roots-p6mm}
 \end{table}

\subsubsection{Bundling strategies}

Let us start with the first two roots \#1 and \#2, where with $\Gamma_{1/2}$, $K_{1/2}$, and $M_{1/2/3/4}$ only one dimensional irreps with real characters are involved. It is easy to check, that the mirrors along the high symmetry lines force us to bundle root \#1 as 
\begin{equation}
    (\Gamma_1+K_1+M_3)\,\oplus\,(\Gamma_2+K_2+M_4).
\end{equation}
We now invoke the usual Chern argument by writing \cite{Fang2012}
\begin{equation}
    e^{\mathrm{i}\pi C/3} = \prod_{i\,\in\,\mathrm{occ.}}(-1)^F \eta_i(\Gamma)\vartheta_i(K)\zeta_i(M),
\end{equation}
where $F=2S$, with $S$ the total spin of the particles, $\eta$, $\vartheta$, and $\zeta$ are the eigenvalues of $\{6^+|(0,0)\}$ and $\{3^+|(0,0)\}$, and $\{2|(0,0)\}$ respectively. In the following, we set $F=0$. For the bands above, this means, the Chern number C of these bands is given by 
\begin{equation}
    C=3+6n\quad\mathrm{with}\quad n \,\in\, \mathbb Z.
\end{equation}
The only way this is compatible with time reversal symmetry is by having a band touching between the sets of bands. The same holds for root \#2.

For roots \#3--\#6, the two-dimensional irrep at $K$ glues the bands together into a bundle of two. Moreover, the mirrors along the lines to $M$ and $\Gamma$ fix the connectivity. For the roots \#7, \#8, \#11, \#12, the 2D irreps at $\Gamma$ do the job, and again, the mirrors to $M$ and $K$ determine the connectivity to their irreps. For the roots \#9 and \#10, the bands are degenerate both at $\Gamma$ and $K$. 

Finally, the types of the roots of $p6mm$ are indicated in the last column of Tab.~\ref{tab:roots-p6mm}.

\subsubsection{Examples}

\begin{figure}[h!bt]
    \centering
    \includegraphics{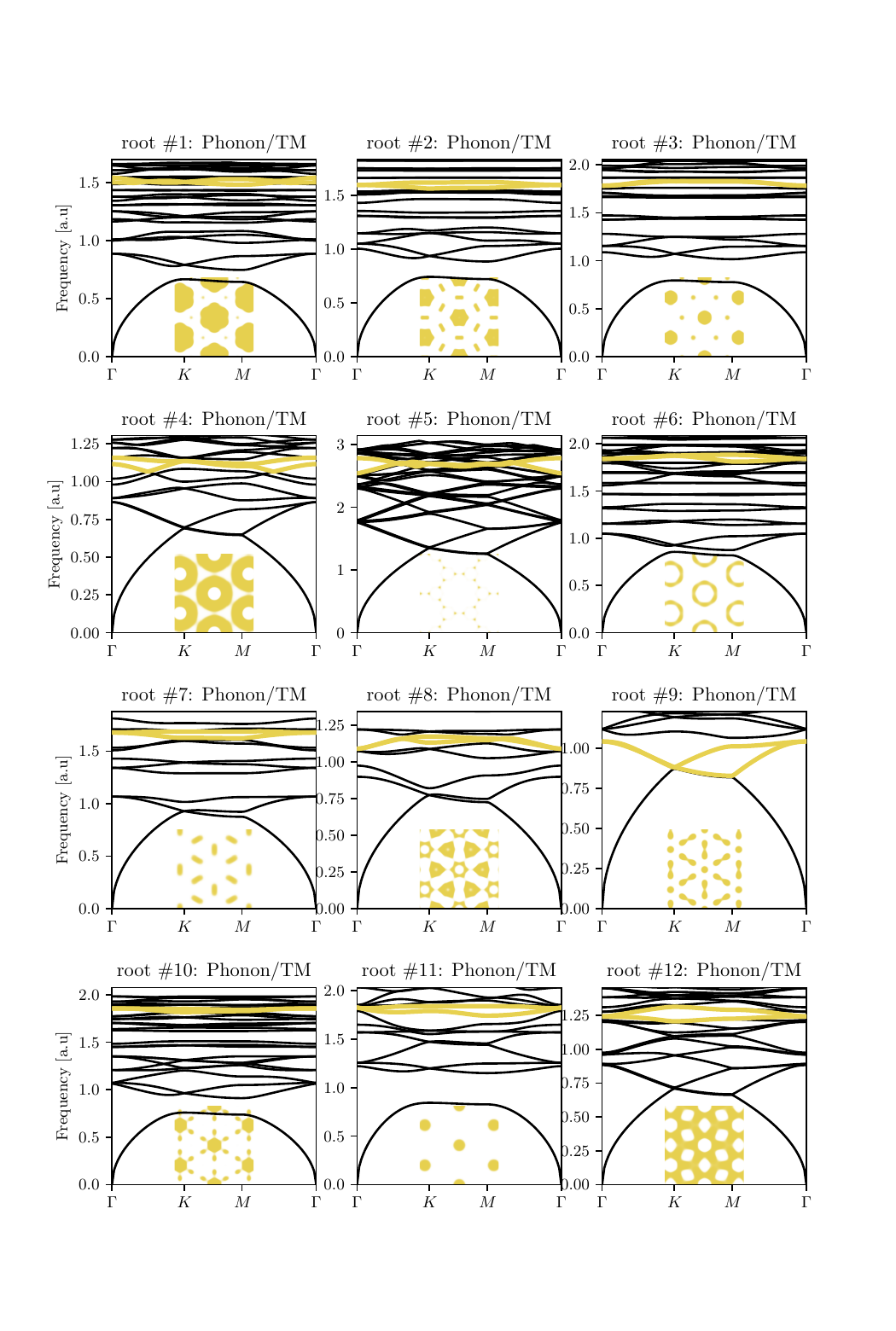}
    \caption[$p6mm$ phonons/TM examples.]{ One sample for each of the roots in $p6mm$ for phonons and TM photons.}
    \label{fig:p6mm-TM-examples}
\end{figure}

\begin{figure}[h!bt]
    \centering
    \includegraphics{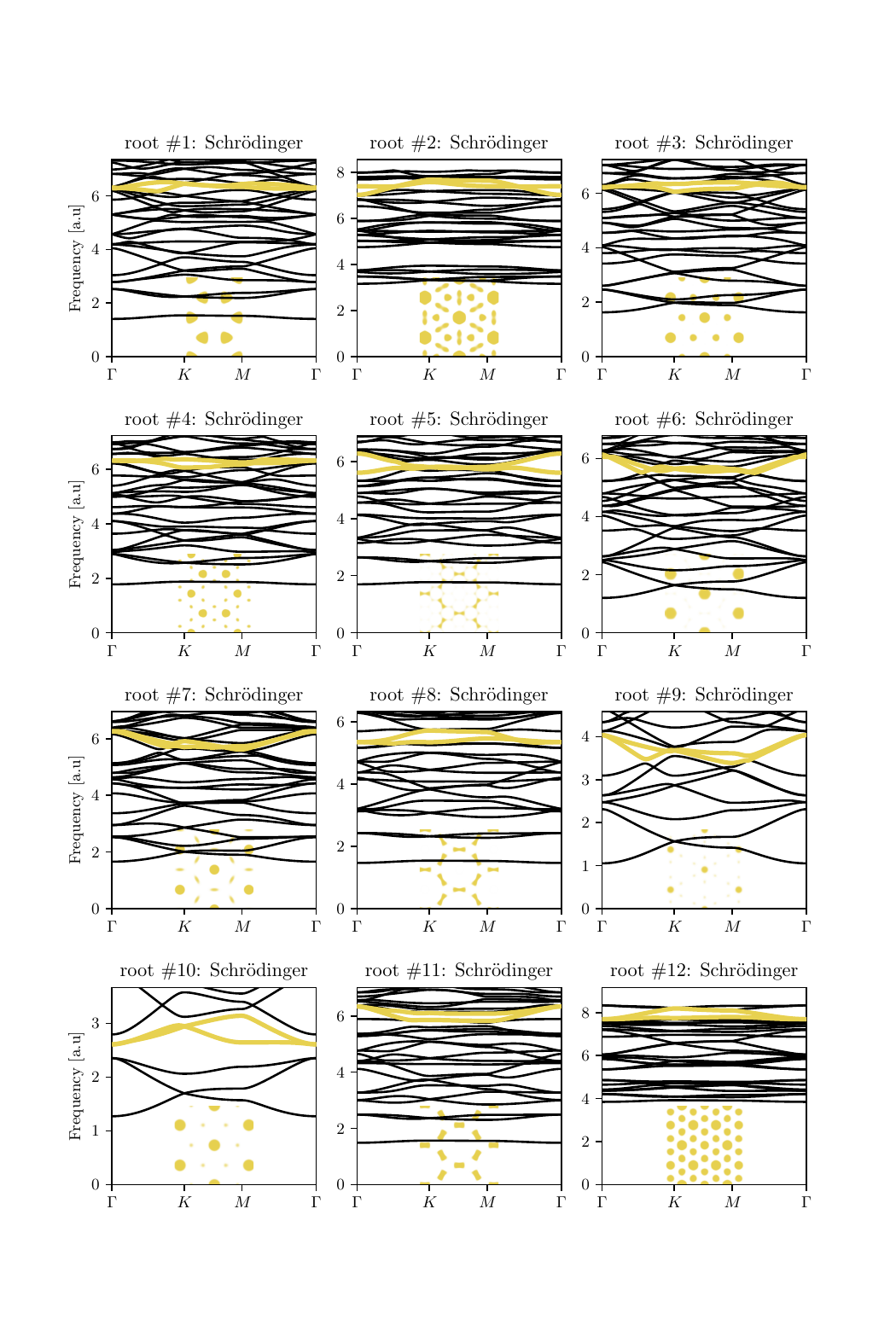}
    \caption[$p6mm$ Schrödinger examples.]{One sample for each of the roots in $p6mm$ for systems described by the Schrödinger equation.}
    \label{fig:p6mm-Q-examples}
\end{figure}

\begin{figure}[h!bt]
    \centering
    \includegraphics{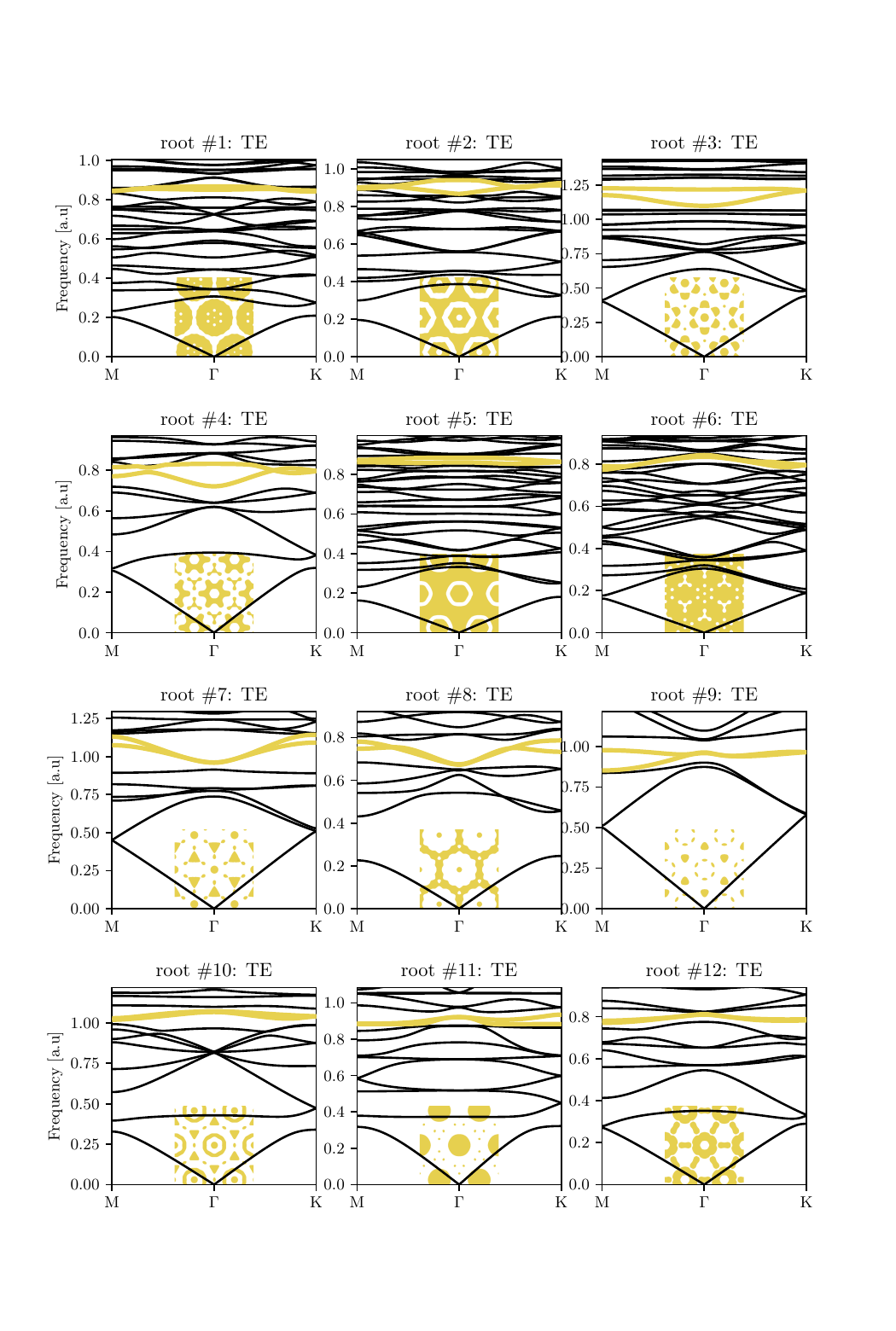}
    \caption[$p6mm$ TE examples.]{ One sample for each of the roots in $p6mm$ for TE photons.}
    \label{fig:p6mm-TE-examples}
\end{figure}

\bibliographystyle{phd-url}
\bibliography{refsi}